\documentclass[%
 aip,
 amsmath,amssymb,
 reprint,%
]{revtex4-1}

\usepackage{graphicx}% Include figure files
\usepackage{dcolumn}% Align table columns on decimal point
\usepackage{bm}% bold math
\usepackage[utf8]{inputenc}
\usepackage[T1]{fontenc}
\usepackage{mathptmx}
\usepackage{etoolbox}

\usepackage{url}
\usepackage{braket}
\usepackage{hyperref}
\hypersetup{
  colorlinks   = true,    % Colours links instead of ugly boxes
  urlcolor     = blue,    % Colour for external hyperlinks
  linkcolor    = blue,    % Colour of internal links
  citecolor    = blue,    % Colour of citations
}
\usepackage{placeins} % added consisting 
\usepackage{amsmath} % split, 	
\usepackage[caption=false]{subfig}
\makeatletter
\def\@email#1#2{%
 \endgroup
 \patchcmd{\titleblock@produce}
  {\frontmatter@RRAPformat}
  {\frontmatter@RRAPformat{\produce@RRAP{*#1\href{mailto:#2}{#2}}}\frontmatter@RRAPformat}
  {}{}
}%
\makeatother
\begin{document}
\onecolumngrid
The following article has been submitted to Journal of Applied Physics. After it is published, it will be found at the following \href{https://pubs.aip.org/aip/jap?gad_source=1&gclid=CjwKCAiA3JCvBhA8EiwA4kujZpdqqarxXPn5ye9PijArvSw9SQwvcT5GwnaJQcb60j9njhX2v1tmGRoCdT8QAvD_BwE}{Link}. 
\twocolumngrid

\preprint{AIP/123-QED}

\title{Nature of point defects in monolayer MoS$_2$ and the MoS$_2$/(111)-Au heterojunction}
% Force line breaks with \\
\author{Roozbeh Anvari}
\author{Wennie Wang} 
%\email{}
\affiliation{ \rm McKetta Department of Chemical Engineering, University of Texas at Austin}

\begin{abstract}
Deposition of $\rm MoS_2$ on (111)-Au alters the electronic properties of $\rm MoS_2$. 
In this study, we investigate the free-standing MoS$_2$ monolayer and the MoS$_2$/(111)-Au heterostructure, with and without strain, as well as defects of interest in memristive and neuromorphic applications.
We focus on so-called atomristor devices based on monolayer materials that achieve resistive switching characteristics with the adsorption and desorption of metal adatoms.
%Our study shows that the electronic structure and transport characteristics of native defects of $\rm MoS_2$ are altered significantly with the amount of biaxial tensile strain that is applied to the heterojuction. 
Our study confirms that the formation of midgap states is the primary mechanism behind the resistive switching.
Our results show that strain lowers the adsorption/desorption energies of Au+defect structures of interest, leading to more favorable switching energies, but simultaneously reduces the switching ratio between states of differing conductivities.  
The presence of the (111)-Au substrate additionally introduces non-uniform amounts of strain throughout and charge transfer to the MoS$_2$ monolayer. 
We propose that the induced strain can contribute to the experimentally observed $n-$ to $p-$type transition and Ohmic to Schottky transition in the MoS$_2$ monolayer.
The charge transfer leads to a permanent polarization at the interface, which can be tuned by strain. 
Our study has important implications on the role of the electrode as being a source of the observed variability in memristive devices and as serving as an additional degree of freedom for tuning the switching characteristics of the memristor device.
\end{abstract}

\maketitle
%===========================================================

\section{Introduction}

Memristors are an emerging technology in low-power and highly integrated devices \cite{chang2011short}, data storage and encryption \cite{lanza2022memristive,bertolazziNonvolatileMemoriesBased2019,chiangMemoryApplications2D2021}, and neuromorphic computing circuits \cite{li2018review}. 
In memristors, the application of an external electric field modulates the electrical resistance between a high resistance state (HRS) and low resistance state (LRS) while retaining a memory of the prior resistive state \cite{chang2011short, chang2011short}. 
%Two-dimensional materials have attracted significant interest in developing next-generation electronic devices \cite{hus2021observation, ge2021library, boschetto2023non_2, boschetto2022ab_1, radisavljevic2011single, radisavljevic2013mobility, kim2012high}.
Non-volatile resistive switching (NVRS) has been observed across several 2D materials such as graphene oxide \cite{tan2015non}, black phosphorous \cite{hao2016liquid}, h-BN \cite{wuThinnestNonvolatileMemory2019}, and various transition metal dichalcogenides (TMDs) \cite{bessonov2015layered, son2016colloidal,ge2021library,ge2018atomristor}.
Atomically-thin monolayer materials exhibiting NVRS, dubbed atomristors \cite{ge2018atomristor}, have been demonstrated to have switching voltages on the order of 1 V or less, making them suitable for low-power, energy-efficient devices \cite{ge2021library}.
The use of (nearly) atomically-thin 2D monolayer materials additionally represents the ultimate material scaling limit for achieving high packing densities. 
Indeed, NVRS in monolayer materials has been observed to occur through the formation and dissolution of point defects, which avoids the stochastic formation of conductive filaments \cite{kwon2010atomic, yao2010resistive} and may help realize energy-efficient devices with resistive switching \cite{ge2018atomristor,ge2021library,hus2021observation}.
While point defects are broadly known to impact the electrical \cite{qiu2013hopping, jena2007enhancement}, optical \cite{tongay2013defects, korn2011low}, and magnetic properties\cite{han2013controlling, zhang2023robust} of 2D materials, the mechanism(s) of how resistive switching arises from the presence of point defects is less well understood.
We consider NVRS in vertically-stacked two-terminal devices consisting of a metal-semiconductor-metal.
In particular, we focus on the interface between MoS$_2$ and gold, which is a commonly used contact metal due to its high electronic conductivity and chemical inertness. \\ 

%MoS$_2$ is a widely-studied 2D material, known for its electric \cite{radisavljevic2011single, radisavljevic2013mobility, kim2012high}, optical \cite{xiao2012coupled, mak2010atomically, mak2013tightly, zeng2012valley}, and mechanical \cite{bertolazzi2011stretching, castellanos2012elastic, zhu2015observation, liu2014strain} properties, and thus serves as a prototypical material for understanding resistive switching.
%The MoS$_2$/Au interface has also attracted significant attention due to potential applications in the exfoliation of monolayers for large-area (~cm) van der Waals heterostructures \cite{velicky2020strain}, flexible chemical sensors \cite{guo2019artificial, zribi2020exfoliated}, lithography patterning of transistors \cite{panasci2021strain, gramling2019spatially}, and field effect transistors \cite{gong2013metal, popov2012designing, chen2013tuning}. \\

Several theoretical and experimental studies have investigated the mechanism of resistive switching mediated by point defects in MoS$_2$ on gold electrodes, but thus far a comprehensive picture is lacking. 
One proposed model from Hus et al.\cite{hus2021observation} attributes the resistive switching behavior to the presence of metal adatoms adsorbed in existing sulfur vacancies and sulfur di-vacancies \cite{hus2021observation,theory_tumino2020nature}. 
By combining scanning tunneling spectroscopy and microscopy with DFT calculations, Hus et al. found such metal adatoms result in an in-gap state near the Fermi level that induces a greater degree of metallicity in the monolayer. 
This model was further refined in the work of Ge et al.\cite{ge2018atomristor,ge2021library}, who proposed a mechanism in which the metal adatom dissociates from the electrode, may diffuse across the monolayer surface, and adsorbs onto an existing vacancy. Ge et al. suggested that a similar mechanism may broadly apply to 2D materials based on the observation of similar resistive switching characteristics in current-voltage measurements across several TMDs. 
These results were corroborated with first-principles calculations based on the Keldysh nonequilibrium Green's function theory from Li and colleagues \cite{li2023resistive} who considered point defects in free-standing MoS$_2$, as well as MoTe$_2$ and WTe$_2$. 
These studies focused mainly on the free-standing monolayer, neglecting the metal electrode(s).
The theoretical work from Boschetto et al.\cite{boschetto2022ab_1,boschetto2023non_2} computed tunneling probabilities of various concentrations of sulfur vacancies and adatoms in a heterostructure of MoS$_2$ and gold. Boschetto et al. proposed that the formation of metal adatom clusters forming a conducting bridge could induce a semiconducting to metallic transition in MoS$_2$.
Nevertheless, several open questions remain: What, if any, is the role of the metal electrode in the resistive switching mechanism? More specifically, what is the role of the metal electrode in the defect energetics and the transport properties?\\ 

The interface between MoS$_2$ and gold metal contacts presents a wealth of physics that has been under-explored in the context of resistive switching.
In building an atomic understanding of the interface, one of the first considerations is the impact of strain.
%In the (pristine) monolayer, it is known that (tensile) strain causes direct to indirect band gap transition, reduces the band gap and induces semiconductor to metal transition, and elevates electrical and thermal conductivity \cite{bhattacharyya2014effect}. 
%As the most stable and chemically inert surface of gold, the (111)-Au surface is known to undergo complex surface reconstructions, often called the "herringbone" pattern.\cite{alerhand1988spontaneous,li2022origin}
When an interface between monolayer MoS$_2$ and (111)-Au is formed, the Au substrate may lead to heterogeneous local strain throughout the MoS$_2$ monolayer. 
The extent to which the heterogeneity exists varies among studies and may vary depending on the method of preparation.
For example, a study from Velick\'{y} and colleagues reported biaxial strains of up to 1.9\% based on the downward shift and broadening of the in-plane $E'$ vibrational mode. 
In contrast, monolayer MoS$_2$ synthesized via chemical vapor deposition on polycrystalline Au has been reported to exhibit only out-of-plane strain and no in-plane strain~\cite{yasuda2017out}.
%In addition to differing reports on the extent of strain in monolayer MoS$_2$ on metal substrates, it is also possible for the monolayer MoS$_2$ to alternate between chemisorption and physisorption depending on the relative offset with the substrate, as found in Moire superstructures of MoS$_2$ on(111)-Au.\cite{silva2022spatial} % is this relevant? no, we do not address this at all
\\ 

Prior theoretical and experimental studies also suggest that variations at the MoS$_2$/Au(111) interface are intimately connected with the wide range of electronic properties reported \cite{mcdonnell2014defect,cook2015influence,carladous2002light,bruix2016single,gong2013metal}.
For example, McDonnell and co-workers \cite{mcdonnell2014defect} used a combination of XPS and STM/STS to correlate the observation of lower Schottky barriers to the presence of sulfur and molybdenum vacancies (i.e., varying stoichiometry in the monolayer).
The surface roughness of the metal electrode, strain between the monolayer and substrate, and hence the interlayer spacing have also been invoked \cite{cook2015influence,gong2013metal,bruix2015situ} to explain the modulation of electronic properties between different metal electrodes or compared to free-standing MoS$_2$. 
However, only a few studies \cite{boschetto2022ab_1,boschetto2023non_2,velickyMechanismGoldAssistedExfoliation2018} have considered the explicit presence of the metal electrode in the context of understanding resistive switching mechanisms.
In these studies, the effects of strain, interlayer spacing, and the presence of defects are only partially considered or omitted. 
Hence, in this study, we seek to further elucidate the impact of the metal electrode and accompanying strain on the transport properties of monolayer MoS$_2$ and connect these insights to resistive switching as mediated by point defects.\\

The paper is organized as follows.
We first discuss our computational methodologies for computing defect energetics, transport properties, and for constructing interfacial models between MoS$_2$ and (111)-Au in Section \ref{section_comp}.
In Sections~\ref{sec_defects_free} and~\ref{sec_strain_free}, we discuss the formation of native point defects with gold adatoms in free-standing MoS$_2$ without strain and with strain. 
Our results confirm that the adsorption of a gold adatom induces large changes in conductivity to explain resistive switching in monolayer MoS$_2$.
Next, we investigate the effects of biaxial tensile strain in the MoS$_2$/(111)-Au interface on the defect formation properties (Section~\ref{sec_defects_in_hetero}), interfacial charge transfer and dipole formation (Section \ref{sec_heterostructure_strain}), Ohmic to Schottky transition and its effect on transport properties (Section~\ref{sec_schottky_ohmic}). 
Finally, we conclude with a discussion of the implications of our calculations on resistive switching mechanisms. \\

% =======================================
\section{Computational Details}\label{section_comp}
% =======================================

First-principles electronic structure calculations based on density functional theory were conducted using the OpenMX software package (version 3.9.2) \cite{opmx_ozaki2003variationally, opmx_ozaki2005efficient, opmx_lejaeghere2016reproducibility}. 
Calculations were performed using norm-conserving pseudopotentials \cite{opmx_morrison1993nonlocal}, the generalized gradient approximation for exchange correlation interactions \cite{opmx_perdew1996generalized}, spin polarization, and pseudo-atomic localized basis functions. 
The standard basis set for the pseudo-atomic orbitals used for Mo was s3p2d2, for S was s2p2d1f1, and for Au was s3p2d2f1, all with cutoff radius of 7.0 a.u. 
A vacuum distance of 50 \AA\, in the direction normal to the surface was used in the calculations. 
The atomic coordinates were relaxed until the interatomic forces were below $10^{-5}$ Hartree/Bohr. 
Brillouin zone sampling was performed using a $12 \times12\times1$ mesh, energy cutoff of 300 Ry and energy-convergence threshold of $10^{-10}$ Hartree. 
The DFT-D3 method is used in calculations involving the heterostruture to account for van der Waals interactions \cite{grimme2011effect}. 
Chemical potentials with respect to the vacuum level were evaluated via the effective screening medium (ESM) method \cite{opmx_otani2006first}, as implemented in OpenMX. 
In the projected density of states (PDOS) reported here, the Gaussian broadening width was set at 0.05 eV. 
Band unfolding procedures were performed with the LCAO method implemented in OpenMX \cite{lee2013unfolding}.\\

The formation energy of defects $E^f$ provides information on the stability of a particular defect and is calculated using 
\begin{equation}
\label{eq_defect_form}
    E^f = E_{def} - E_{host} - \sum_i n_i \mu_i  + q(E_v + E_F) + E_{corr}.
\end{equation}
$E_{def}$ and $E_{host}$ correspond to the DFT total energies of the system with and without the defect, $\mu_i$ is the chemical potential of species $i$, and $n_i$ represents the number of species $i$ added ($n_i>0$) or removed ($n_i<0$). 
$E_v$ and $E_F$ are the valence band maximum (VBM) and Fermi level position, respectively. 
$E_{corr}$ indicates the energy correction term corresponding to the spurious interactions between images of the defects within the periodic supercell approach and is included using the effective screening method \cite{otani2006first} as implemented in the openMX code. 
We consider only neutral defects in this paper ($q=0$), namely the gold adatom adsorbed to sulfur vacancies, sulfur divacancies, and molybdenum vacancies as shown in Figure~\ref{simple_defect_struct}. 
The chemical potential of S is referenced to the solid, monoclinic phase; as such, $\rm \mu_{Mo} = \mu_{MoS_2} - 2\mu_{S}$ and $\rm \mu_{Au} = (\mu_{Au_2S} - \mu_{S})/2$. 
The calculated heat of formation of monolayer MoS$_2$ is $-0.988$ eV/atom, which compares favorably to that on the Materials Project~\cite{jainCommentaryMaterialsProject2013} ($-0.964$ eV/atom). 

We consider the energetics of an Au adatom interacting with monolayer MoS$_2$.
The adsorption energy is computed using  $E_{ads} =  E_{\text{host+adatom}} - E_{\text{host}} - E_{\text{adatom}}$, where $E$ represents the computed DFT total energy and $E_{\text{adatom}}$ is referenced to bulk gold.
We also compute the binding energy of Au adatoms to existing native point defects (e.g., sulfur vacancies), which is computed as~\cite{freysoldtFirstprinciplesCalculationsPoint2014} $E^{b} =  E^f_{\text{vacancy+adatom}} - E^f_{\text{vacancy}} - E^f_{\text{adatom}}$, where $E^f$ represents a formation energy as defined in Eq.~\ref{eq_defect_form}.  \\ 

%Comparison of the PBE-D and HSE calculated formation energies at 0 K performed by Komsa et al. indicates that the transition levels of the neutral $V_S$ and $V_{Mo}$ with respect to the band edges of freestanding monolayer MoS2 are only shifted marginally \cite{komsa2015native}. Moreover, previous study shows that the strong Coulombic screening by metal Au(111) slab with infinite dielectric constant minimizes the many body effect in MoS2, justifies the application of GGA functional in MoS2/(111)-Au heterojunction calculations \cite{gong2014unusual,  }. \\

For transport calculations of free-standing MoS$_2$, the linearized semi-classical Boltzmann transport equation is solved using the Boltztrap code \cite{madsen2006boltztrap} to compute the conductivity tensor  
\begin{equation}
\label{eq_BTE}
            \sigma_{\alpha \beta }(T, \mu) = \frac{e^2}{V} \int  \tau(i,\textbf{k})  \, v_\alpha(i,\textbf{k})  v_\beta(i,\textbf{k}) \left [   -  \frac{\partial f_\mu (T, \epsilon)}{\partial \epsilon}  \right ] d \epsilon,           
\end{equation}
where $e$, $T$, $V$, $\tau$  and $f_\mu$ are the electronic charge, temperature, the volume of the unit cell, the relaxation time, and the Fermi-Dirac distribution respectively. 
The band velocity of $i$th energy band  $\epsilon(i,\textbf{k})$ is given by $v_\alpha(i,\textbf{k}) = \frac{1}{\hbar} \frac{\partial \epsilon(i,\textbf{k})}{\partial \textbf{k}_\alpha}$ where $\textbf{k}_\alpha$ is the $\alpha$ component of the wavevector $\textbf{k}$. 
The Brillouin zone was sampled with a $k$-point mesh of $32 \times 32 \times 1$. 
The electrical conductivity is obtained within the constant relaxation time approximation (CRTA). 
The CRTA is based on the assumption that the relaxation time does not change significantly with energy within the relevant range of the Fermi level and is thus taken to be approximately a constant ($\tau(i,\textbf{k}) \approx \tau$). 
In this study, we compute a normalized conductivity ($\bar{\sigma} = \sigma/\tau$).
This approximation scheme is computationally efficient and has been widely applied to study TMDs and other 2D materials \cite{jin2015revisit, guo2013high}. 
However, its predictive power of the absolute conductivity depends on the empirical choice of scattering rate \cite{ponce2021first, ganose2021efficient, bhattacharyya2014effect}.
%We obtain a normalized conductivity of $\sigma/\tau$ = 1.61 $\times 10^8$ under the CRTA using the Fermi level for free-standing MoS$_2$ with one sulfur vacancy.
%The experimental carrier lifetime corresponding to single layer of MoS$_2$ is $\tau =$  40 ps which yields a conductivity value of 6.45 $\times 10^{-3}$ $ \rm S/cm$. 
%The electrical conductivity of MoS$_2$ is approximately 6.39 $\times 10^{-2}$  $ \rm S/cm$  at room temperature \cite{el1977temperature}. 
%Calculated normalized electrical conductivities for free standing MoS2 structures are depicted in Figure \ref{} and are tabulated in Table \ref{STAL_strain_conductivity}. 
We emphasize that this study on resistive switching is focused on the ratios of the conductivity between the low- and high-resistance states. 
As such, we expect ratios involving $\bar{\sigma}$ to be weakly dependent on the details of the relaxation time $\tau$.
Thus, trends in the transport properties of the defects investigated here are expected to be retained.\\
%We note that the application of the CRTA approximation here is similar to that in the calculation of the Seebeck factor, which displays a weaker dependence on electron lifetimes than mobility and conductivity due to error cancellation \cite{ganose2021efficient, ricci2017ab}. \\ 

We next describe our rationale for constructing various heterostructures of monolayer MoS$_2$ on (111)-Au.
We focus on the MoS$_2$/(111)-Au heterostructure and find this structural model captures many features of the Au/MoS$_2$/Au vertical stack, as detailed in the Supplementary Information (Figure~\ref{hetero_coordinates}).
Our calculations indicate that the formation energy of sulfur vacancies are converged to within 1.3 meV for a (6x6) supercell with a vacuum of 55 \AA, which corresponds to a defect density of  $\rm 3.14 \times 10^{13} \,\, cm^{-2}$. 
Details of convergence are given in the SI. 
$\rm MoS_2$ is known to go through a ($\rm \sqrt{3} \times \sqrt{3}$) reconstruction when deposited on (111)-Au \cite{min2016reduction, kang2014computational, gong2014unusual}. 
Our calculations yield a lattice mismatch of $\rm 5.79 \%$ between relaxed lattice constant of the top reconstructed layer $\rm (2\sqrt{3} \times 2\sqrt{3})-MoS_2$ (11.06~\AA) and that of the (111)-Au substrate (11.75~\AA). 
Previous experimental studies indicate that different levels of biaxial strain are present in the MoS$_2$ top layer after deposition on (111)-Au \cite{cook2015influence,gong2013metal,bruix2015situ}. 
Thus, we study the effect of strain in two extreme scenarios in which MoS$_2$ goes through a biaxial tensile strain to adapt to the lattice vectors of Au and vice versa (i.e., 0\% to 6.23\% strain of MoS$_2$); four additional intermediate structures are chosen in which both MoS$_2$ and (111)-Au are strained, as depicted in Figure~\ref{fig_subs_all_a}. 
%The defect density corresponding to the heterostructure with MoS2 at zero and 5.87$\%$ tensile strain is $\rm 9.43 \times 10^{13} \,\, cm^{-2}$ and $\rm 8.35 \times 10^{13} \,\, cm^{-2}$, respectively. [**To discuss with Wennie: changes of defect density with strain**]\\ 

% STM studies of Silva et al. reveal that an (11x11)MoS2 on (12x12)Au reconstruction is the closest commensurate model of hexagonal reconstruction of MoS2-on-(111)Au moire superstructure.  \cite{silva2022spatial} .

% ============================================
\section{Results}\label{sec_results}
% ============================================

\subsection{Point defects in free-standing monolayer $\rm MoS_2$ } \label{sec_defects_free}

It has been suggested that the adsorption of gold on monolayer MoS$_2$ is responsible for the resistive switching~\cite{hus2021observation,li2023resistive}.
In order to understand the influence of point defects, we consider the sulfur vacancy ($\rm V_S$), sulfur divacancy ($\rm V_{2S}$), molybdenum vacancy ($\rm V_{Mo}$); this is in addition to the defect relevant to resistive switching, i.e., a gold adatom adsorbed on the pristine monolayer $\rm Au+pr$ and on each of the vacancies considered ($\rm Au_S, \, Au_{2S}, \,Au_{Mo}$).
To begin, we report on the defect formation energies of the vacancies without an Au adatom (see also Figure~\ref{simple_defect_energ_vac} and Table~\ref{table_stal_energies} in the Supplementary Information). 
The molybdenum vacancy $\rm V_{Mo}$ is predicted to be the least stable vacancy with a formation energy of at least 4 eV (in sulfur-rich conditions).
The sulfur divacancy $\rm V_{2S}$ is the second most stable defect over a wide range of the chemical potentials of sulfur, and only becomes less favorable than the molybdenum vacancy ($\rm V_{Mo}$) in extremely sulfur-rich environments. 
Our calculations confirm that the single sulfur vacancy ($\rm V_S$) is the most energetically favorable vacancy for all ranges of chemical potential of S, in agreement with prior computational studies \cite{komsa2015native}. 
Our observations of the relative stability of vacancies in MoS$_2$ are also in line with experimental observations based on STM/STS \cite{hus2021observation}.
%Experimental observations yield a ratio of 1:10 for $\rm V_{2S}$ to $\rm V_{S}$ \cite{hus2021observation}.
%The calculated formation energy for neutral defects are in agreement with previous theoretical and experimental studies, as tabulated in Table \ref{table_stal_energies} of the SI. \\

% FiG 1 standalone MoS2 
%-----------------------
\begin{figure}[h!]
 \centering
     \subfloat[\label{simple_defect_energ_Au}]{  
    \includegraphics[width=.9\columnwidth]
      {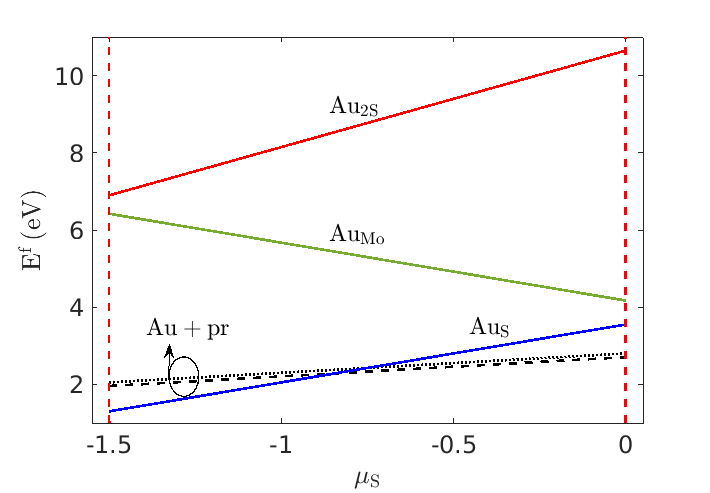}
      } 
      
 \subfloat[\label{native_defect_band_dos_cond_b}]{  
    \includegraphics[width=1.05\columnwidth]
      {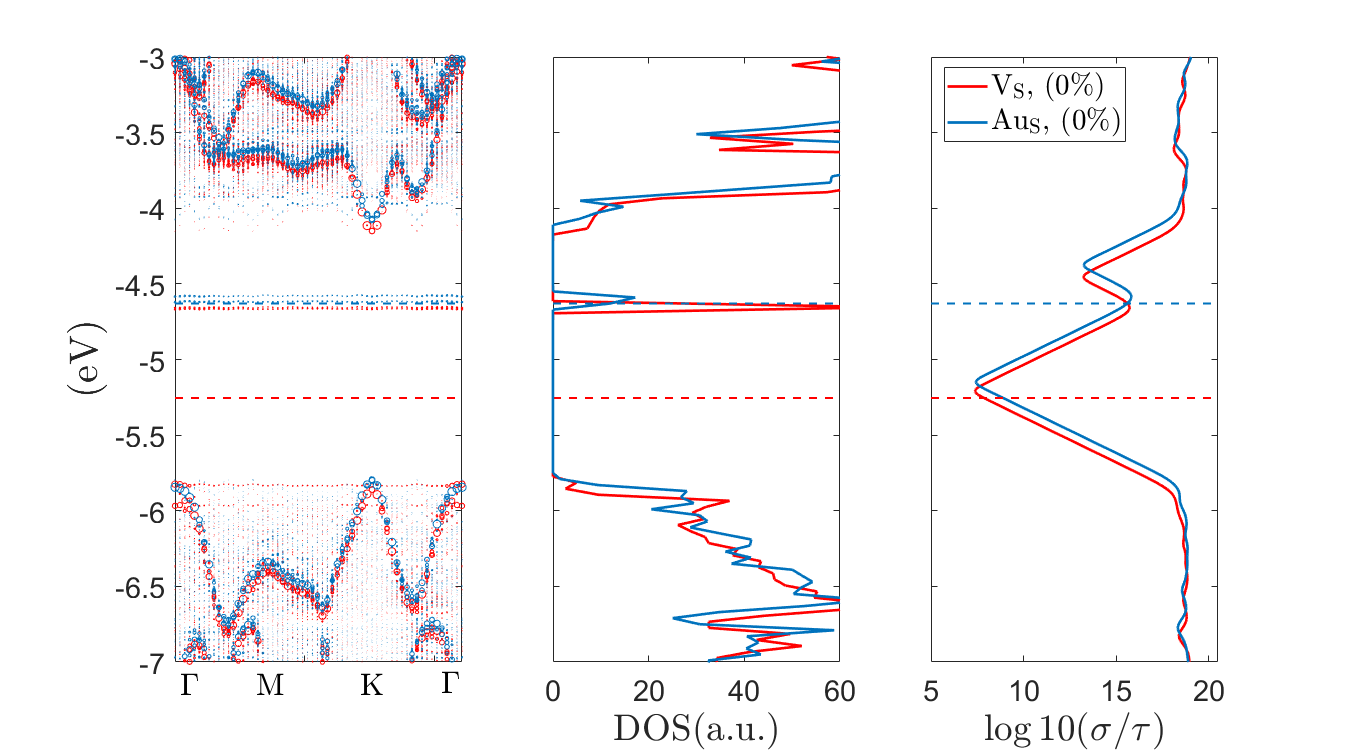}
     }
     \caption{(a) Calculated formation energies of a number of defect complexes involving an Au adatom and native vacancy in standalone MoS$_2$. (b) Unfolded bandstructure (left), DOS (middle) and normalized electrical conductivity (right) of (6x6) supercell of $\rm MoS_2$ with Au adatom adsorbed on various cation and anion vacancies. The electronic structure and corresponding normalized conductivity of the sulfur divacancy and molybdenum vacancy with and without an Au adatom are provided in Figure~\ref{native_defect_band_dos_cond_SI}.   
     }
     \label{defect_free_standing}
\end{figure} 

We next consider the incorporation of an Au adatom on a sulfur vacancy site $\rm Au_S$, a sulfur divacancy site $\rm Au_{2S}$, and a molybdenum vacancy site $\rm Au_{Mo}$ (corresponding structures in Figure~\ref{simple_defect_struct}). 
The defect formation energy phase diagram of the cation and anion vacancies of $\rm MoS_2$ containing an Au adatom is depicted in Figure~\ref{simple_defect_energ_Au}. 
Figure~\ref{simple_defect_energ_Au} shows that the formation energy of an Au-adatom adsorbed onto the pristine structure at the Mo, the S, or the hollow sites are within several meV of each other and comparable to the defect formation energy of the sulfur vacancy under sulfur-rich conditions. 
%more favorable than the adsorption of Au into sulfur ($\rm Au_{S}$, $\rm Au_{2S}$) and molybdenum ($\rm Au_{Mo}$) vacancies. 
We find that the formation energy for the direct substitution of Au for a sulfur site across relevant ranges of the sulfur chemical potential is unfavorable ($\sim$1.6 eV to 3.5 eV) compared to the adsorption energy of Au onto a pre-existing sulfur vacancy (0.95 eV). 
These results indicate that the adsorption of an Au adatom onto a pre-existing anion vacancy is a primary mechanism for resistive switching in atomristor devices, in alignment with prior studies \cite{hus2021observation,li2022conductive,li2023resistive}.\\

In addition to the stability of the defect complexes considered, we consider the impact of the various defect complexes involving an Au adatom and vacancy on the electronic structure and the change in conductivity.
The electronic structure (unfolded band structure and DOS) and normalized electrical conductivity of $\rm MoS_2$ structures with a vacancy and Au+vacancy defect complex are depicted in Figures \ref{defect_free_standing} and \ref{native_defect_band_dos_cond_SI}.
%The energy levels corresponding to the band edges and defect levels at high symmetry points are summarized in Figure \ref{simple_defect_band_edges}. 
Analysis of the band structure indicates that all the vacancy defects and their complexes with Au have a direct band gap at the high-symmetry K-point. 
Both $\rm V_{S}$ and $\rm V_{2S}$ defects induce midgap states, approximately 0.5 eV below the conduction band minimum (CBM), while $\rm V_{Mo}$ induces several mid-gap states that are located farther from the CBM. 
From our analysis of the unfolded band structure, we find minimal changes of the electron velocity near the band edges (K-point). 
As a result, the difference in normalized electrical conductivity is determined by the density of states (see Eq.~\ref{eq_BTE}).\\ 

Based on a density of states argument, one would expect a high (normalized) conductivity when the Fermi level is resonant with a partially filled electronic state, either of the defect or of the host monolayer. 
We focus here on $\rm Au_S$ and present the same data in Figure~\ref{native_defect_band_dos_cond_SI} for the remaining defect configurations.
As observed in Figure~\ref{native_defect_band_dos_cond_a}, the pristine monolayer indeed has a negligible electrical conductivity in the vicinity of the computed Fermi level , which lies within the bandgap. 
When introducing a sulfur vacancy (Figure~\ref{native_defect_band_dos_cond_b}), unoccupied defects emerge and are energetically separated from the computed Fermi level, leading to a moderate ($4-5$ orders of magnitude) increase in the normalized conductivity.
By contrast, the introduction of a molybdenum vacancy (Figure~\ref{native_defect_band_dos_cond_SI}) leads to defect states resonant with the computed Fermi level, resulting in the normalized conductivity increasing by nearly ten orders of magnitude compared to the pristine monolayer case.
As such, the adsorption of Au on the molybdenum vacancy site leads to a minimal change in the computed normalized conductivity.
However, the formation energy of molybdenum vacancies high and thus would not be expected to contribute significantly to the resistive switching of the overall device.
Notably, the Fermi level of configurations involving the gold adatom (e.g., $\rm Au_S$ and $\rm Au+pr$ structures) are located at the energy level resonant with the energy of the midgap defect state. 
Thus, based on the formation energy and transport properties of these defects, one would expect the resistive switching to be mediated predominantly by the presence of sulfur vacancies with some but limited contribution from divacancies, in line with experimental observations\cite{hus2021observation}. 
Indeed, the normalized conductivity increases by several orders of magnitude when considering the adsorption of an Au adatom onto the sulfur vacancy (around ten orders of magnitude) and divacancy sites (around  five orders of magnitude).
Previous experimental work shows that the ON/OFF ratio, or the ratio of resistance at LRS to that of HRS, can reach $10^8$ \cite{ge2018atomristor,gong2014unusual}.
Calculated normalized conductivities corresponding to native defects complexes are tabulated in Table \ref{table_point_def_cond}.\\

Thus far, we have a basic understanding of a resistive switching mechanism in atomristors to be mediated by the formation of point defect complexes involving the adsorption of Au adatoms. 
We investigate next the effect of strain and the presence of the metal electrode/substrate, separately (Section~\ref{sec_strain_free}) and in conjunction (Sections~\ref{sec_defects_in_hetero} and \ref{sec_heterostructure_strain}).

% =================================================================
\subsection{Effect of strain in free-standing monolayer MoS$_2$}\label{sec_strain_free}

\begin{figure}[!htb]
 \centering
     \subfloat[\label{fig_stal_pdos_strain_b}]{%
    \includegraphics[width=1.05\linewidth] 
      {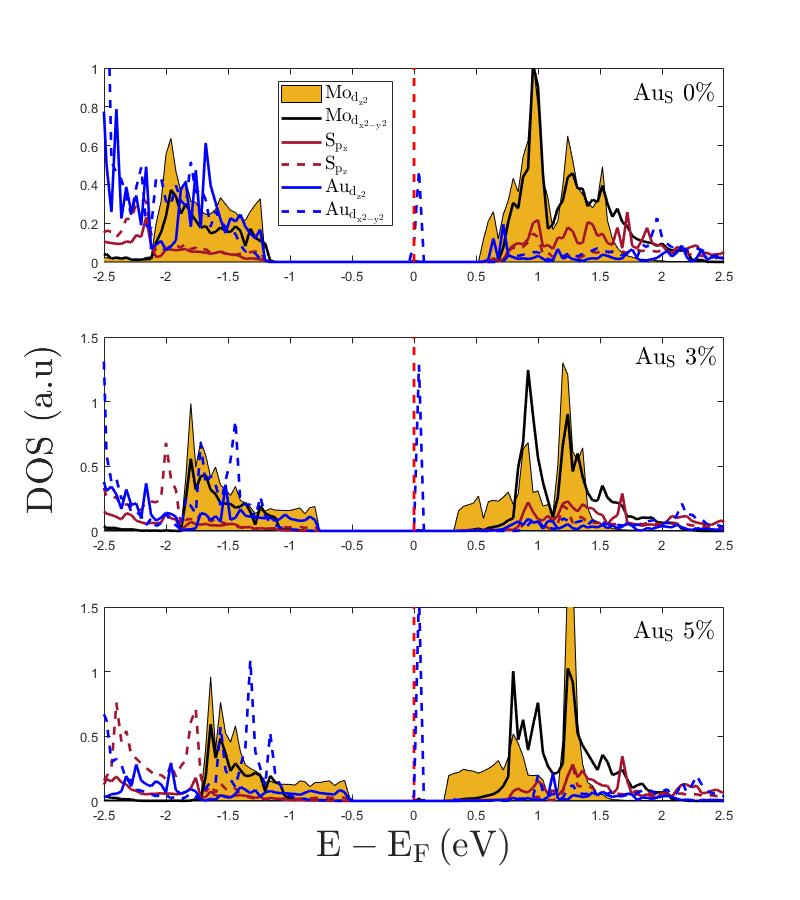}
      } 
      
     \subfloat[\label{stal_strain_bands_d}]{  
    \includegraphics[width=1.05\columnwidth]
      {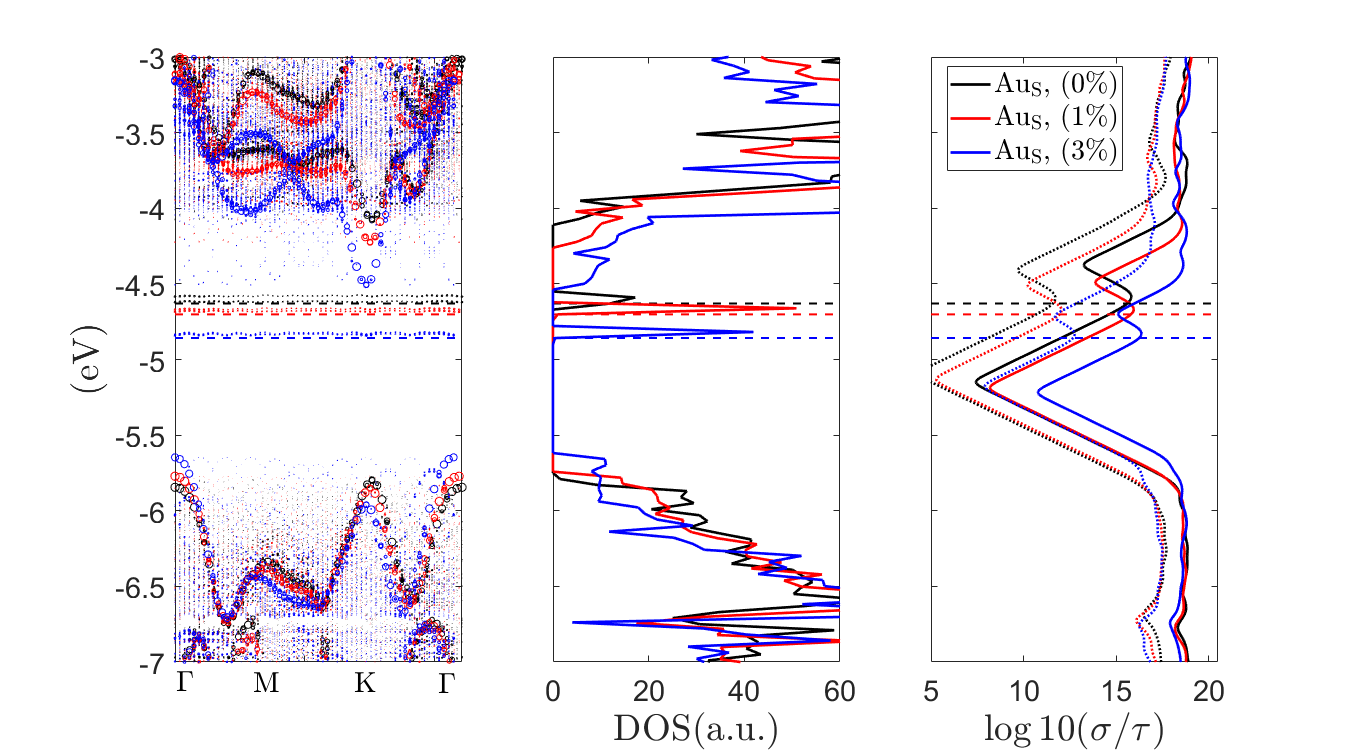}
     }
     \caption{ (a) Effect of biaxial tensile strain of the projected density of states. (b) Effect of biaxial tensile strain on the electronic structure, in-plane conductivity (thick solid lines), and out-of-plane conductivity (dotted thin lines) of defective MoS$_2$ containing an Au adatom adsorbed on a sulfur vacancy. }
     \label{stal_strain_bands}
\end{figure} 

In this section, we explore the effect of biaxial tensile strain on free-standing MoS$_2$.
The changes of the band edges of the pristine MoS$_2$ structure with biaxial tensile strain are depicted in Figures~\ref{stal_strain_bands_SI}, \ref{fig_stal_pdos_strain}, and \ref{fig_stal_unfold_pr_projected}. 
At zero strain, the band gap is direct with dominant character from the out-of-plane Mo ($d_{z^2}$) states at the CBM located at the high-symmetry $\rm K$-point; the VB edge at the $\rm K$ contains contributions from the in-plane ($d_{x^2-y^2}$) and the VB edge at $\rm \Gamma$ contains contributions from out-of-plane ($d_{z^2}$) orbitals of Mo.
As the tensile strain increases (0 to 5$\%$), the states with Mo:$d_{z^2}$ and S:$p_z$ character shift up relative to $E_f$ by approximately 0.54 eV and the VBM mainly consists of Mo:$d_{z^2}$ at higher tensile strain (Figure~\ref{fig_stal_unfold_pr_projected}). 
As a result, the band gap energy of the strained structure shrinks and becomes indirect. 
Moreover, the band velocity increases at both CBM (K-point) and VBM ($\Gamma$-point).
Therefore, it is expected that changes of group velocity play a pronounced role in the conductivity of strained structures (either with or without defects). \\
%Further details of band unfolding of free standing structures with defects are given in the SI (Figure \ref{fig_stal_unfold_pr_proj_def}). 
%With increasing tensile strain, the CBM shifts down from approximately 0.8 eV$- E_f$ at 0$\%$ strain to 0.32 eV$-E_f$ at 5$\%$ strain. 
%The CBM remains mainly of Mo:$d_{z^2}$ nature at higher strain.
%The band gap at 0 $\%$ strain is direct and at the VBM, one observes states of Mo:$d_{x^2-y^2}$ and Mo:$d_{z^2}$ character. 
%When introducing a sulfur vacancy to free-standing $\rm V_S$, the CBM shifts down fron 1.08-$\rm E_f$ to 0.48-$\rm E_f$ eV, similar to the pristine monolayer. 
%The sulfur vacancy introduces defect levels that are located 510 meV below CBM, 95 meV resonant within the CBs and few meV above VBM at zero strain. As the strain increases from 0 to 5 $\%$, both Mo:$\rm d_{z^2}$+S:$\rm p_Z$ shift from Ef-.6 eV to Ef-.24 eV. Interestingly midgap defect level, mainly consists of Mo:$\rm d_{z^2}$ components [**absurd observation, discuss with Wennie**].\\ 

With this in mind, we consider the effect of biaxial tensile strain on the electronic structure and conductivity of free-standing monolayer containing defects.
We consider the pristine monolayer, the monolayer with a sulfur vacancy and pristine monolayer with Au adatom as relevant defects to resistive switching; in particular, we focus on the Au adatom adsorbed on a sulfur vacancy in Figure~\ref{stal_strain_bands}.
Figure~\ref{stal_strain_bands_SI} shows the corresponding results for the pristine monolayer, MoS$_2$ with a sulfur vacancy only, and pristine MoS$_2$ with an Au adatom adsorbed. \\

When including the effect of strain with the adsorption of an Au adatom on the pristine surface, we again observe similar changes in the band edges of free-standing monolayer MoS$_2$.
The inclusion of an Au adatom on the pristine monolayer introduces a defect level that is 759 meV below the CBM, which is resonant with the Fermi level. 
As the strain increases from 0 to 5$\%$, the defect level gets closer to the CBM but remains resonant with the Fermi level. 
Notably, the midgap defect state consists of mainly of Au:$d_{z^2}$ character followed by Mo:$d_{z^2}$ character.% plus a negligible (**use better word ?) contribution from Au:$\rm d_{x^2-y^2}$.  
As tensile strain increases, the contribution of Au:$d_{z^2}$ character to the defect level increases and arises from the relaxation of the Au adatom minimizing its distance to the plane of the monolayer, leading to a greater interaction between the Au adatom and monolayer (see Tables~\ref{table_stal_strain_charge_Au} and ~\ref{table_stal_strain_charge_AuS} for changes in measured distance and occupations with strain). \\
%VBM mainly consists of Mo:$\rm d_{x^2-y^2}$ and Au:$\rm d_{z^2}$. However as the strain increases, Mo orbitals dominate the VBM with a smaller contribution from Au:$\rm d_{x^2-y^2}$. CBM is always dominated by Mo:$\rm d_{z^2}$. CBM shifts down from 0.72 at 0$\%$ to 0.32 and .08 eV above $\rm E_f$ at 3 and 5$\%$, respectively. Band edges of unoccupied Mo:$\rm d_{x^2-y^2}$ and Au also shift down closer to $\rm E_f$ but are always above the CBM. \\

%=================================================================
% FiG 3
%-----------------------
\begin{figure}[h!]
    \centering
    \subfloat[\label{fig_str_ads_energ}]{  
    \includegraphics[width=.9\columnwidth]
      {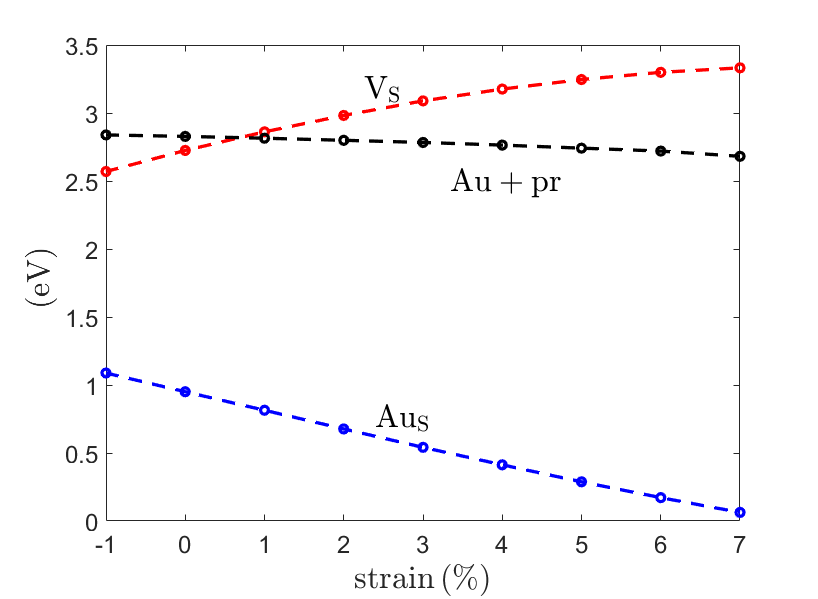}
      } 
      
    \subfloat[\label{fig_str_band}]{%
    \includegraphics[width=.9\columnwidth] 
      {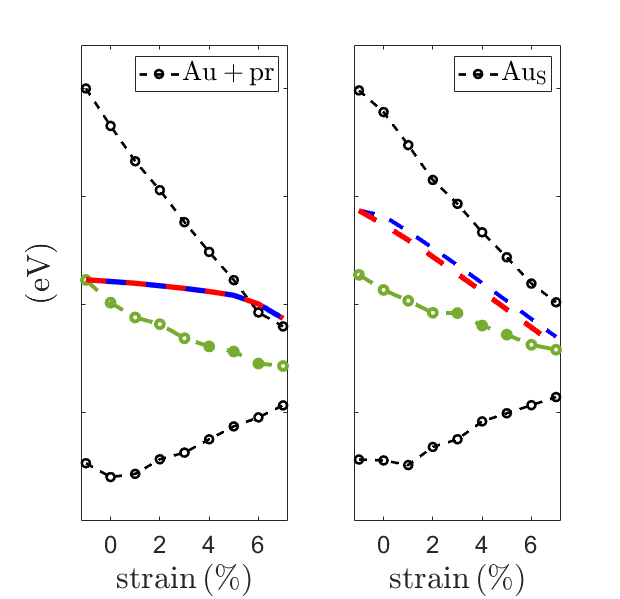}
      }
     \hfill
 %     \subfloat[\label{fig_str_band}]{%
 %   \includegraphics[width=.9\columnwidth] 
 %     {"./fig_draft3/stal_band_edges".png}
 %     }
     \caption{ (a) Effect of strain on formation energy of $\rm V_S$ and adsorption energy of the Au adatom on the pristine structure $\rm Au+pr$ and sulfur vacancy $\rm Au_S$. (b) Effect of strain on the band edges of $\rm Au_S$ and $\rm Au+pr$ structures. Dashed blue lines correspond to defect level and red lines correspond to the Fermi-level.  Center of the d-band is shown by green. Effect of strain on the band edges of pristine and  $\rm V_S$ are given in \ref{fig_SI_strain_band_edge_2}. }
     \label{fig_str_band}
\end{figure} 

We also present in Figure~\ref{stal_strain_bands} the results for an Au adatom adsorbed on a sulfur vacancy in free-standing MoS$_2$.
The midgap defect state in $\rm Au_S$ is resonant with the Fermi level.
Unlike the defect level in the Au+pr configuration, the defect level in the $\rm Au_S$ configuration interestingly consists mainly of Au:$d_{x^2 - y^2}$, illustrating the different degree of hybridization of the Au adatom with the monolayer when sulfur vacancies are present.
An analysis of the orbital decomposition of the unfolded bands in Figure~\ref{fig_stal_pdos_strain_b} indicates that strain leads to the following changes in the electronic structure for MoS$_2$ containing an Au adatom at a sulfur vacancy.
%Mo:$\rm d_{z^2}$ at the CBM at all strain levels. As the tensile strain increase the CBM shifts down from 0.52 eV to 0.24 eV relative to $\rm E_f$. 
%Moreover, the magnitude of shift is lower than that of Au+pr. VBM at 0$\%$ is dominated by Mo:$\rm d_{x^2-y^2}$. Similar to Au+pr, Mo:$\rm d_{x^2-y^2}$ does not shift considerably with strain. As a result, as the strain increases it becomes dominated by Mo:$\rm d_{z^2}$ with a smaller contribution with Au:$\rm d_{z^2}$.[** double check proportion of Mo/Au, ** to discuss : DOS is a good measure here, ] \\
\\

Figure~\ref{fig_str_ads_energ} shows that the adsorption energy  of the Au adatom on the pristine and defective structures decreases with increasing tensile strain. 
We can rationalize this trend based on the $d$-band center theory \cite{ruban1997surface,  schnur2010strain}. 
Here, the $d$-band center is considered to be the average of the lowest energy level of the unoccupied and highest energy level of the occupied Mo$_{d_{z^2}}$ bands.
Tensile strain leads to a reduction of the wavefunction overlap in the vicinity of the adsorption site and consequently a narrowing of the $d$-bands. 
In the case of an early transition metal such as Mo, the narrowing of the $d$-bands leads to a decreased population of the $d$-bands. 
Due to charge conservation, the $d$-bands will shift downward, leading to to a higher occupation of the anti-bonding orbitals.
This downward shift of the $d$-bands corresponds to a weakened interaction between the adsorbate and the surface. 
Figure~\ref{fig_str_band} shows the computed $d$-band center with respect to the band edges and Fermi level as a function of tensile strain, in agreement with the trend in computed adsorption energies. \\

\begin{figure}[h!]
    \centering
    \subfloat[\label{fig_str_ads_charg}]{%
    \includegraphics[width=.9\columnwidth] 
      {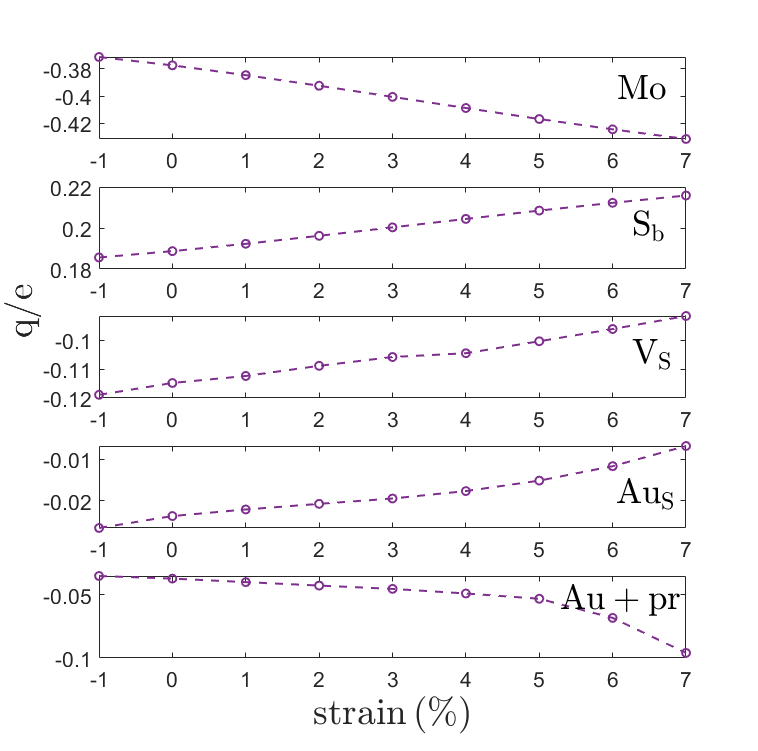}
       }
       \hfill
    \subfloat[\label{fig_str_ads_dip}]{  
    \includegraphics[width=1.05\columnwidth]
      {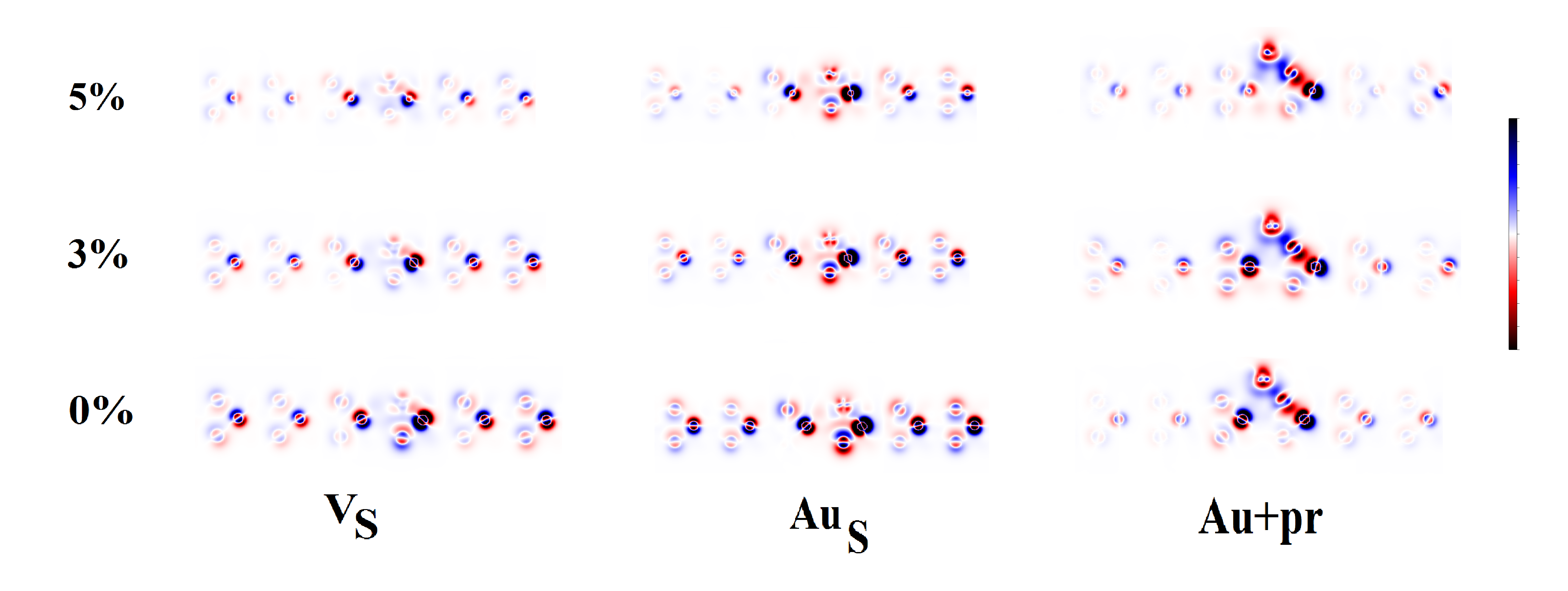}
      } 
     \caption{ (a) Effect of biaxial strain on the Mulliken charge on Mo and S of pristine structure (demonstrating an overall charge delocalization), S-vacancy of the $\rm v_S$ structure, and Au-adatom adsorbed on $\rm Au_S$, and $\rm Au+pr$ structures. Accumulation and depletion of charge is represented by positive and negative signs, respectively. (b) 2D-cross section of the charge-differenpositive and negative numbers, respectivel.ce density. The charge difference is obtained by $\rm \rho_{defect} - \rm \rho_{pr} \mp \rho_{adsorbate}$  (positive for vacancy and negative for adsorbate). Here red and blue colors correspond to loss and gain of charge density, respectively.}
     \label{fig_str_ads_}
\end{figure}

Accompanying these shifts in the $d$-band is a redistribution of charge in the vicinity of the defect site, as shown in Figure~\ref{fig_str_ads_}.
We observe that the vicinity of the $\rm Au_S$ defect experiences a net loss of charge with increasing tensile strain whereas the vicinity of $\rm Au+pr$ undergoes a net gain of charge with increasing tensile strain.
From Figures~\ref{fig_str_band} and ~\ref{fig_stal_pdos_strain}, we observe that the center of $d_{z^2}$ band gets closer toward the Fermi level with increasing strain, while that of $d_{x^2-y^2}$ shifts away from the Fermi level in all defect configurations considered.
Therefore, it is expected that the $d_{z^2}$ orbital becomes more populated while the $d_{x^2-y^2}$ orbital becomes less populated as the tensile strain increases.
Indeed, as shown in Tables~\ref{table_stal_strain_charge_Au} and \ref{table_stal_strain_charge_AuS}, the orbital-decomposed charges of $\rm Au+pr$ and $\rm Au_S$ structures confirms this trend. 
We revisit these trends when considering the heterostructure of Mo$_2$ and gold.\\   

% FiG , STAL , Fig 4 
%-----------------------
 
We briefly comment on the apparent Fermi level pinning of the $\rm Au+pr$ defect configurations.
Several experimental studies have reported that Fermi level pinning occurs for MoS$_2$ in contact with metals \cite{gong2013metal,gong2014unusual}. 
The mechanism of the observed Fermi level pinning at the MoS$_2$/Au(111) interface has been attributed to formation of interfacial dipoles (as discussed in section \ref{sec_heterostructure_strain} and \ref{sec_schottky_ohmic}).
Notably, this mechanism is distinct from well-known Fermi level pinning mechanisms that are applicable to common metal-semiconductor junctions\cite{gong2014unusual}: (1) Bardeen \cite{bardeen1947surface, darling1991defect}, (2) metal induced gap states  \cite{heine1965theory, louie1975self}, (3) defect induced gap-states  \cite{hasegawa1983electrical}.
Interestingly, the Fermi level in the $\rm Au_S$ defect configuration is resonant with the defect level introduced by the Au adatom at all strain levels considered.
Furthermore, our calculations (see Figure~\ref{fig_str_band}) show that strain in free-standing monolayer Mo$_2$ can induce an $n$-type to more $p$-type-like transition for defective structures, as reported in previous theoretical reports \cite{bhattacharyya2014effect, choi2018strain}. 
In section \ref{sec_schottky_ohmic} we connect this discussion of the changes in electronic structure with strain and defects to understand experimental observations of $n$- to $p$-type-like transition in MoS$_2$/Au(111) structures \cite{cook2015influence, mcdonnell2014defect}. \\% can/may be attributed to the local increase of tensile strain. \\  

\begin{figure}[h!]
    \centering 
    \subfloat[\label{fig_str_cond_x}]{%
    \includegraphics[width=.99\columnwidth] 
      {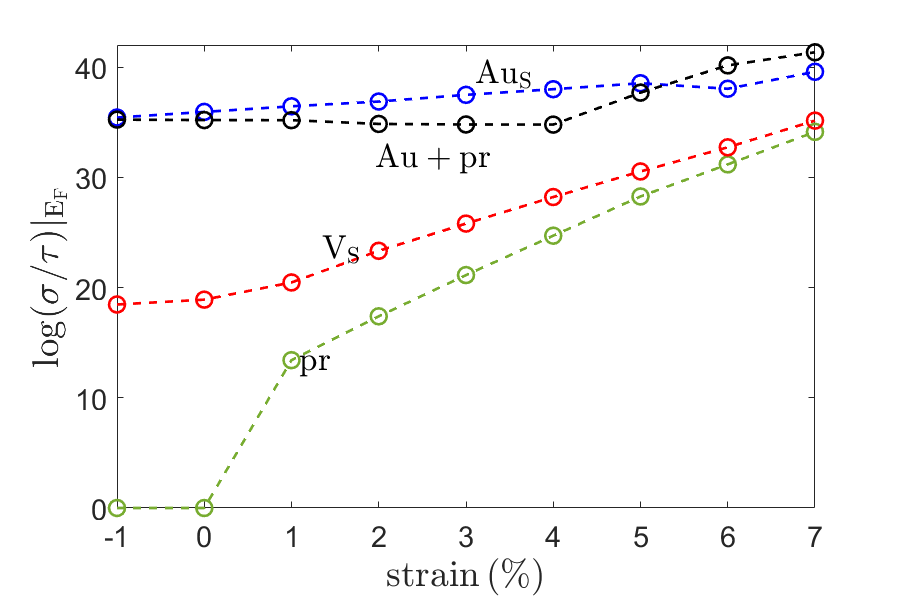}
}

    \subfloat[\label{fig_str_cond_y}]{  
    \includegraphics[width=.99\columnwidth]
         {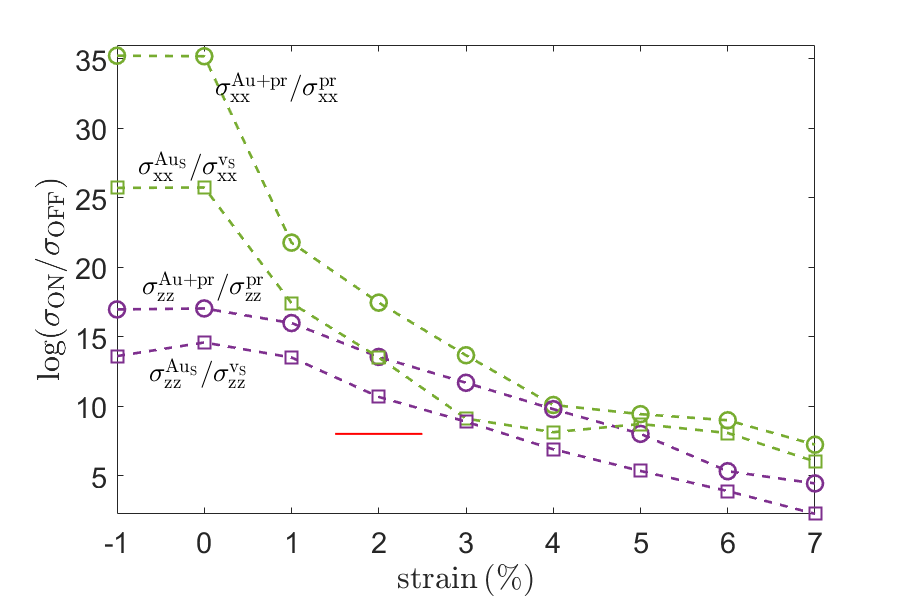}
}   
     \caption{(a) Effect of biaxial strain on the normalized electrical conductivity of stand alone pristine $\rm MoS_2$ structure (green), $\rm V_S$ (red), $\rm Au_S$ (blue) and $\rm Au+pr$ (black) structures at Fermi-level at 300 K in units of 1/$\rm \Omega \cdot cm \cdot s$. (b) Calculated switching ratio of $\rm Au_S/V_S$ (squares) and  $\rm Au+pr/pr$ (circles) combinations of transport in the lateral (dashed) and perpendicular direction (solid) based on normalized conductivities. The red line corresponds to experimental value of the switching ratio \cite{gong2014unusual}.}
     \label{fig_str_cond_}
\end{figure}

Finally, we consider the effect of strain on the normalized conductivity.
Figure~\ref{fig_str_cond_} shows the effect of biaxial strain on the normalized tensor elements of electrical conductivity of the free-standing pristine and defective structures.
In all cases considered (i.e., the pristine structure, $\rm Au+pr$, $\rm V_s$, $\rm Au_S$), tensile strain leads to an increase in the normalized conductivity. 
As shown in Figure \ref{stal_strain_bands}, the increase of the magnitude of the normalized conductivity in strained structures is due to the increase of the band velocity, which is apparent in the band edges for both the in-plane and out-of-plane conductivities.
As shown in Figure~\ref{stal_strain_bands_d}, the normalized conductivity reaches a local maximum when the Fermi level is resonant with the energy level of defect. 
Thus, the normalized conductivity of defects with an Au adatom are generally higher than those without an Au adatom due to the Fermi level being resonant with the corresponding defect level in the former case.
As the monolayer is strained, we also observe the computed Fermi level follow the downward shift of the defect level.
We find a similar trend for the pristine monolayer, the monolayer containing a sulfur vacancy, and the pristine monolayer containing an Au adatom (Figure~\ref{stal_strain_bands_SI}).\\

In the context of resistive switching, it is the ratio of conductivity between the high-resistance and low-resistance states that is relevant.
Figure~\ref{fig_str_cond_y} provides the ratio of the normalized conductivity at ON/OFF states for in in-plane ($\rm xx$) and out-of-plane ($\rm zz$) components.
We focus on the ratio between involving the resistive switching event between the sulfur vacancy alone and the Au adatom adsorbed on the sulfur vacancy ($\rm \sigma_{Au_S}/\sigma_{v_S}$).
For $\rm \sigma_{Au_S}/\sigma_{v_S}$ the ratio of the normalized conductivity across the monolayer diminishes by around five orders of magnitude going from 0\% to 2$\%$ strain in the MoS$_2$ monolayer.
For reference, the switching ratios reported in the literature are in the range of 6-8 orders of magnitude \cite{hus2021observation} (red line in Figure~\ref{fig_str_cond_}).
While some degree of resistive switching via point defects is preserved up to large strains, we anticipate that the consideration of strain in resistive switching devices based on 2D materials can explain some of variability observed within and among samples and will be an important factor in comparing the performance (in this case, the switching ratio) among materials platforms and sample preparation methods.\\

%WW: I don't think we can meaningfully comment on this. 
%[**still working on this**] Now we explain why conductivity at midgap defect level of AuS is higher than that of Au+pr structure. The tunneling current in the direction normal to the surface is proportional to $\rm I(z) \approx \exp{(-2 z  \kappa_{\parallel})}$, where the decay constant is proportional to $\rm \kappa=\sqrt{2m\phi /\hbar^2 +k_{\parallel}^2}$ \cite{}. It was discussed that the defect level in Au+pr consists mainly of $\rm d_{z^2}$ orbitals, while that of AuS is mainly of $\rm d_{x^2-y^2}$ nature. As a result Au+pr has a higher decay constant and a lower conductivity compared to that of AuS. [** 1. assuming constant velocity, or effective mass in this representation**] [** 2. I am not convinced this approximation can be applied to trap level, as the concept of effective mass is not clear here. perhaps BTE is more appropriate]. \\ 

%-----------------------%-----------------------
% FiG  STAL, conductivity , Fig 5 
%-----------------------%-----------------------

% ==========================================================================
% ========================    HETEROSTRUCCTURE         =====================
% ==========================================================================

% =======================================
\subsection{Point defects and strain in the MoS$_2$/(111)-Au heterostructure}\label{sec_defects_in_hetero}
% =======================================

\begin{figure}[h!]
    \centering
    \subfloat[\label{fig_het_ads}]{%
    \includegraphics[width=1.0\columnwidth] 
      {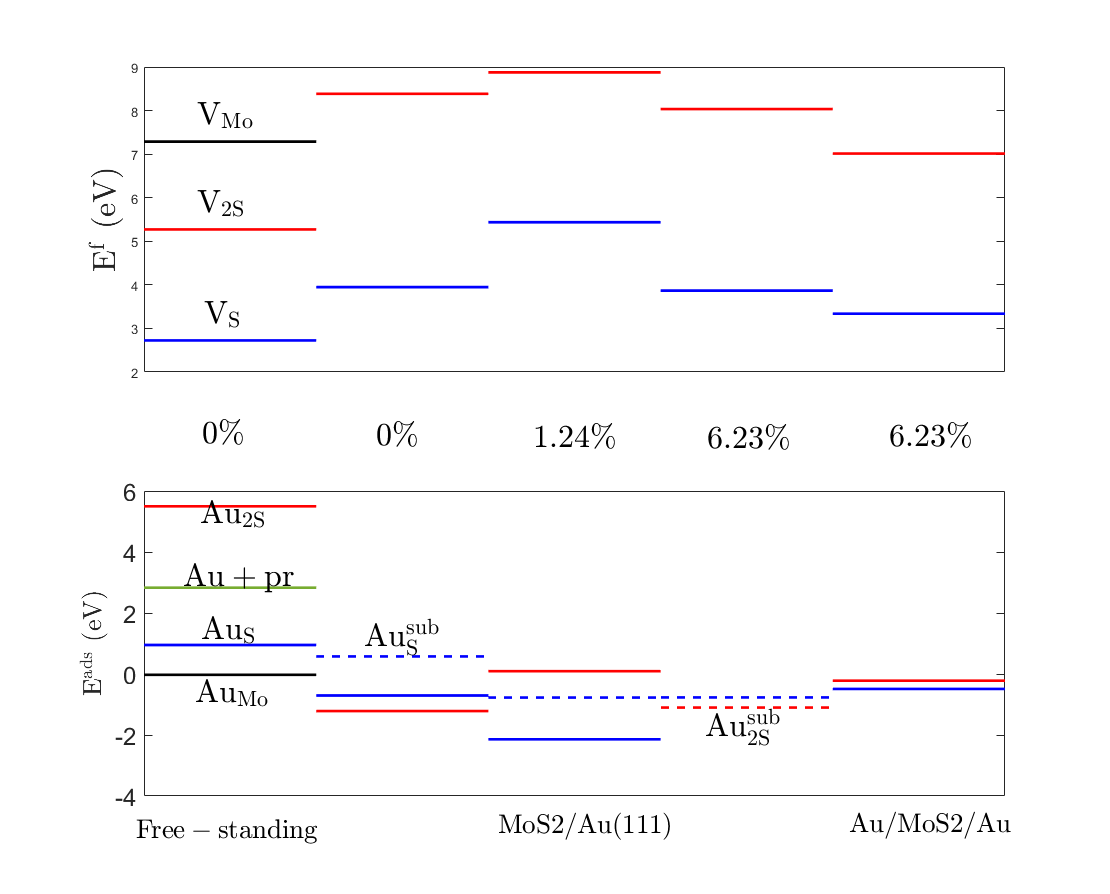}
} 

    \subfloat[\label{fig_hetero_defect_combined}]{%
    \includegraphics[width=.99\columnwidth] 
      {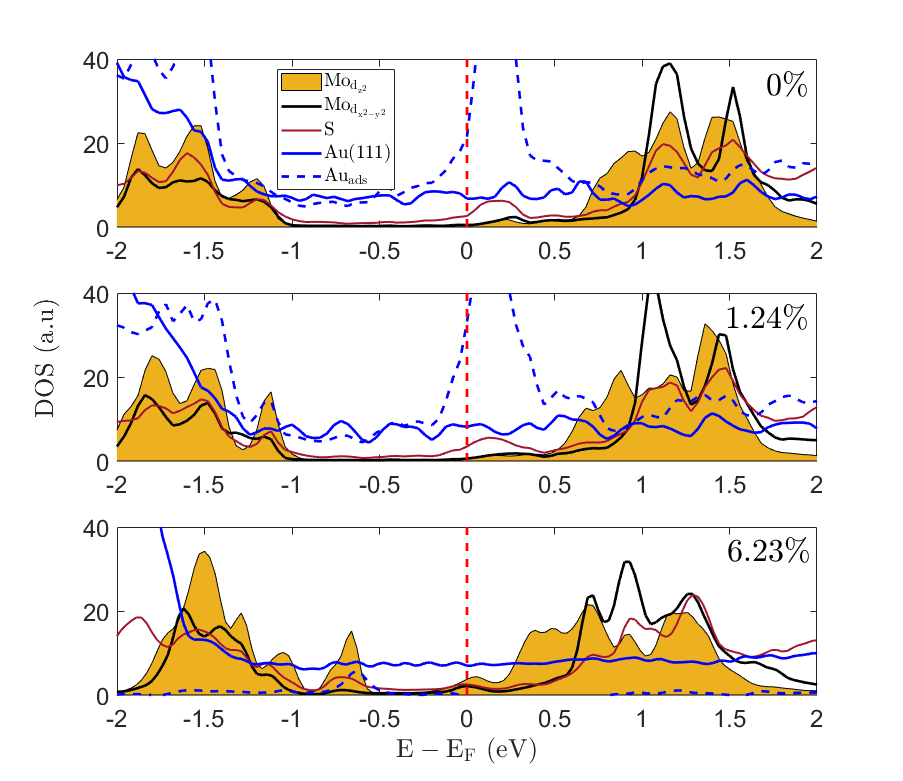}
} 
     \caption{ (a) Effect of tensile strain on the formation energy of vacancies (top) and adsorption energy of Au+vacancy structures. Columns from left to right correspond to Free-standing MoS$_2$, MoS$_2$/Au(111) heterostructure under 0$\%$, 1.24$\%$ and 6.23$\%$ tensile strain, and double electrode structure under 6.23$\%$ tensile strain, respectively. Adsorption energies represented by solid (dashed) line correspond to Au-adatom provided by an external reservoir (substrate). Calculated energies are tabulated in Table \ref{table_hetero_energies} of SI. (b)  Electronic structure (DOS) of $\rm Au_{S}$+$\rm V_S$ defect complex in the $\rm MoS_2/(111)-Au$ heterostructure. A more complete picture is given in Figures~\ref{fig_pdos_het_0_defect_}, \ref{fig_pdos_het_1p2_defect_},  and\ref{fig_pdos_het_5_defect_}.}
     \label{fig_het_ads_}
\end{figure}

We now discuss the influence of the metal substrate using the MoS$_2$/(111)-Au heterostructure.
The presence of the metal substrate is analyzed from the perspective of introducing strain into the monolayer and in driving charge transfer to the monolayer.
As with the discussion of the free-standing monolayer, we consider the impact of each of these effects in the presence of defects relevant to memristor devices.
With the formation of a heterostructure, we distinguish between the sulfur in the vicinity of the defect that is closest to the substrate $\rm S_b$ and farthest from the substrate $\rm S_t$. 
Figure \ref{fig_het_ads} depicts that the presence of the substrate leads to an increase of the formation energy of the sulfur vacancy and divacancy (we omit the Mo vacancy as it has already a high defect formation energy without the substrate).
In contrast, the presence of a gold substrate leads to a decrease of adsorption energies compared to the free-standing monolayer (calculated energies are tabulated in Table \ref{table_hetero_energies} of SI), including with tensile strain included. 
As a result, we conclude that the adsorption of an Au adatom on a pre-existing vacancy is the dominant defect responsible for switching, rather than direct substitution of a Au adatom for sulfur, consistent with the conclusions of the free-standing monolayer.\\

% FiG , hetero , adsorption , Fig 6 
%-----------------------

Next, we examine the electronic structure of the MoS$_2$/Au(111) heterostructure under strain to understand the trends observed in the computed ON/OFF ratio. 
The PDOS of defect complexes in the MoS$_2$/Au(111) heterostructure under 0, 1.24 and 6.23$\rm\%$ biaxial strain is depicted in Figure~\ref{fig_hetero_defect_combined} (see also Figures~\ref{fig_pdos_het_0_defect_}, \ref{fig_pdos_het_1p2_defect_} and \ref{fig_pdos_het_5_defect_}).  
%In the heterostructure with a relaxed top layer ( 0$\rm \%$ strain), the band gap of MoS2 is approximately 1.5 eV, with VBM and CBM located approximately at -1 and 0.5 eV, respectively.
For the heterostructure with 1.24\% strain, a typical strain found in experimental samples, the band gap decrease to 1.32 eV and the  Mo$:{d_{z^2}}$ orbitals have shifted towards the VBM, as also observed in the free-standing monolayer.
%The peak of VBM is now located 1.16 eV below the Fermi level. 
% The defect level of $\rm V_S$ structure consists of S and Au character and to a lower degree by Mo orbitals. 
 Adsorption of Au into the vacancy ($\rm Au_S$) still leads to a defect energy level, which is within 0.16 eV of the Fermi level.
 This behavior is also observed for the $\rm Au_{S2}$ defect configuration.
%The peak of the $\rm V_{2S}$ structure is also located 0.16 eV above the Fermi level. However, it has a more significant  contribution from S orbitals. The band gap of this structure is approximately the same as that of $\rm V_S$.  
The $\rm Au_{2S}$ structure sees two additional peaks located at 0.04 and 0.68 eV relative to the Fermi level.
% The defect level with the highest energy is located inside the CBM. As a result, formation of $\rm Au_{2S}$ can cause Schottky to Ohmic transition [** move to that section**].
% The trend of $\rm Au_{2S}$ + $\rm v_{Au}$ is the same of that of$ Au_{2S}$, however the defect level located in the vicinity of Fermi level splits, such that 2 peaks at 0.04 and .24 eV relative to the Fermi level are created.
A comparison between the studied strained structures indicates that at very high strain (6.23\%) the electronic states at the band edges and Fermi level are then dominated by  Mo$:{d_{z^2}}$ orbitals. 
Consequently, high levels of strain lead to a suppression of the defect state induced by the adsorption of a Au adatom.
Therefore, although high levels of strain significantly lower the adsorption/desorption energy levels, we expect that strain concurrently leads to a suppression of the expected switching ratio, as similarly found for the stand-alone monolayer.
While beyond the scope of this paper, a rigorous study of the transport properties across the vertical stack architecture would quantify the impact of the presence of the MoS$_2$/(111)-Au interface and is a topic of future study.\\

% =======================================
\subsection{Strain and charge transfer in the Mo$_2$/(111)-Au heterostructure}\label{sec_heterostructure_strain}
% =======================================
  
\begin{figure}[h!]
    \centering
      \subfloat[\label{fig_subs_all_a}]{%
    \includegraphics[width=.99\columnwidth] 
      {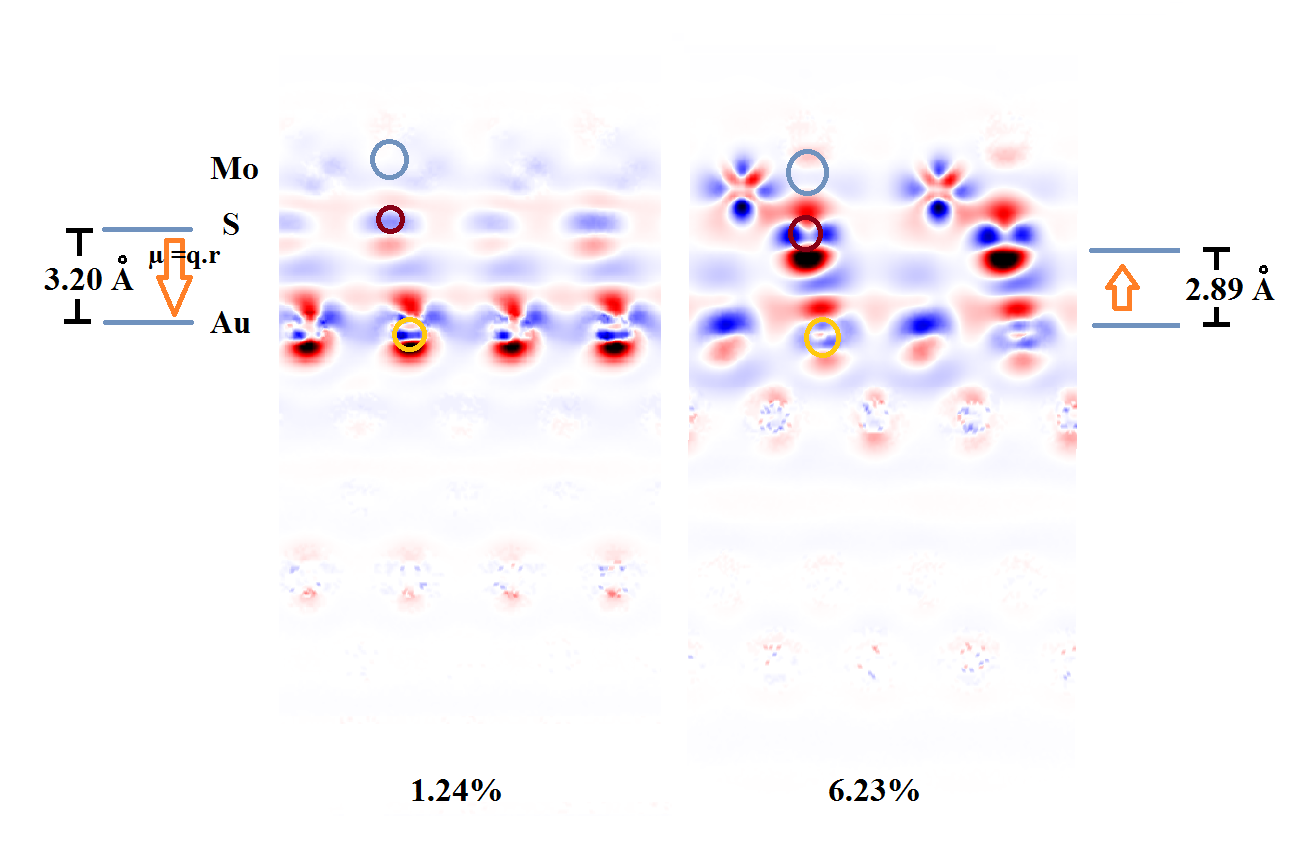}
    } 
    
 %   \subfloat[\label{fig_subs_all_a}]{%
 %   \includegraphics[width=.99\columnwidth] 
 %     {"./fig_draft4/all_hetero_2".png}
 %   } 
 
%\subfloat[\label{fig_subs_all_b}]{%
%    \includegraphics[width=.99\columnwidth] 
%      {"./fig_draft4/subst_2d_charge_3".png}
%    }
 
%    \subfloat[\label{fig_subs_all_c}]{%
%    \includegraphics[width=.45\columnwidth] 
%      {"./fig_draft2/substrate/subs_strain_charge".png}
%}
\subfloat[\label{fig_subs_all_d}]{%
    \includegraphics[width=.99\columnwidth] 
      {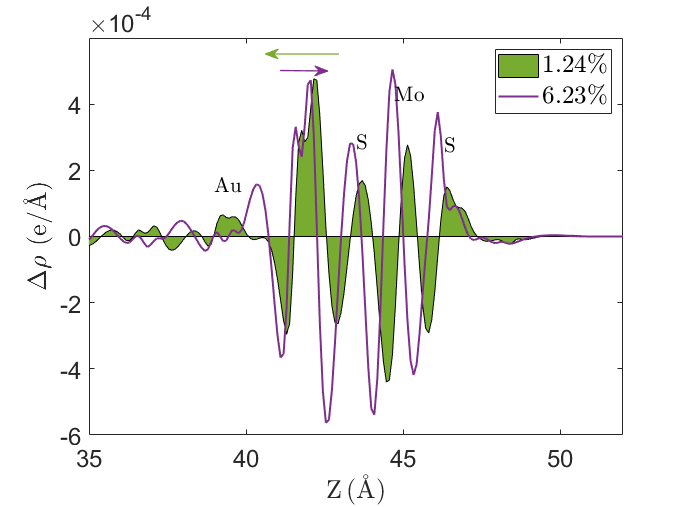}
    }

\caption{(a) Charge difference plot and interlayer distances due to biaxial strain caused by lattice mismatch of the top $\rm MoS_2$ layer with the (111)-Au substrate. The charge difference is obtained by $\rm \rho_{heterostructure}$-$\rm \rho_{Au(111)}$-$\rm \rho_{MoS_2}$. Gold, red and blue circles correspond to the location of the surface of (111)-Au, S and Mo atoms, respectively. 
The blue bars and the arrow in between represent the interlayer distance and the magnitude and direction of the interfacial dipole moment, respectively. 
Lattice constants, interlayer distances and the corresponding charge difference plot of the other strained heterostructures are given in \ref{SI_fig_subs_all}. (b) Cross section of the charge difference density along the direction normal to the surface.} 
     \label{fig_subs_all}
\end{figure}

In this section we discuss how charge transfer across the heterostructure varies with increasing tensile strain and leads to different interfacial dipole moments. 
%Figure \ref{fig_subs_all} shows the effect of strain on the charge transfer characteristics of the MoS$_2$/Au(1111) heterostructure.
Figure~\ref{fig_subs_all} shows that increasing biaxial tensile strain leads to a reduction of the interlayer spacing between MoS$_2$ and the gold substrate (see also Figure~\ref{fig_subs_all_a}).
Accompanying the change in interlayer spacing with strain are changes in the charge transfer between MoS$_2$ and gold electrode. 
Charge difference plots (\ref{fig_subs_all_b}) show that more charge is accumulated on both the $\rm S_{b}$ atom and the surface of the electrode with increasing strain and decreasing interlayer spacing. 
As a result, a space charge region forms in the interfacial area and forms permanent dipole moments. 
Moreover, the surface integral of the charge depletion region, corresponding to $\rm \Delta \rho < 0$ regions in Figure~\ref{fig_subs_all_d}, is larger at the boundary of the interfacial region near the electrode at low strain regimes and near MoS$_2$ at high strain regimes.
This results in a permanent interfacial dipole moments to be pointing toward the electrode at low strains and pointing toward MoS$_2$ at high strains.
Although the magnitude of the charge accumulation at both sides of the heterostructure is larger for higher levels of strain  (corresponding to darker colors in Figure~\ref{fig_subs_all_b}), the decrease of the interlayer distance lowers the magnitude of the dipole moment. 
%Accumulation and depletion of charge at the top layer, which is consequently manifested in modulation of the direction and magnitude of the interfacial dipole moment affects the barrier height at the interface, and band bending at the semiconductor side. 
These phenomena are expected to contribute to the observed the $n-$ to $p-$type transition and Ohmic to Schottky transition in experimental results, as discussed in the next section.
We also expect this dipole to affect the dynamics of the resistive switching and to be an additional degree of freedom with which to modulate the resistive switching characteristics, a topic we leave to future study. \\ 

%=====================================================================
%=====================================================================
%=====================================================================

\subsection{Strain and defects in the heterostructure and the Ohmic to Schottky transition}\label{sec_schottky_ohmic}

% ==================================================
% Effect of interlayer distancing and strain on n--> p and/or 
%  ohmic to schottky transition 

% fig 8   
%-----------------------
\begin{figure}[h!]
    \centering
    \subfloat[\label{fig_sub_dist_a}]{%
    \includegraphics[width=.99\columnwidth] 
      {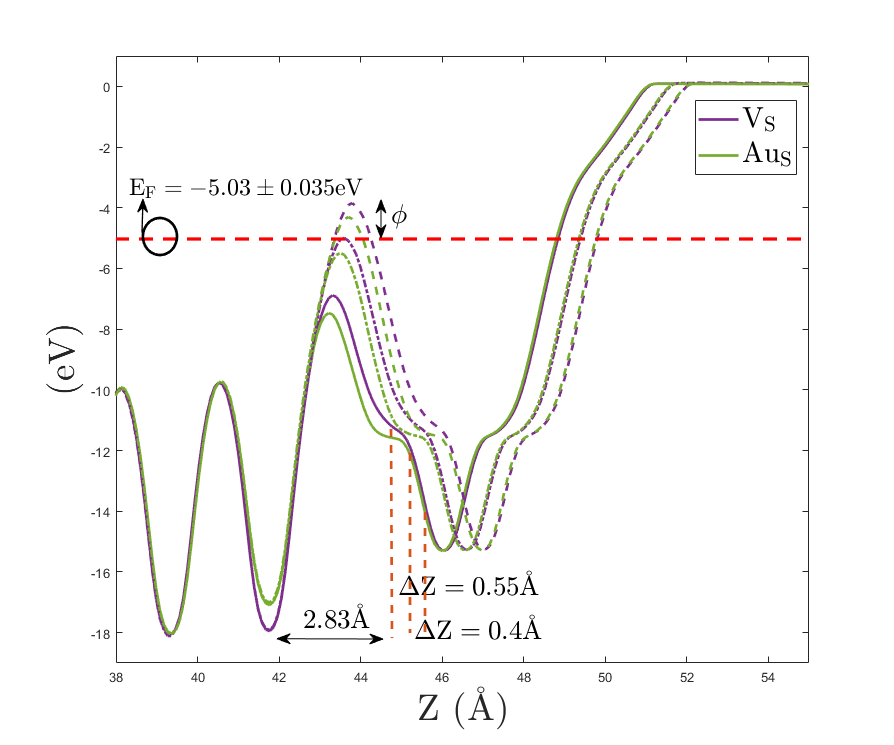}
} 

   \subfloat[\label{fig_sub_dist_b}]{%
    \includegraphics[width=.99\columnwidth] 
      {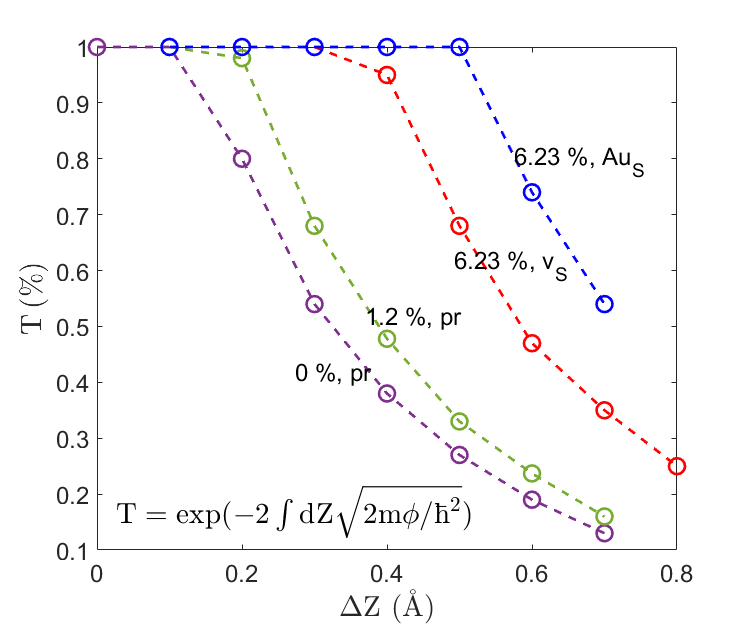}
} 

%    \subfloat[\label{fig_sub_dist_c}]{%
%    \includegraphics[width=.45\columnwidth] 
%      {"./fig_draft2/substrate/subs_dist_charge_analysis".png}
%}
     \caption{(a) Changes of the electrostatic potential of atoms with interlayer distance of the MoS$_2$/(111)Au heterostructure under 6.23 $\%$ tensile strain. (b) Changes of tunneling probability with interlayer distance for select MoS$_2$/(111)-Au heterostructures. Changes of the corresponding Mulliken charges are given in Figure~\ref{subs_shift_charge_5p8}.   
     }
     \label{fig_sub_dist}
\end{figure} 

In this section, we present the relationships among strain, the interlayer spacing, and changes of the Schottky barrier height $\phi$ at the interface. 
This correspondence is first illustrated with the heavily strained (6.23\%) and initially Ohmic heterostructure in Figure~\ref{fig_sub_dist_a}, which shows the changes in the average electrostatic potential along the direction normal to the surface with variable interlayer distance. 
%As demonstrated in Figure \ref{fig_sub_dist_b}, the Mulliken charges with displacement of the MoS$_2$ in the surface-normal direction. 
Our results indicate that an increase of 0.55 \AA\, in the interlayer distance brings the heavily strained and initially Ohmic heterosutructure  (6.23$\%$) to the verge of transitioning to a Schottky interface; a further increase of the interlayer distance results in additional elevation of the barrier height above the Fermi level into the Schottky regime. 
Contrary to case of free-standing $\rm Au_S$ and $\rm Au+pr$, where the adsorbed Au-adatom donates charge to the host MoS$_2$, the adsorbed Au-adatom formed in the $\rm Au_S$/(111)-Au structure obtains extra electronic charge. 
As shown in Figure~\ref{subs_shift_charge_5p8}, the relative changes of the Mulliken charges of the Au(111) electrode and $\rm S_b$ also show an extremum at $\rm \Delta Z$=0.5 \AA, which is the onset of Schottky to Ohmic transiton and is an indicator of the change of the direction of the interfacial dipole moment. \\

We repeat the calculation of the average electrostatic potential at different representative strains for a number of interlayer spacings and select defect configurations.
Figure \ref{fig_sub_dist_b} demonstrates the effect of strain on the tunneling probability $T$ and the corresponding interlayer spacing at which the Ohmic to Schottky transition occurs.
Biaxial tensile strain lowers the barrier height, which is equivalent to an increase in the onset (i.e., required shift of the interlayer distance) of the Ohmic to Schottky transition. 
Moreover, diffusion of the Au-adatom from electrode into a sulfur vacancy in the MoS$_2$ layer lowers the barrier and increases the tunneling probability. 
Therefore, the change of barrier height is an additional degree of freedom that contributes to switching mechanism, as influenced by strain, interlayer distance, and defects. 
This parameter is clearly a function of biaxial tensile strain caused by the deformation potential; we further suggest that the variation of the interlayer spacing is a descriptor of the varying surface roughness of the underlying Au(111) electrode. 
These results suggest that variability in the metal electrodes is transferred to the electronic and transport properties of the monolayer. \\

There are several implications to memristor devices that our results have.
There is a plethora of theoretical and experimental evidence that hint at such variations in the deformation potential of MoS$_2$/(111)-Au interface. 
For example, experimental results have reported seemingly random $n-$type and $p-$type behavior with Fermi level fluctuations up to 1 eV \cite{cook2015influence, mcdonnell2014defect}. 
Moreover, a Schottky to Ohmic transition has been observed for samples grown under the same conditions \cite{cook2015influence, carladous2002light}. 
%A combined STS/DFT study of Silva et al. shows a periodic displacement of MoS$_2$ layer in the direction normal to the surface with a magnitude between 0.5 (DFT) and 1.0 $\rm \AA$ (STS) \cite{silva2022spatial}. 
%Another DFT study on different stacking of S and bottom Au of the electrode yields an interlayer distance of 2.51, 3.03 and 3.20 $\rm \AA$, and a binding energy of -.25, -.03 and -.04 eV respectively \cite{bruix2016single}. 
Ballistic electron emission microscopy experiments, which studied the growth of Au on Mo$_2$, also reveal that the roughness and deformation of Au grown on MoS$_2$ increases thickness of the Au layer \cite{cook2015influence, carladous2002light}. 
%Topographic images of Au(111) grown on MoS$_2$ indicate that the surface morphology of Au transitions from terraced triangles to a mix of terraced hexagonal and irregular-shaped structures as Au(111) film thickness exceeded 16 nm. Here, the increasing strain can be attributed to Volmer-Weber growth mode through which bulk properties of Au begin to dominate at higher thicknesses and cause the deformation of interfacial morphology \cite{cook2015influence, carladous2002light, gong2013metal}. 
Raman spectra also reveal the emergence of tensile strain in the MoS$_2$ with thicker Au films. 
Here, with increasing thickness of Au, the interlayer distance and Schottky barrier increase whereas the transmission exponentially decreases through the (111)-Au/MoS$_2$ junction, though controversy of the exact nature of the transition remains~\cite{carladous2002light}. 
Interestingly, experimental studies report that the strength of interaction between monolayer MoS$_2$ and gold can be tuned when varying the surface roughness of the gold substrate \cite{velicky2020strain, michail2016optical}.\\
 
 %Similar to prior work\cite{michail2016optical}, Velick\'{y} et al. \cite{velicky2020strain,14,15} reported biaxial strains of up to 1.9\% based on the downward shift and broadening of the in-plane $E'$ mode. the splitting of the out-of-plane $A1$ to n-type doping up to concentrations of $2.6 \times 10^{13}$ cm$^{-2}$ in locations of the MoS$_2$ monolayer in close contact with Au. 
 
Thus, our results show that the barrier height is affected by the deformation potential of the electrode and the resulting interfacial dipole moments. 
As an example of the importance of this concept in memristors, Hus et al.\cite{hus2021observation} observed in scanning tunneling spectroscopy studies both an Ohmic-like and Schottky-like behavior in the I-V characteristics of sulfur vacancies in different locations on monolayer MoS$_2$. 
Their experiment suggest that the local environment of sulfur vacancies may differ and further impact the transport properties when a metal adatom is introduced. 
Our results suggest that the Ohmic and Schottky behavior correspond to either regions with high and low strain levels, or regions with different interlayer spacings due to surface morphology of the electrode. \\ 
 
% ============================================
% ============================================
\section{Conclusion and future work}\label{sec_concl }

In this study, we explored the role of strain and the presence of the substrate for point defect complexes relevant to resistive switching applications for free-standing MoS$_2$ and the (111)Au-MoS$_2$ junction. 
Specifically, we inspected the characteristics of neutral sulfur vacancies, sulfur divacancies, and molybdenum vacancies of $\rm MoS_2$ and their interaction with an Au adatom in biaxially strained configurations of the free-standing monolayer and the interface with (111)-Au. \\

For the free-standing monolayer, our results indicated that formation of $\rm Au_S$ structure is not energetically as favorable as that of $\rm Au+pr$ structure, suggesting that Au adatoms adsorb onto a pre-existing anion vacancy and that $\rm Au+pr$ structure serves as an intermediate state.
Our calculations also showed that mechanical strain lowers the adsorption energy of the Au adatom, making the resistive switching energetically more favorable. 
The lowering of the adsorption energy may be rationalized using the $d$-band center theory from electrocatalysis.
%The adsorption of Au on a sulfur vacancy site also leads to Fermi level pinning, which in part aids in the change in conductivity and in part limits the tunability of the type of contact between MoS$_2$ and substrate.
The change in the normalized conductivity ($\sigma/\tau$) is found to arise from the Fermi level being resonant with the defect level induced by the Au adatom; changes of band velocities with tensile strain is a secondary effect of strain in free-standing structures. 
However, increasing tensile strain is found to lower the switching ratios of the ON/OFF states for all of the defect configurations considered. 
Consequently, any use of strain engineering in the 2D monolayer must balance the advantages of lowering the adsorption/desorption energies, i.e., switching energies, and the disadvantages of reducing the ON/OFF switching ratio in the design of memristor devices. \\

For the heterostructure of MoS$_2$/(111)-Au, we consider the effects of strain and charge transfer across the interface.
The lattice mismatch between the substrate and MoS$_2$ causes tensile strain in MoS$_2$. 
Our study showed that the formation energies of all vacancies in the heterostructure increase and adsorption/desorption energies of Au+defect complexes of interest decrease with increasing tensile strain. 
Thus, the formation of an Au-adatom on a pre-existing vacancy is still more favorable, as it is in the free-standing monolayer. 
We find that moderate levels of strain (1.9$\%$) bring the defect level of adsorbed Au-adatom closer to the Fermi level, suggesting a slight improvement in switching ratio. 
In contrast, very high levels of strain lead to a suppression of the electronic states from Au-adtom in the band-gap region, resulting in a diminished change in conductivity in the ON state,  which is not favorable for resistive switching. \\  

A second effect due to the presence of the electrode is the interfacial charge transfer and formation of interfacial dipole moments.
Our calculations show that the magnitude and direction of the interfacial dipole moments change with tensile strain and interlayer distance between the electrode and monolayer. 
Thus, we suggest that a primary mechanism behind the experimental observation of $n-$ to $p-$type behavior and the Ohmic to Schottky transition is related to the deformation potential of the Au(111) electrode. 
Indeed, our results suggest that the variability in performance often observed within and across memristor devices can be explained through this picture of strain, charge transfer, and interlayer spacing. 
This phenomenon can be affected and potentially controlled through engineering processing conditions, chemical doping, strain, and electric field. 
Simultaneously, the presence of the electrode introduces additional degrees of freedom (e.g., the dipole moment) that can be used to alter the resistive switching response of the device.
We leave this and the study of electric field and transport of the heterostructure (i.e., device) to future work. \\ 

In summary, we find that there are quantitative impacts that the presence of strain and the substrate have on the defect energetics related to the switching energy and transport properties related to the switching ratio of the atomristor. 
Therefore, the interaction of the monolayer and electrode is an important consideration in the search of the ideal atomristor.  \\

\section{Supplementary Material}
The following supplementary information is provided: (1) Section S1: structural models of defect configurations and heterostructure, defect energetics, details electronic structure and transport properties of remaining defect configurations; (2) Section S2: effect of biaxial tensile strain in free-standing MoS$_2$ on the electronic structure, transport properties, orbtial occupations, and charge density; (3) Section S3: effect of biaxial tensile strain in the MoS$_2$/(111)-Au heterostructure on charge density, site occupations, defect energetics, and electronic structure.

\begin{acknowledgments}
We thank Deji Akinwande for valuable discussions.
The work was supported in by part by The Welch Foundation (grant F-2172-20230405).
The authors acknowledge the Texas Advanced Computing Center (TACC) at The University of Texas at Austin for providing HPC resources that have contributed to the research results reported within this paper. URL: http://www.tacc.utexas.edu 
\end{acknowledgments}

%=============================================================
\section{References}
\bibliography{memristor}

\begin{thebibliography}{65}%
\makeatletter
\providecommand \@ifxundefined [1]{%
 \@ifx{#1\undefined}
}%
\providecommand \@ifnum [1]{%
 \ifnum #1\expandafter \@firstoftwo
 \else \expandafter \@secondoftwo
 \fi
}%
\providecommand \@ifx [1]{%
 \ifx #1\expandafter \@firstoftwo
 \else \expandafter \@secondoftwo
 \fi
}%
\providecommand \natexlab [1]{#1}%
\providecommand \enquote  [1]{``#1''}%
\providecommand \bibnamefont  [1]{#1}%
\providecommand \bibfnamefont [1]{#1}%
\providecommand \citenamefont [1]{#1}%
\providecommand \href@noop [0]{\@secondoftwo}%
\providecommand \href [0]{\begingroup \@sanitize@url \@href}%
\providecommand \@href[1]{\@@startlink{#1}\@@href}%
\providecommand \@@href[1]{\endgroup#1\@@endlink}%
\providecommand \@sanitize@url [0]{\catcode `\\12\catcode `\$12\catcode `\&12\catcode `\#12\catcode `\^12\catcode `\_12\catcode `\%12\relax}%
\providecommand \@@startlink[1]{}%
\providecommand \@@endlink[0]{}%
\providecommand \url  [0]{\begingroup\@sanitize@url \@url }%
\providecommand \@url [1]{\endgroup\@href {#1}{\urlprefix }}%
\providecommand \urlprefix  [0]{URL }%
\providecommand \Eprint [0]{\href }%
\providecommand \doibase [0]{http://dx.doi.org/}%
\providecommand \selectlanguage [0]{\@gobble}%
\providecommand \bibinfo  [0]{\@secondoftwo}%
\providecommand \bibfield  [0]{\@secondoftwo}%
\providecommand \translation [1]{[#1]}%
\providecommand \BibitemOpen [0]{}%
\providecommand \bibitemStop [0]{}%
\providecommand \bibitemNoStop [0]{.\EOS\space}%
\providecommand \EOS [0]{\spacefactor3000\relax}%
\providecommand \BibitemShut  [1]{\csname bibitem#1\endcsname}%
\let\auto@bib@innerbib\@empty
%</preamble>
\bibitem [{\citenamefont {Chang}, \citenamefont {Jo},\ and\ \citenamefont {Lu}(2011)}]{chang2011short}%
  \BibitemOpen
  \bibfield  {author} {\bibinfo {author} {\bibfnamefont {T.}~\bibnamefont {Chang}}, \bibinfo {author} {\bibfnamefont {S.-H.}\ \bibnamefont {Jo}}, \ and\ \bibinfo {author} {\bibfnamefont {W.}~\bibnamefont {Lu}},\ }\bibfield  {title} {\enquote {\bibinfo {title} {Short-term memory to long-term memory transition in a nanoscale memristor},}\ }\href@noop {} {\bibfield  {journal} {\bibinfo  {journal} {ACS nano}\ }\textbf {\bibinfo {volume} {5}},\ \bibinfo {pages} {7669--7676} (\bibinfo {year} {2011})}\BibitemShut {NoStop}%
\bibitem [{\citenamefont {Lanza}\ \emph {et~al.}(2022)\citenamefont {Lanza}, \citenamefont {Sebastian}, \citenamefont {Lu}, \citenamefont {Le~Gallo}, \citenamefont {Chang}, \citenamefont {Akinwande}, \citenamefont {Puglisi}, \citenamefont {Alshareef}, \citenamefont {Liu},\ and\ \citenamefont {Roldan}}]{lanza2022memristive}%
  \BibitemOpen
  \bibfield  {author} {\bibinfo {author} {\bibfnamefont {M.}~\bibnamefont {Lanza}}, \bibinfo {author} {\bibfnamefont {A.}~\bibnamefont {Sebastian}}, \bibinfo {author} {\bibfnamefont {W.~D.}\ \bibnamefont {Lu}}, \bibinfo {author} {\bibfnamefont {M.}~\bibnamefont {Le~Gallo}}, \bibinfo {author} {\bibfnamefont {M.-F.}\ \bibnamefont {Chang}}, \bibinfo {author} {\bibfnamefont {D.}~\bibnamefont {Akinwande}}, \bibinfo {author} {\bibfnamefont {F.~M.}\ \bibnamefont {Puglisi}}, \bibinfo {author} {\bibfnamefont {H.~N.}\ \bibnamefont {Alshareef}}, \bibinfo {author} {\bibfnamefont {M.}~\bibnamefont {Liu}}, \ and\ \bibinfo {author} {\bibfnamefont {J.~B.}\ \bibnamefont {Roldan}},\ }\bibfield  {title} {\enquote {\bibinfo {title} {Memristive technologies for data storage, computation, encryption, and radio-frequency communication},}\ }\href@noop {} {\bibfield  {journal} {\bibinfo  {journal} {Science}\ }\textbf {\bibinfo {volume} {376}},\ \bibinfo {pages} {eabj9979} (\bibinfo {year} {2022})}\BibitemShut {NoStop}%
\bibitem [{\citenamefont {Bertolazzi}\ \emph {et~al.}()\citenamefont {Bertolazzi}, \citenamefont {Bondavalli}, \citenamefont {Roche}, \citenamefont {San}, \citenamefont {Choi}, \citenamefont {Colombo}, \citenamefont {Bonaccorso},\ and\ \citenamefont {Samorì}}]{bertolazziNonvolatileMemoriesBased2019}%
  \BibitemOpen
  \bibfield  {author} {\bibinfo {author} {\bibfnamefont {S.}~\bibnamefont {Bertolazzi}}, \bibinfo {author} {\bibfnamefont {P.}~\bibnamefont {Bondavalli}}, \bibinfo {author} {\bibfnamefont {S.}~\bibnamefont {Roche}}, \bibinfo {author} {\bibfnamefont {T.}~\bibnamefont {San}}, \bibinfo {author} {\bibfnamefont {S.}~\bibnamefont {Choi}}, \bibinfo {author} {\bibfnamefont {L.}~\bibnamefont {Colombo}}, \bibinfo {author} {\bibfnamefont {F.}~\bibnamefont {Bonaccorso}}, \ and\ \bibinfo {author} {\bibfnamefont {P.}~\bibnamefont {Samorì}},\ }\bibfield  {title} {\enquote {\bibinfo {title} {Nonvolatile {{Memories Based}} on {{Graphene}} and {{Related 2D Materials}}},}\ }\href {\doibase 10.1002/adma.201806663} {\ \textbf {\bibinfo {volume} {31}},\ \bibinfo {pages} {1806663}}\BibitemShut {NoStop}%
\bibitem [{\citenamefont {Chiang}\ \emph {et~al.}()\citenamefont {Chiang}, \citenamefont {Ostwal}, \citenamefont {Wu}, \citenamefont {Pang}, \citenamefont {Zhang}, \citenamefont {Chen},\ and\ \citenamefont {Appenzeller}}]{chiangMemoryApplications2D2021}%
  \BibitemOpen
  \bibfield  {author} {\bibinfo {author} {\bibfnamefont {C.-C.}\ \bibnamefont {Chiang}}, \bibinfo {author} {\bibfnamefont {V.}~\bibnamefont {Ostwal}}, \bibinfo {author} {\bibfnamefont {P.}~\bibnamefont {Wu}}, \bibinfo {author} {\bibfnamefont {C.-S.}\ \bibnamefont {Pang}}, \bibinfo {author} {\bibfnamefont {F.}~\bibnamefont {Zhang}}, \bibinfo {author} {\bibfnamefont {Z.}~\bibnamefont {Chen}}, \ and\ \bibinfo {author} {\bibfnamefont {J.}~\bibnamefont {Appenzeller}},\ }\bibfield  {title} {\enquote {\bibinfo {title} {Memory applications from {{2D}} materials},}\ }\href {\doibase 10.1063/5.0038013} {\ \textbf {\bibinfo {volume} {8}},\ \bibinfo {pages} {021306}}\BibitemShut {NoStop}%
\bibitem [{\citenamefont {Li}\ \emph {et~al.}(2018)\citenamefont {Li}, \citenamefont {Wang}, \citenamefont {Midya}, \citenamefont {Xia},\ and\ \citenamefont {Yang}}]{li2018review}%
  \BibitemOpen
  \bibfield  {author} {\bibinfo {author} {\bibfnamefont {Y.}~\bibnamefont {Li}}, \bibinfo {author} {\bibfnamefont {Z.}~\bibnamefont {Wang}}, \bibinfo {author} {\bibfnamefont {R.}~\bibnamefont {Midya}}, \bibinfo {author} {\bibfnamefont {Q.}~\bibnamefont {Xia}}, \ and\ \bibinfo {author} {\bibfnamefont {J.~J.}\ \bibnamefont {Yang}},\ }\bibfield  {title} {\enquote {\bibinfo {title} {Review of memristor devices in neuromorphic computing: materials sciences and device challenges},}\ }\href@noop {} {\bibfield  {journal} {\bibinfo  {journal} {Journal of Physics D: Applied Physics}\ }\textbf {\bibinfo {volume} {51}},\ \bibinfo {pages} {503002} (\bibinfo {year} {2018})}\BibitemShut {NoStop}%
\bibitem [{\citenamefont {Tan}\ \emph {et~al.}(2015)\citenamefont {Tan}, \citenamefont {Liu}, \citenamefont {Huang},\ and\ \citenamefont {Zhang}}]{tan2015non}%
  \BibitemOpen
  \bibfield  {author} {\bibinfo {author} {\bibfnamefont {C.}~\bibnamefont {Tan}}, \bibinfo {author} {\bibfnamefont {Z.}~\bibnamefont {Liu}}, \bibinfo {author} {\bibfnamefont {W.}~\bibnamefont {Huang}}, \ and\ \bibinfo {author} {\bibfnamefont {H.}~\bibnamefont {Zhang}},\ }\bibfield  {title} {\enquote {\bibinfo {title} {Non-volatile resistive memory devices based on solution-processed ultrathin two-dimensional nanomaterials},}\ }\href@noop {} {\bibfield  {journal} {\bibinfo  {journal} {Chemical society reviews}\ }\textbf {\bibinfo {volume} {44}},\ \bibinfo {pages} {2615--2628} (\bibinfo {year} {2015})}\BibitemShut {NoStop}%
\bibitem [{\citenamefont {Hao}\ \emph {et~al.}(2016)\citenamefont {Hao}, \citenamefont {Wen}, \citenamefont {Xiang}, \citenamefont {Yuan}, \citenamefont {Yang}, \citenamefont {Li}, \citenamefont {Wang}, \citenamefont {Zeng}, \citenamefont {Wang}, \citenamefont {Liu} \emph {et~al.}}]{hao2016liquid}%
  \BibitemOpen
  \bibfield  {author} {\bibinfo {author} {\bibfnamefont {C.}~\bibnamefont {Hao}}, \bibinfo {author} {\bibfnamefont {F.}~\bibnamefont {Wen}}, \bibinfo {author} {\bibfnamefont {J.}~\bibnamefont {Xiang}}, \bibinfo {author} {\bibfnamefont {S.}~\bibnamefont {Yuan}}, \bibinfo {author} {\bibfnamefont {B.}~\bibnamefont {Yang}}, \bibinfo {author} {\bibfnamefont {L.}~\bibnamefont {Li}}, \bibinfo {author} {\bibfnamefont {W.}~\bibnamefont {Wang}}, \bibinfo {author} {\bibfnamefont {Z.}~\bibnamefont {Zeng}}, \bibinfo {author} {\bibfnamefont {L.}~\bibnamefont {Wang}}, \bibinfo {author} {\bibfnamefont {Z.}~\bibnamefont {Liu}},  \emph {et~al.},\ }\bibfield  {title} {\enquote {\bibinfo {title} {Liquid-exfoliated black phosphorous nanosheet thin films for flexible resistive random access memory applications},}\ }\href@noop {} {\bibfield  {journal} {\bibinfo  {journal} {Advanced Functional Materials}\ }\textbf {\bibinfo {volume} {26}},\ \bibinfo {pages} {2016--2024} (\bibinfo {year} {2016})}\BibitemShut {NoStop}%
\bibitem [{\citenamefont {Wu}\ \emph {et~al.}()\citenamefont {Wu}, \citenamefont {Ge}, \citenamefont {Chen}, \citenamefont {Chou}, \citenamefont {Zhang}, \citenamefont {Zhang}, \citenamefont {Banerjee}, \citenamefont {Chiang}, \citenamefont {Lee},\ and\ \citenamefont {Akinwande}}]{wuThinnestNonvolatileMemory2019}%
  \BibitemOpen
  \bibfield  {author} {\bibinfo {author} {\bibfnamefont {X.}~\bibnamefont {Wu}}, \bibinfo {author} {\bibfnamefont {R.}~\bibnamefont {Ge}}, \bibinfo {author} {\bibfnamefont {P.-A.}\ \bibnamefont {Chen}}, \bibinfo {author} {\bibfnamefont {H.}~\bibnamefont {Chou}}, \bibinfo {author} {\bibfnamefont {Z.}~\bibnamefont {Zhang}}, \bibinfo {author} {\bibfnamefont {Y.}~\bibnamefont {Zhang}}, \bibinfo {author} {\bibfnamefont {S.}~\bibnamefont {Banerjee}}, \bibinfo {author} {\bibfnamefont {M.-H.}\ \bibnamefont {Chiang}}, \bibinfo {author} {\bibfnamefont {J.~C.}\ \bibnamefont {Lee}}, \ and\ \bibinfo {author} {\bibfnamefont {D.}~\bibnamefont {Akinwande}},\ }\bibfield  {title} {\enquote {\bibinfo {title} {Thinnest {{Nonvolatile Memory Based}} on {{Monolayer}} h-{{BN}}},}\ }\href {\doibase 10.1002/adma.201806790} {\ \textbf {\bibinfo {volume} {31}},\ \bibinfo {pages} {1806790}}\BibitemShut {NoStop}%
\bibitem [{\citenamefont {Bessonov}\ \emph {et~al.}(2015)\citenamefont {Bessonov}, \citenamefont {Kirikova}, \citenamefont {Petukhov}, \citenamefont {Allen}, \citenamefont {Ryh{\"a}nen},\ and\ \citenamefont {Bailey}}]{bessonov2015layered}%
  \BibitemOpen
  \bibfield  {author} {\bibinfo {author} {\bibfnamefont {A.~A.}\ \bibnamefont {Bessonov}}, \bibinfo {author} {\bibfnamefont {M.~N.}\ \bibnamefont {Kirikova}}, \bibinfo {author} {\bibfnamefont {D.~I.}\ \bibnamefont {Petukhov}}, \bibinfo {author} {\bibfnamefont {M.}~\bibnamefont {Allen}}, \bibinfo {author} {\bibfnamefont {T.}~\bibnamefont {Ryh{\"a}nen}}, \ and\ \bibinfo {author} {\bibfnamefont {M.~J.}\ \bibnamefont {Bailey}},\ }\bibfield  {title} {\enquote {\bibinfo {title} {Layered memristive and memcapacitive switches for printable electronics},}\ }\href@noop {} {\bibfield  {journal} {\bibinfo  {journal} {Nature materials}\ }\textbf {\bibinfo {volume} {14}},\ \bibinfo {pages} {199--204} (\bibinfo {year} {2015})}\BibitemShut {NoStop}%
\bibitem [{\citenamefont {Son}\ \emph {et~al.}(2016)\citenamefont {Son}, \citenamefont {Chae}, \citenamefont {Kim}, \citenamefont {Choi}, \citenamefont {Yang}, \citenamefont {Park}, \citenamefont {Kale}, \citenamefont {Koo}, \citenamefont {Choi}, \citenamefont {Lee} \emph {et~al.}}]{son2016colloidal}%
  \BibitemOpen
  \bibfield  {author} {\bibinfo {author} {\bibfnamefont {D.}~\bibnamefont {Son}}, \bibinfo {author} {\bibfnamefont {S.~I.}\ \bibnamefont {Chae}}, \bibinfo {author} {\bibfnamefont {M.}~\bibnamefont {Kim}}, \bibinfo {author} {\bibfnamefont {M.~K.}\ \bibnamefont {Choi}}, \bibinfo {author} {\bibfnamefont {J.}~\bibnamefont {Yang}}, \bibinfo {author} {\bibfnamefont {K.}~\bibnamefont {Park}}, \bibinfo {author} {\bibfnamefont {V.~S.}\ \bibnamefont {Kale}}, \bibinfo {author} {\bibfnamefont {J.~H.}\ \bibnamefont {Koo}}, \bibinfo {author} {\bibfnamefont {C.}~\bibnamefont {Choi}}, \bibinfo {author} {\bibfnamefont {M.}~\bibnamefont {Lee}},  \emph {et~al.},\ }\bibfield  {title} {\enquote {\bibinfo {title} {Colloidal synthesis of uniform-sized molybdenum disulfide nanosheets for wafer-scale flexible nonvolatile memory},}\ }\href@noop {} {\bibfield  {journal} {\bibinfo  {journal} {Advanced materials}\ }\textbf {\bibinfo {volume} {28}},\ \bibinfo {pages} {9326--9332} (\bibinfo {year} {2016})}\BibitemShut {NoStop}%
\bibitem [{\citenamefont {Ge}\ \emph {et~al.}(2021)\citenamefont {Ge}, \citenamefont {Wu}, \citenamefont {Liang}, \citenamefont {Hus}, \citenamefont {Gu}, \citenamefont {Okogbue}, \citenamefont {Chou}, \citenamefont {Shi}, \citenamefont {Zhang}, \citenamefont {Banerjee} \emph {et~al.}}]{ge2021library}%
  \BibitemOpen
  \bibfield  {author} {\bibinfo {author} {\bibfnamefont {R.}~\bibnamefont {Ge}}, \bibinfo {author} {\bibfnamefont {X.}~\bibnamefont {Wu}}, \bibinfo {author} {\bibfnamefont {L.}~\bibnamefont {Liang}}, \bibinfo {author} {\bibfnamefont {S.~M.}\ \bibnamefont {Hus}}, \bibinfo {author} {\bibfnamefont {Y.}~\bibnamefont {Gu}}, \bibinfo {author} {\bibfnamefont {E.}~\bibnamefont {Okogbue}}, \bibinfo {author} {\bibfnamefont {H.}~\bibnamefont {Chou}}, \bibinfo {author} {\bibfnamefont {J.}~\bibnamefont {Shi}}, \bibinfo {author} {\bibfnamefont {Y.}~\bibnamefont {Zhang}}, \bibinfo {author} {\bibfnamefont {S.~K.}\ \bibnamefont {Banerjee}},  \emph {et~al.},\ }\bibfield  {title} {\enquote {\bibinfo {title} {A library of atomically thin 2d materials featuring the conductive-point resistive switching phenomenon},}\ }\href@noop {} {\bibfield  {journal} {\bibinfo  {journal} {Advanced Materials}\ }\textbf {\bibinfo {volume} {33}},\ \bibinfo {pages} {2007792} (\bibinfo {year} {2021})}\BibitemShut {NoStop}%
\bibitem [{\citenamefont {Ge}\ \emph {et~al.}(2018)\citenamefont {Ge}, \citenamefont {Wu}, \citenamefont {Kim}, \citenamefont {Shi}, \citenamefont {Sonde}, \citenamefont {Tao}, \citenamefont {Zhang}, \citenamefont {Lee},\ and\ \citenamefont {Akinwande}}]{ge2018atomristor}%
  \BibitemOpen
  \bibfield  {author} {\bibinfo {author} {\bibfnamefont {R.}~\bibnamefont {Ge}}, \bibinfo {author} {\bibfnamefont {X.}~\bibnamefont {Wu}}, \bibinfo {author} {\bibfnamefont {M.}~\bibnamefont {Kim}}, \bibinfo {author} {\bibfnamefont {J.}~\bibnamefont {Shi}}, \bibinfo {author} {\bibfnamefont {S.}~\bibnamefont {Sonde}}, \bibinfo {author} {\bibfnamefont {L.}~\bibnamefont {Tao}}, \bibinfo {author} {\bibfnamefont {Y.}~\bibnamefont {Zhang}}, \bibinfo {author} {\bibfnamefont {J.~C.}\ \bibnamefont {Lee}}, \ and\ \bibinfo {author} {\bibfnamefont {D.}~\bibnamefont {Akinwande}},\ }\bibfield  {title} {\enquote {\bibinfo {title} {Atomristor: nonvolatile resistance switching in atomic sheets of transition metal dichalcogenides},}\ }\href@noop {} {\bibfield  {journal} {\bibinfo  {journal} {Nano letters}\ }\textbf {\bibinfo {volume} {18}},\ \bibinfo {pages} {434--441} (\bibinfo {year} {2018})}\BibitemShut {NoStop}%
\bibitem [{\citenamefont {Kwon}\ \emph {et~al.}(2010)\citenamefont {Kwon}, \citenamefont {Kim}, \citenamefont {Jang}, \citenamefont {Jeon}, \citenamefont {Lee}, \citenamefont {Kim}, \citenamefont {Li}, \citenamefont {Park}, \citenamefont {Lee}, \citenamefont {Han} \emph {et~al.}}]{kwon2010atomic}%
  \BibitemOpen
  \bibfield  {author} {\bibinfo {author} {\bibfnamefont {D.-H.}\ \bibnamefont {Kwon}}, \bibinfo {author} {\bibfnamefont {K.~M.}\ \bibnamefont {Kim}}, \bibinfo {author} {\bibfnamefont {J.~H.}\ \bibnamefont {Jang}}, \bibinfo {author} {\bibfnamefont {J.~M.}\ \bibnamefont {Jeon}}, \bibinfo {author} {\bibfnamefont {M.~H.}\ \bibnamefont {Lee}}, \bibinfo {author} {\bibfnamefont {G.~H.}\ \bibnamefont {Kim}}, \bibinfo {author} {\bibfnamefont {X.-S.}\ \bibnamefont {Li}}, \bibinfo {author} {\bibfnamefont {G.-S.}\ \bibnamefont {Park}}, \bibinfo {author} {\bibfnamefont {B.}~\bibnamefont {Lee}}, \bibinfo {author} {\bibfnamefont {S.}~\bibnamefont {Han}},  \emph {et~al.},\ }\bibfield  {title} {\enquote {\bibinfo {title} {Atomic structure of conducting nanofilaments in tio2 resistive switching memory},}\ }\href@noop {} {\bibfield  {journal} {\bibinfo  {journal} {Nature nanotechnology}\ }\textbf {\bibinfo {volume} {5}},\ \bibinfo {pages} {148--153} (\bibinfo {year} {2010})}\BibitemShut {NoStop}%
\bibitem [{\citenamefont {Yao}\ \emph {et~al.}(2010)\citenamefont {Yao}, \citenamefont {Sun}, \citenamefont {Zhong}, \citenamefont {Natelson},\ and\ \citenamefont {Tour}}]{yao2010resistive}%
  \BibitemOpen
  \bibfield  {author} {\bibinfo {author} {\bibfnamefont {J.}~\bibnamefont {Yao}}, \bibinfo {author} {\bibfnamefont {Z.}~\bibnamefont {Sun}}, \bibinfo {author} {\bibfnamefont {L.}~\bibnamefont {Zhong}}, \bibinfo {author} {\bibfnamefont {D.}~\bibnamefont {Natelson}}, \ and\ \bibinfo {author} {\bibfnamefont {J.~M.}\ \bibnamefont {Tour}},\ }\bibfield  {title} {\enquote {\bibinfo {title} {Resistive switches and memories from silicon oxide},}\ }\href@noop {} {\bibfield  {journal} {\bibinfo  {journal} {Nano letters}\ }\textbf {\bibinfo {volume} {10}},\ \bibinfo {pages} {4105--4110} (\bibinfo {year} {2010})}\BibitemShut {NoStop}%
\bibitem [{\citenamefont {Hus}\ \emph {et~al.}(2021)\citenamefont {Hus}, \citenamefont {Ge}, \citenamefont {Chen}, \citenamefont {Liang}, \citenamefont {Donnelly}, \citenamefont {Ko}, \citenamefont {Huang}, \citenamefont {Chiang}, \citenamefont {Li},\ and\ \citenamefont {Akinwande}}]{hus2021observation}%
  \BibitemOpen
  \bibfield  {author} {\bibinfo {author} {\bibfnamefont {S.~M.}\ \bibnamefont {Hus}}, \bibinfo {author} {\bibfnamefont {R.}~\bibnamefont {Ge}}, \bibinfo {author} {\bibfnamefont {P.-A.}\ \bibnamefont {Chen}}, \bibinfo {author} {\bibfnamefont {L.}~\bibnamefont {Liang}}, \bibinfo {author} {\bibfnamefont {G.~E.}\ \bibnamefont {Donnelly}}, \bibinfo {author} {\bibfnamefont {W.}~\bibnamefont {Ko}}, \bibinfo {author} {\bibfnamefont {F.}~\bibnamefont {Huang}}, \bibinfo {author} {\bibfnamefont {M.-H.}\ \bibnamefont {Chiang}}, \bibinfo {author} {\bibfnamefont {A.-P.}\ \bibnamefont {Li}}, \ and\ \bibinfo {author} {\bibfnamefont {D.}~\bibnamefont {Akinwande}},\ }\bibfield  {title} {\enquote {\bibinfo {title} {Observation of single-defect memristor in an mos2 atomic sheet},}\ }\href {https://www.nature.com/articles/s41565-020-00789-w} {\bibfield  {journal} {\bibinfo  {journal} {Nature Nanotechnology}\ }\textbf {\bibinfo {volume} {16}},\ \bibinfo {pages} {58--62} (\bibinfo {year} {2021})}\BibitemShut {NoStop}%
\bibitem [{\citenamefont {Qiu}\ \emph {et~al.}(2013)\citenamefont {Qiu}, \citenamefont {Xu}, \citenamefont {Wang}, \citenamefont {Ren}, \citenamefont {Nan}, \citenamefont {Ni}, \citenamefont {Chen}, \citenamefont {Yuan}, \citenamefont {Miao}, \citenamefont {Song} \emph {et~al.}}]{qiu2013hopping}%
  \BibitemOpen
  \bibfield  {author} {\bibinfo {author} {\bibfnamefont {H.}~\bibnamefont {Qiu}}, \bibinfo {author} {\bibfnamefont {T.}~\bibnamefont {Xu}}, \bibinfo {author} {\bibfnamefont {Z.}~\bibnamefont {Wang}}, \bibinfo {author} {\bibfnamefont {W.}~\bibnamefont {Ren}}, \bibinfo {author} {\bibfnamefont {H.}~\bibnamefont {Nan}}, \bibinfo {author} {\bibfnamefont {Z.}~\bibnamefont {Ni}}, \bibinfo {author} {\bibfnamefont {Q.}~\bibnamefont {Chen}}, \bibinfo {author} {\bibfnamefont {S.}~\bibnamefont {Yuan}}, \bibinfo {author} {\bibfnamefont {F.}~\bibnamefont {Miao}}, \bibinfo {author} {\bibfnamefont {F.}~\bibnamefont {Song}},  \emph {et~al.},\ }\bibfield  {title} {\enquote {\bibinfo {title} {Hopping transport through defect-induced localized states in molybdenum disulphide},}\ }\href@noop {} {\bibfield  {journal} {\bibinfo  {journal} {Nature communications}\ }\textbf {\bibinfo {volume} {4}},\ \bibinfo {pages} {2642} (\bibinfo {year} {2013})}\BibitemShut {NoStop}%
\bibitem [{\citenamefont {Jena}\ and\ \citenamefont {Konar}(2007)}]{jena2007enhancement}%
  \BibitemOpen
  \bibfield  {author} {\bibinfo {author} {\bibfnamefont {D.}~\bibnamefont {Jena}}\ and\ \bibinfo {author} {\bibfnamefont {A.}~\bibnamefont {Konar}},\ }\bibfield  {title} {\enquote {\bibinfo {title} {Enhancement of carrier mobility in semiconductor nanostructures by dielectric engineering},}\ }\href@noop {} {\bibfield  {journal} {\bibinfo  {journal} {Physical review letters}\ }\textbf {\bibinfo {volume} {98}},\ \bibinfo {pages} {136805} (\bibinfo {year} {2007})}\BibitemShut {NoStop}%
\bibitem [{\citenamefont {Tongay}\ \emph {et~al.}(2013)\citenamefont {Tongay}, \citenamefont {Suh}, \citenamefont {Ataca}, \citenamefont {Fan}, \citenamefont {Luce}, \citenamefont {Kang}, \citenamefont {Liu}, \citenamefont {Ko}, \citenamefont {Raghunathanan}, \citenamefont {Zhou} \emph {et~al.}}]{tongay2013defects}%
  \BibitemOpen
  \bibfield  {author} {\bibinfo {author} {\bibfnamefont {S.}~\bibnamefont {Tongay}}, \bibinfo {author} {\bibfnamefont {J.}~\bibnamefont {Suh}}, \bibinfo {author} {\bibfnamefont {C.}~\bibnamefont {Ataca}}, \bibinfo {author} {\bibfnamefont {W.}~\bibnamefont {Fan}}, \bibinfo {author} {\bibfnamefont {A.}~\bibnamefont {Luce}}, \bibinfo {author} {\bibfnamefont {J.~S.}\ \bibnamefont {Kang}}, \bibinfo {author} {\bibfnamefont {J.}~\bibnamefont {Liu}}, \bibinfo {author} {\bibfnamefont {C.}~\bibnamefont {Ko}}, \bibinfo {author} {\bibfnamefont {R.}~\bibnamefont {Raghunathanan}}, \bibinfo {author} {\bibfnamefont {J.}~\bibnamefont {Zhou}},  \emph {et~al.},\ }\bibfield  {title} {\enquote {\bibinfo {title} {Defects activated photoluminescence in two-dimensional semiconductors: interplay between bound, charged and free excitons},}\ }\href@noop {} {\bibfield  {journal} {\bibinfo  {journal} {Scientific reports}\ }\textbf {\bibinfo {volume} {3}},\ \bibinfo {pages} {2657} (\bibinfo {year} {2013})}\BibitemShut {NoStop}%
\bibitem [{\citenamefont {Korn}\ \emph {et~al.}(2011)\citenamefont {Korn}, \citenamefont {Heydrich}, \citenamefont {Hirmer}, \citenamefont {Schmutzler},\ and\ \citenamefont {Sch{\"u}ller}}]{korn2011low}%
  \BibitemOpen
  \bibfield  {author} {\bibinfo {author} {\bibfnamefont {T.}~\bibnamefont {Korn}}, \bibinfo {author} {\bibfnamefont {S.}~\bibnamefont {Heydrich}}, \bibinfo {author} {\bibfnamefont {M.}~\bibnamefont {Hirmer}}, \bibinfo {author} {\bibfnamefont {J.}~\bibnamefont {Schmutzler}}, \ and\ \bibinfo {author} {\bibfnamefont {C.}~\bibnamefont {Sch{\"u}ller}},\ }\bibfield  {title} {\enquote {\bibinfo {title} {Low-temperature photocarrier dynamics in monolayer mos2},}\ }\href@noop {} {\bibfield  {journal} {\bibinfo  {journal} {Applied Physics Letters}\ }\textbf {\bibinfo {volume} {99}} (\bibinfo {year} {2011})}\BibitemShut {NoStop}%
\bibitem [{\citenamefont {Han}\ \emph {et~al.}(2013)\citenamefont {Han}, \citenamefont {Hwang}, \citenamefont {Kim}, \citenamefont {Yun}, \citenamefont {Lee}, \citenamefont {Park}, \citenamefont {Ryu}, \citenamefont {Park}, \citenamefont {Yoo}, \citenamefont {Yoon} \emph {et~al.}}]{han2013controlling}%
  \BibitemOpen
  \bibfield  {author} {\bibinfo {author} {\bibfnamefont {S.~W.}\ \bibnamefont {Han}}, \bibinfo {author} {\bibfnamefont {Y.~H.}\ \bibnamefont {Hwang}}, \bibinfo {author} {\bibfnamefont {S.-H.}\ \bibnamefont {Kim}}, \bibinfo {author} {\bibfnamefont {W.~S.}\ \bibnamefont {Yun}}, \bibinfo {author} {\bibfnamefont {J.~D.}\ \bibnamefont {Lee}}, \bibinfo {author} {\bibfnamefont {M.~G.}\ \bibnamefont {Park}}, \bibinfo {author} {\bibfnamefont {S.}~\bibnamefont {Ryu}}, \bibinfo {author} {\bibfnamefont {J.~S.}\ \bibnamefont {Park}}, \bibinfo {author} {\bibfnamefont {D.-H.}\ \bibnamefont {Yoo}}, \bibinfo {author} {\bibfnamefont {S.-P.}\ \bibnamefont {Yoon}},  \emph {et~al.},\ }\bibfield  {title} {\enquote {\bibinfo {title} {Controlling ferromagnetic easy axis in a layered mos 2 single crystal},}\ }\href@noop {} {\bibfield  {journal} {\bibinfo  {journal} {Physical review letters}\ }\textbf {\bibinfo {volume} {110}},\ \bibinfo {pages} {247201} (\bibinfo {year} {2013})}\BibitemShut {NoStop}%
\bibitem [{\citenamefont {Zhang}\ \emph {et~al.}(2023)\citenamefont {Zhang}, \citenamefont {Li}, \citenamefont {Cheng}, \citenamefont {Gao}, \citenamefont {Liao},\ and\ \citenamefont {Ying}}]{zhang2023robust}%
  \BibitemOpen
  \bibfield  {author} {\bibinfo {author} {\bibfnamefont {M.}~\bibnamefont {Zhang}}, \bibinfo {author} {\bibfnamefont {Q.}~\bibnamefont {Li}}, \bibinfo {author} {\bibfnamefont {W.}~\bibnamefont {Cheng}}, \bibinfo {author} {\bibfnamefont {Y.}~\bibnamefont {Gao}}, \bibinfo {author} {\bibfnamefont {B.}~\bibnamefont {Liao}}, \ and\ \bibinfo {author} {\bibfnamefont {M.}~\bibnamefont {Ying}},\ }\bibfield  {title} {\enquote {\bibinfo {title} {Robust room-temperature ferromagnetism induced by defect engineering in monolayer mos2},}\ }\href@noop {} {\bibfield  {journal} {\bibinfo  {journal} {Applied Surface Science}\ }\textbf {\bibinfo {volume} {608}},\ \bibinfo {pages} {155220} (\bibinfo {year} {2023})}\BibitemShut {NoStop}%
\bibitem [{\citenamefont {Tumino}\ \emph {et~al.}(2020)\citenamefont {Tumino}, \citenamefont {Casari}, \citenamefont {Li~Bassi},\ and\ \citenamefont {Tosoni}}]{theory_tumino2020nature}%
  \BibitemOpen
  \bibfield  {author} {\bibinfo {author} {\bibfnamefont {F.}~\bibnamefont {Tumino}}, \bibinfo {author} {\bibfnamefont {C.~S.}\ \bibnamefont {Casari}}, \bibinfo {author} {\bibfnamefont {A.}~\bibnamefont {Li~Bassi}}, \ and\ \bibinfo {author} {\bibfnamefont {S.}~\bibnamefont {Tosoni}},\ }\bibfield  {title} {\enquote {\bibinfo {title} {Nature of point defects in single-layer mos2 supported on au (111)},}\ }\href@noop {} {\bibfield  {journal} {\bibinfo  {journal} {The Journal of Physical Chemistry C}\ }\textbf {\bibinfo {volume} {124}},\ \bibinfo {pages} {12424--12431} (\bibinfo {year} {2020})}\BibitemShut {NoStop}%
\bibitem [{\citenamefont {Li}\ \emph {et~al.}(2023)\citenamefont {Li}, \citenamefont {Wang}, \citenamefont {Chen},\ and\ \citenamefont {Li}}]{li2023resistive}%
  \BibitemOpen
  \bibfield  {author} {\bibinfo {author} {\bibfnamefont {X.-D.}\ \bibnamefont {Li}}, \bibinfo {author} {\bibfnamefont {B.-Q.}\ \bibnamefont {Wang}}, \bibinfo {author} {\bibfnamefont {N.-K.}\ \bibnamefont {Chen}}, \ and\ \bibinfo {author} {\bibfnamefont {X.-B.}\ \bibnamefont {Li}},\ }\bibfield  {title} {\enquote {\bibinfo {title} {Resistive switching mechanism of mos2 based atomristor},}\ }\href@noop {} {\bibfield  {journal} {\bibinfo  {journal} {Nanotechnology}\ }\textbf {\bibinfo {volume} {34}},\ \bibinfo {pages} {205201} (\bibinfo {year} {2023})}\BibitemShut {NoStop}%
\bibitem [{\citenamefont {Boschetto}\ \emph {et~al.}(2022)\citenamefont {Boschetto}, \citenamefont {Carapezzi}, \citenamefont {Delacour}, \citenamefont {Abernot}, \citenamefont {Gil},\ and\ \citenamefont {Todri-Sanial}}]{boschetto2022ab_1}%
  \BibitemOpen
  \bibfield  {author} {\bibinfo {author} {\bibfnamefont {G.}~\bibnamefont {Boschetto}}, \bibinfo {author} {\bibfnamefont {S.}~\bibnamefont {Carapezzi}}, \bibinfo {author} {\bibfnamefont {C.}~\bibnamefont {Delacour}}, \bibinfo {author} {\bibfnamefont {M.}~\bibnamefont {Abernot}}, \bibinfo {author} {\bibfnamefont {T.}~\bibnamefont {Gil}}, \ and\ \bibinfo {author} {\bibfnamefont {A.}~\bibnamefont {Todri-Sanial}},\ }\bibfield  {title} {\enquote {\bibinfo {title} {Ab initio computer simulations on interfacial properties of single-layer mos2 and au contacts for two-dimensional nanodevices},}\ }\href {https://pubs.acs.org/doi/epdf/10.1021/acsanm.2c00995} {\bibfield  {journal} {\bibinfo  {journal} {ACS Applied Nano Materials}\ }\textbf {\bibinfo {volume} {5}},\ \bibinfo {pages} {10192--10202} (\bibinfo {year} {2022})}\BibitemShut {NoStop}%
\bibitem [{\citenamefont {Boschetto}, \citenamefont {Carapezzi},\ and\ \citenamefont {Todri-Sanial}(2023)}]{boschetto2023non_2}%
  \BibitemOpen
  \bibfield  {author} {\bibinfo {author} {\bibfnamefont {G.}~\bibnamefont {Boschetto}}, \bibinfo {author} {\bibfnamefont {S.}~\bibnamefont {Carapezzi}}, \ and\ \bibinfo {author} {\bibfnamefont {A.}~\bibnamefont {Todri-Sanial}},\ }\bibfield  {title} {\enquote {\bibinfo {title} {Non-volatile resistive switching mechanism in single-layer mos 2 memristors: Insights from ab initio modelling of au and mos 2 interfaces},}\ }\href {https://pubs.rsc.org/en/content/articlepdf/2023/na/d3na00045a} {\bibfield  {journal} {\bibinfo  {journal} {Nanoscale Advances}\ } (\bibinfo {year} {2023})}\BibitemShut {NoStop}%
\bibitem [{\citenamefont {Yasuda}\ \emph {et~al.}(2017)\citenamefont {Yasuda}, \citenamefont {Takahashi}, \citenamefont {Osaka}, \citenamefont {Kumagai}, \citenamefont {Miyata}, \citenamefont {Okada}, \citenamefont {Hayamizu},\ and\ \citenamefont {Murakoshi}}]{yasuda2017out}%
  \BibitemOpen
  \bibfield  {author} {\bibinfo {author} {\bibfnamefont {S.}~\bibnamefont {Yasuda}}, \bibinfo {author} {\bibfnamefont {R.}~\bibnamefont {Takahashi}}, \bibinfo {author} {\bibfnamefont {R.}~\bibnamefont {Osaka}}, \bibinfo {author} {\bibfnamefont {R.}~\bibnamefont {Kumagai}}, \bibinfo {author} {\bibfnamefont {Y.}~\bibnamefont {Miyata}}, \bibinfo {author} {\bibfnamefont {S.}~\bibnamefont {Okada}}, \bibinfo {author} {\bibfnamefont {Y.}~\bibnamefont {Hayamizu}}, \ and\ \bibinfo {author} {\bibfnamefont {K.}~\bibnamefont {Murakoshi}},\ }\bibfield  {title} {\enquote {\bibinfo {title} {Out-of-plane strain induced in a moir{\'e} superstructure of monolayer mos2 and mose2 on au (111)},}\ }\href@noop {} {\bibfield  {journal} {\bibinfo  {journal} {Small}\ }\textbf {\bibinfo {volume} {13}},\ \bibinfo {pages} {1700748} (\bibinfo {year} {2017})}\BibitemShut {NoStop}%
\bibitem [{\citenamefont {McDonnell}\ \emph {et~al.}(2014)\citenamefont {McDonnell}, \citenamefont {Addou}, \citenamefont {Buie}, \citenamefont {Wallace},\ and\ \citenamefont {Hinkle}}]{mcdonnell2014defect}%
  \BibitemOpen
  \bibfield  {author} {\bibinfo {author} {\bibfnamefont {S.}~\bibnamefont {McDonnell}}, \bibinfo {author} {\bibfnamefont {R.}~\bibnamefont {Addou}}, \bibinfo {author} {\bibfnamefont {C.}~\bibnamefont {Buie}}, \bibinfo {author} {\bibfnamefont {R.~M.}\ \bibnamefont {Wallace}}, \ and\ \bibinfo {author} {\bibfnamefont {C.~L.}\ \bibnamefont {Hinkle}},\ }\bibfield  {title} {\enquote {\bibinfo {title} {Defect-dominated doping and contact resistance in mos2},}\ }\href@noop {} {\bibfield  {journal} {\bibinfo  {journal} {ACS nano}\ }\textbf {\bibinfo {volume} {8}},\ \bibinfo {pages} {2880--2888} (\bibinfo {year} {2014})}\BibitemShut {NoStop}%
\bibitem [{\citenamefont {Cook}\ \emph {et~al.}(2015)\citenamefont {Cook}, \citenamefont {Palandech}, \citenamefont {Doore}, \citenamefont {Ye}, \citenamefont {Ye}, \citenamefont {He},\ and\ \citenamefont {Stollenwerk}}]{cook2015influence}%
  \BibitemOpen
  \bibfield  {author} {\bibinfo {author} {\bibfnamefont {M.}~\bibnamefont {Cook}}, \bibinfo {author} {\bibfnamefont {R.}~\bibnamefont {Palandech}}, \bibinfo {author} {\bibfnamefont {K.}~\bibnamefont {Doore}}, \bibinfo {author} {\bibfnamefont {Z.}~\bibnamefont {Ye}}, \bibinfo {author} {\bibfnamefont {G.}~\bibnamefont {Ye}}, \bibinfo {author} {\bibfnamefont {R.}~\bibnamefont {He}}, \ and\ \bibinfo {author} {\bibfnamefont {A.~J.}\ \bibnamefont {Stollenwerk}},\ }\bibfield  {title} {\enquote {\bibinfo {title} {Influence of interface coupling on the electronic properties of the au/mos 2 junction},}\ }\href@noop {} {\bibfield  {journal} {\bibinfo  {journal} {Physical Review B}\ }\textbf {\bibinfo {volume} {92}},\ \bibinfo {pages} {201302} (\bibinfo {year} {2015})}\BibitemShut {NoStop}%
\bibitem [{\citenamefont {Carladous}\ \emph {et~al.}(2002)\citenamefont {Carladous}, \citenamefont {Coratger}, \citenamefont {Ajustron}, \citenamefont {Seine}, \citenamefont {P{\'e}chou},\ and\ \citenamefont {Beauvillain}}]{carladous2002light}%
  \BibitemOpen
  \bibfield  {author} {\bibinfo {author} {\bibfnamefont {A.}~\bibnamefont {Carladous}}, \bibinfo {author} {\bibfnamefont {R.}~\bibnamefont {Coratger}}, \bibinfo {author} {\bibfnamefont {F.}~\bibnamefont {Ajustron}}, \bibinfo {author} {\bibfnamefont {G.}~\bibnamefont {Seine}}, \bibinfo {author} {\bibfnamefont {R.}~\bibnamefont {P{\'e}chou}}, \ and\ \bibinfo {author} {\bibfnamefont {J.}~\bibnamefont {Beauvillain}},\ }\bibfield  {title} {\enquote {\bibinfo {title} {Light emission from spectral analysis of au/mos 2 nanocontacts stimulated by scanning tunneling microscopy},}\ }\href@noop {} {\bibfield  {journal} {\bibinfo  {journal} {Physical Review B}\ }\textbf {\bibinfo {volume} {66}},\ \bibinfo {pages} {045401} (\bibinfo {year} {2002})}\BibitemShut {NoStop}%
\bibitem [{\citenamefont {Bruix}\ \emph {et~al.}(2016)\citenamefont {Bruix}, \citenamefont {Miwa}, \citenamefont {Hauptmann}, \citenamefont {Wegner}, \citenamefont {Ulstrup}, \citenamefont {Gr{\o}nborg}, \citenamefont {Sanders}, \citenamefont {Dendzik}, \citenamefont {{\v{C}}abo}, \citenamefont {Bianchi} \emph {et~al.}}]{bruix2016single}%
  \BibitemOpen
  \bibfield  {author} {\bibinfo {author} {\bibfnamefont {A.}~\bibnamefont {Bruix}}, \bibinfo {author} {\bibfnamefont {J.~A.}\ \bibnamefont {Miwa}}, \bibinfo {author} {\bibfnamefont {N.}~\bibnamefont {Hauptmann}}, \bibinfo {author} {\bibfnamefont {D.}~\bibnamefont {Wegner}}, \bibinfo {author} {\bibfnamefont {S.}~\bibnamefont {Ulstrup}}, \bibinfo {author} {\bibfnamefont {S.~S.}\ \bibnamefont {Gr{\o}nborg}}, \bibinfo {author} {\bibfnamefont {C.~E.}\ \bibnamefont {Sanders}}, \bibinfo {author} {\bibfnamefont {M.}~\bibnamefont {Dendzik}}, \bibinfo {author} {\bibfnamefont {A.~G.}\ \bibnamefont {{\v{C}}abo}}, \bibinfo {author} {\bibfnamefont {M.}~\bibnamefont {Bianchi}},  \emph {et~al.},\ }\bibfield  {title} {\enquote {\bibinfo {title} {Single-layer mos 2 on au (111): Band gap renormalization and substrate interaction},}\ }\href@noop {} {\bibfield  {journal} {\bibinfo  {journal} {Physical Review B}\ }\textbf {\bibinfo {volume} {93}},\ \bibinfo {pages} {165422} (\bibinfo {year} {2016})}\BibitemShut {NoStop}%
\bibitem [{\citenamefont {Gong}\ \emph {et~al.}(2013)\citenamefont {Gong}, \citenamefont {Huang}, \citenamefont {Miller}, \citenamefont {Cheng}, \citenamefont {Hao}, \citenamefont {Cobden}, \citenamefont {Kim}, \citenamefont {Ruoff}, \citenamefont {Wallace}, \citenamefont {Cho} \emph {et~al.}}]{gong2013metal}%
  \BibitemOpen
  \bibfield  {author} {\bibinfo {author} {\bibfnamefont {C.}~\bibnamefont {Gong}}, \bibinfo {author} {\bibfnamefont {C.}~\bibnamefont {Huang}}, \bibinfo {author} {\bibfnamefont {J.}~\bibnamefont {Miller}}, \bibinfo {author} {\bibfnamefont {L.}~\bibnamefont {Cheng}}, \bibinfo {author} {\bibfnamefont {Y.}~\bibnamefont {Hao}}, \bibinfo {author} {\bibfnamefont {D.}~\bibnamefont {Cobden}}, \bibinfo {author} {\bibfnamefont {J.}~\bibnamefont {Kim}}, \bibinfo {author} {\bibfnamefont {R.~S.}\ \bibnamefont {Ruoff}}, \bibinfo {author} {\bibfnamefont {R.~M.}\ \bibnamefont {Wallace}}, \bibinfo {author} {\bibfnamefont {K.}~\bibnamefont {Cho}},  \emph {et~al.},\ }\bibfield  {title} {\enquote {\bibinfo {title} {Metal contacts on physical vapor deposited monolayer mos2},}\ }\href {https://pubs.acs.org/doi/epdf/10.1021/nn4052138} {\bibfield  {journal} {\bibinfo  {journal} {ACS nano}\ }\textbf {\bibinfo {volume} {7}},\ \bibinfo {pages} {11350--11357} (\bibinfo {year} {2013})}\BibitemShut {NoStop}%
\bibitem [{\citenamefont {Bruix}\ \emph {et~al.}(2015)\citenamefont {Bruix}, \citenamefont {Fuchtbauer}, \citenamefont {Tuxen}, \citenamefont {Walton}, \citenamefont {Andersen}, \citenamefont {Porsgaard}, \citenamefont {Besenbacher}, \citenamefont {Hammer},\ and\ \citenamefont {Lauritsen}}]{bruix2015situ}%
  \BibitemOpen
  \bibfield  {author} {\bibinfo {author} {\bibfnamefont {A.}~\bibnamefont {Bruix}}, \bibinfo {author} {\bibfnamefont {H.~G.}\ \bibnamefont {Fuchtbauer}}, \bibinfo {author} {\bibfnamefont {A.~K.}\ \bibnamefont {Tuxen}}, \bibinfo {author} {\bibfnamefont {A.~S.}\ \bibnamefont {Walton}}, \bibinfo {author} {\bibfnamefont {M.}~\bibnamefont {Andersen}}, \bibinfo {author} {\bibfnamefont {S.}~\bibnamefont {Porsgaard}}, \bibinfo {author} {\bibfnamefont {F.}~\bibnamefont {Besenbacher}}, \bibinfo {author} {\bibfnamefont {B.}~\bibnamefont {Hammer}}, \ and\ \bibinfo {author} {\bibfnamefont {J.~V.}\ \bibnamefont {Lauritsen}},\ }\bibfield  {title} {\enquote {\bibinfo {title} {In situ detection of active edge sites in single-layer mos2 catalysts},}\ }\href@noop {} {\bibfield  {journal} {\bibinfo  {journal} {ACS nano}\ }\textbf {\bibinfo {volume} {9}},\ \bibinfo {pages} {9322--9330} (\bibinfo {year} {2015})}\BibitemShut {NoStop}%
\bibitem [{\citenamefont {Velický}\ \emph {et~al.}()\citenamefont {Velický}, \citenamefont {Donnelly}, \citenamefont {Hendren}, \citenamefont {McFarland}, \citenamefont {Scullion}, \citenamefont {DeBenedetti}, \citenamefont {Correa}, \citenamefont {Han}, \citenamefont {Wain}, \citenamefont {Hines}, \citenamefont {Muller}, \citenamefont {Novoselov}, \citenamefont {Abruña}, \citenamefont {Bowman}, \citenamefont {Santos},\ and\ \citenamefont {Huang}}]{velickyMechanismGoldAssistedExfoliation2018}%
  \BibitemOpen
  \bibfield  {author} {\bibinfo {author} {\bibfnamefont {M.}~\bibnamefont {Velický}}, \bibinfo {author} {\bibfnamefont {G.~E.}\ \bibnamefont {Donnelly}}, \bibinfo {author} {\bibfnamefont {W.~R.}\ \bibnamefont {Hendren}}, \bibinfo {author} {\bibfnamefont {S.}~\bibnamefont {McFarland}}, \bibinfo {author} {\bibfnamefont {D.}~\bibnamefont {Scullion}}, \bibinfo {author} {\bibfnamefont {W.~J.~I.}\ \bibnamefont {DeBenedetti}}, \bibinfo {author} {\bibfnamefont {G.~C.}\ \bibnamefont {Correa}}, \bibinfo {author} {\bibfnamefont {Y.}~\bibnamefont {Han}}, \bibinfo {author} {\bibfnamefont {A.~J.}\ \bibnamefont {Wain}}, \bibinfo {author} {\bibfnamefont {M.~A.}\ \bibnamefont {Hines}}, \bibinfo {author} {\bibfnamefont {D.~A.}\ \bibnamefont {Muller}}, \bibinfo {author} {\bibfnamefont {K.~S.}\ \bibnamefont {Novoselov}}, \bibinfo {author} {\bibfnamefont {H.~D.}\ \bibnamefont {Abruña}}, \bibinfo {author} {\bibfnamefont {R.~M.}\ \bibnamefont {Bowman}}, \bibinfo {author} {\bibfnamefont {E.~J.~G.}\ \bibnamefont {Santos}}, \ and\
  \bibinfo {author} {\bibfnamefont {F.}~\bibnamefont {Huang}},\ }\bibfield  {title} {\enquote {\bibinfo {title} {Mechanism of {{Gold-Assisted Exfoliation}} of {{Centimeter-Sized Transition-Metal Dichalcogenide Monolayers}}},}\ }\href {\doibase 10.1021/acsnano.8b06101} {\ \textbf {\bibinfo {volume} {12}},\ \bibinfo {pages} {10463--10472}}\BibitemShut {NoStop}%
\bibitem [{\citenamefont {Ozaki}(2003)}]{opmx_ozaki2003variationally}%
  \BibitemOpen
  \bibfield  {author} {\bibinfo {author} {\bibfnamefont {T.}~\bibnamefont {Ozaki}},\ }\bibfield  {title} {\enquote {\bibinfo {title} {Variationally optimized atomic orbitals for large-scale electronic structures},}\ }\href@noop {} {\bibfield  {journal} {\bibinfo  {journal} {Physical Review B}\ }\textbf {\bibinfo {volume} {67}},\ \bibinfo {pages} {155108} (\bibinfo {year} {2003})}\BibitemShut {NoStop}%
\bibitem [{\citenamefont {Ozaki}\ and\ \citenamefont {Kino}(2005)}]{opmx_ozaki2005efficient}%
  \BibitemOpen
  \bibfield  {author} {\bibinfo {author} {\bibfnamefont {T.}~\bibnamefont {Ozaki}}\ and\ \bibinfo {author} {\bibfnamefont {H.}~\bibnamefont {Kino}},\ }\bibfield  {title} {\enquote {\bibinfo {title} {Efficient projector expansion for the ab initio lcao method},}\ }\href@noop {} {\bibfield  {journal} {\bibinfo  {journal} {Physical Review B}\ }\textbf {\bibinfo {volume} {72}},\ \bibinfo {pages} {045121} (\bibinfo {year} {2005})}\BibitemShut {NoStop}%
\bibitem [{\citenamefont {Lejaeghere}\ \emph {et~al.}(2016)\citenamefont {Lejaeghere}, \citenamefont {Bihlmayer}, \citenamefont {Bj{\"o}rkman}, \citenamefont {Blaha}, \citenamefont {Bl{\"u}gel}, \citenamefont {Blum}, \citenamefont {Caliste}, \citenamefont {Castelli}, \citenamefont {Clark}, \citenamefont {Dal~Corso} \emph {et~al.}}]{opmx_lejaeghere2016reproducibility}%
  \BibitemOpen
  \bibfield  {author} {\bibinfo {author} {\bibfnamefont {K.}~\bibnamefont {Lejaeghere}}, \bibinfo {author} {\bibfnamefont {G.}~\bibnamefont {Bihlmayer}}, \bibinfo {author} {\bibfnamefont {T.}~\bibnamefont {Bj{\"o}rkman}}, \bibinfo {author} {\bibfnamefont {P.}~\bibnamefont {Blaha}}, \bibinfo {author} {\bibfnamefont {S.}~\bibnamefont {Bl{\"u}gel}}, \bibinfo {author} {\bibfnamefont {V.}~\bibnamefont {Blum}}, \bibinfo {author} {\bibfnamefont {D.}~\bibnamefont {Caliste}}, \bibinfo {author} {\bibfnamefont {I.~E.}\ \bibnamefont {Castelli}}, \bibinfo {author} {\bibfnamefont {S.~J.}\ \bibnamefont {Clark}}, \bibinfo {author} {\bibfnamefont {A.}~\bibnamefont {Dal~Corso}},  \emph {et~al.},\ }\bibfield  {title} {\enquote {\bibinfo {title} {Reproducibility in density functional theory calculations of solids},}\ }\href@noop {} {\bibfield  {journal} {\bibinfo  {journal} {Science}\ }\textbf {\bibinfo {volume} {351}},\ \bibinfo {pages} {aad3000} (\bibinfo {year} {2016})}\BibitemShut {NoStop}%
\bibitem [{\citenamefont {Morrison}, \citenamefont {Bylander},\ and\ \citenamefont {Kleinman}(1993)}]{opmx_morrison1993nonlocal}%
  \BibitemOpen
  \bibfield  {author} {\bibinfo {author} {\bibfnamefont {I.}~\bibnamefont {Morrison}}, \bibinfo {author} {\bibfnamefont {D.}~\bibnamefont {Bylander}}, \ and\ \bibinfo {author} {\bibfnamefont {L.}~\bibnamefont {Kleinman}},\ }\bibfield  {title} {\enquote {\bibinfo {title} {Nonlocal hermitian norm-conserving vanderbilt pseudopotential},}\ }\href@noop {} {\bibfield  {journal} {\bibinfo  {journal} {Physical Review B}\ }\textbf {\bibinfo {volume} {47}},\ \bibinfo {pages} {6728} (\bibinfo {year} {1993})}\BibitemShut {NoStop}%
\bibitem [{\citenamefont {Perdew}, \citenamefont {Burke},\ and\ \citenamefont {Ernzerhof}(1996)}]{opmx_perdew1996generalized}%
  \BibitemOpen
  \bibfield  {author} {\bibinfo {author} {\bibfnamefont {J.~P.}\ \bibnamefont {Perdew}}, \bibinfo {author} {\bibfnamefont {K.}~\bibnamefont {Burke}}, \ and\ \bibinfo {author} {\bibfnamefont {M.}~\bibnamefont {Ernzerhof}},\ }\bibfield  {title} {\enquote {\bibinfo {title} {Generalized gradient approximation made simple},}\ }\href@noop {} {\bibfield  {journal} {\bibinfo  {journal} {Physical review letters}\ }\textbf {\bibinfo {volume} {77}},\ \bibinfo {pages} {3865} (\bibinfo {year} {1996})}\BibitemShut {NoStop}%
\bibitem [{\citenamefont {Grimme}, \citenamefont {Ehrlich},\ and\ \citenamefont {Goerigk}(2011)}]{grimme2011effect}%
  \BibitemOpen
  \bibfield  {author} {\bibinfo {author} {\bibfnamefont {S.}~\bibnamefont {Grimme}}, \bibinfo {author} {\bibfnamefont {S.}~\bibnamefont {Ehrlich}}, \ and\ \bibinfo {author} {\bibfnamefont {L.}~\bibnamefont {Goerigk}},\ }\bibfield  {title} {\enquote {\bibinfo {title} {Effect of the damping function in dispersion corrected density functional theory},}\ }\href@noop {} {\bibfield  {journal} {\bibinfo  {journal} {Journal of computational chemistry}\ }\textbf {\bibinfo {volume} {32}},\ \bibinfo {pages} {1456--1465} (\bibinfo {year} {2011})}\BibitemShut {NoStop}%
\bibitem [{\citenamefont {Otani}\ and\ \citenamefont {Sugino}(2006{\natexlab{a}})}]{opmx_otani2006first}%
  \BibitemOpen
  \bibfield  {author} {\bibinfo {author} {\bibfnamefont {M.}~\bibnamefont {Otani}}\ and\ \bibinfo {author} {\bibfnamefont {O.}~\bibnamefont {Sugino}},\ }\bibfield  {title} {\enquote {\bibinfo {title} {First-principles calculations of charged surfaces and interfaces: A plane-wave nonrepeated slab approach},}\ }\href@noop {} {\bibfield  {journal} {\bibinfo  {journal} {Physical Review B}\ }\textbf {\bibinfo {volume} {73}},\ \bibinfo {pages} {115407} (\bibinfo {year} {2006}{\natexlab{a}})}\BibitemShut {NoStop}%
\bibitem [{\citenamefont {Lee}, \citenamefont {Yamada-Takamura},\ and\ \citenamefont {Ozaki}(2013)}]{lee2013unfolding}%
  \BibitemOpen
  \bibfield  {author} {\bibinfo {author} {\bibfnamefont {C.-C.}\ \bibnamefont {Lee}}, \bibinfo {author} {\bibfnamefont {Y.}~\bibnamefont {Yamada-Takamura}}, \ and\ \bibinfo {author} {\bibfnamefont {T.}~\bibnamefont {Ozaki}},\ }\bibfield  {title} {\enquote {\bibinfo {title} {Unfolding method for first-principles lcao electronic structure calculations},}\ }\href@noop {} {\bibfield  {journal} {\bibinfo  {journal} {Journal of Physics: Condensed Matter}\ }\textbf {\bibinfo {volume} {25}},\ \bibinfo {pages} {345501} (\bibinfo {year} {2013})}\BibitemShut {NoStop}%
\bibitem [{\citenamefont {Otani}\ and\ \citenamefont {Sugino}(2006{\natexlab{b}})}]{otani2006first}%
  \BibitemOpen
  \bibfield  {author} {\bibinfo {author} {\bibfnamefont {M.}~\bibnamefont {Otani}}\ and\ \bibinfo {author} {\bibfnamefont {O.}~\bibnamefont {Sugino}},\ }\bibfield  {title} {\enquote {\bibinfo {title} {First-principles calculations of charged surfaces and interfaces: A plane-wave nonrepeated slab approach},}\ }\href@noop {} {\bibfield  {journal} {\bibinfo  {journal} {Physical Review B}\ }\textbf {\bibinfo {volume} {73}},\ \bibinfo {pages} {115407} (\bibinfo {year} {2006}{\natexlab{b}})}\BibitemShut {NoStop}%
\bibitem [{\citenamefont {Jain}\ \emph {et~al.}()\citenamefont {Jain}, \citenamefont {Ong}, \citenamefont {Hautier}, \citenamefont {Chen}, \citenamefont {Richards}, \citenamefont {Dacek}, \citenamefont {Cholia}, \citenamefont {Gunter}, \citenamefont {Skinner}, \citenamefont {Ceder},\ and\ \citenamefont {Persson}}]{jainCommentaryMaterialsProject2013}%
  \BibitemOpen
  \bibfield  {author} {\bibinfo {author} {\bibfnamefont {A.}~\bibnamefont {Jain}}, \bibinfo {author} {\bibfnamefont {S.~P.}\ \bibnamefont {Ong}}, \bibinfo {author} {\bibfnamefont {G.}~\bibnamefont {Hautier}}, \bibinfo {author} {\bibfnamefont {W.}~\bibnamefont {Chen}}, \bibinfo {author} {\bibfnamefont {W.~D.}\ \bibnamefont {Richards}}, \bibinfo {author} {\bibfnamefont {S.}~\bibnamefont {Dacek}}, \bibinfo {author} {\bibfnamefont {S.}~\bibnamefont {Cholia}}, \bibinfo {author} {\bibfnamefont {D.}~\bibnamefont {Gunter}}, \bibinfo {author} {\bibfnamefont {D.}~\bibnamefont {Skinner}}, \bibinfo {author} {\bibfnamefont {G.}~\bibnamefont {Ceder}}, \ and\ \bibinfo {author} {\bibfnamefont {K.~A.}\ \bibnamefont {Persson}},\ }\bibfield  {title} {\enquote {\bibinfo {title} {Commentary: {{The Materials Project}}: {{A}} materials genome approach to accelerating materials innovation},}\ }\href {\doibase 10.1063/1.4812323} {\ \textbf {\bibinfo {volume} {1}},\ \bibinfo {pages} {011002}}\BibitemShut {NoStop}%
\bibitem [{\citenamefont {Freysoldt}\ \emph {et~al.}()\citenamefont {Freysoldt}, \citenamefont {Grabowski}, \citenamefont {Hickel}, \citenamefont {Neugebauer}, \citenamefont {Kresse}, \citenamefont {Janotti},\ and\ \citenamefont {Van~de Walle}}]{freysoldtFirstprinciplesCalculationsPoint2014}%
  \BibitemOpen
  \bibfield  {author} {\bibinfo {author} {\bibfnamefont {C.}~\bibnamefont {Freysoldt}}, \bibinfo {author} {\bibfnamefont {B.}~\bibnamefont {Grabowski}}, \bibinfo {author} {\bibfnamefont {T.}~\bibnamefont {Hickel}}, \bibinfo {author} {\bibfnamefont {J.}~\bibnamefont {Neugebauer}}, \bibinfo {author} {\bibfnamefont {G.}~\bibnamefont {Kresse}}, \bibinfo {author} {\bibfnamefont {A.}~\bibnamefont {Janotti}}, \ and\ \bibinfo {author} {\bibfnamefont {C.~G.}\ \bibnamefont {Van~de Walle}},\ }\bibfield  {title} {\enquote {\bibinfo {title} {First-principles calculations for point defects in solids},}\ }\href {\doibase 10.1103/RevModPhys.86.253} {\ \textbf {\bibinfo {volume} {86}},\ \bibinfo {pages} {253--305}}\BibitemShut {NoStop}%
\bibitem [{\citenamefont {Madsen}\ and\ \citenamefont {Singh}(2006)}]{madsen2006boltztrap}%
  \BibitemOpen
  \bibfield  {author} {\bibinfo {author} {\bibfnamefont {G.~K.}\ \bibnamefont {Madsen}}\ and\ \bibinfo {author} {\bibfnamefont {D.~J.}\ \bibnamefont {Singh}},\ }\bibfield  {title} {\enquote {\bibinfo {title} {Boltztrap. a code for calculating band-structure dependent quantities},}\ }\href@noop {} {\bibfield  {journal} {\bibinfo  {journal} {Computer Physics Communications}\ }\textbf {\bibinfo {volume} {175}},\ \bibinfo {pages} {67--71} (\bibinfo {year} {2006})}\BibitemShut {NoStop}%
\bibitem [{\citenamefont {Jin}\ \emph {et~al.}(2015)\citenamefont {Jin}, \citenamefont {Liao}, \citenamefont {Fang}, \citenamefont {Liu}, \citenamefont {Liu}, \citenamefont {Ding}, \citenamefont {Luo},\ and\ \citenamefont {Yang}}]{jin2015revisit}%
  \BibitemOpen
  \bibfield  {author} {\bibinfo {author} {\bibfnamefont {Z.}~\bibnamefont {Jin}}, \bibinfo {author} {\bibfnamefont {Q.}~\bibnamefont {Liao}}, \bibinfo {author} {\bibfnamefont {H.}~\bibnamefont {Fang}}, \bibinfo {author} {\bibfnamefont {Z.}~\bibnamefont {Liu}}, \bibinfo {author} {\bibfnamefont {W.}~\bibnamefont {Liu}}, \bibinfo {author} {\bibfnamefont {Z.}~\bibnamefont {Ding}}, \bibinfo {author} {\bibfnamefont {T.}~\bibnamefont {Luo}}, \ and\ \bibinfo {author} {\bibfnamefont {N.}~\bibnamefont {Yang}},\ }\bibfield  {title} {\enquote {\bibinfo {title} {A revisit to high thermoelectric performance of single-layer mos 2},}\ }\href@noop {} {\bibfield  {journal} {\bibinfo  {journal} {Scientific reports}\ }\textbf {\bibinfo {volume} {5}},\ \bibinfo {pages} {18342} (\bibinfo {year} {2015})}\BibitemShut {NoStop}%
\bibitem [{\citenamefont {Guo}\ \emph {et~al.}(2013)\citenamefont {Guo}, \citenamefont {Yang}, \citenamefont {Tao}, \citenamefont {Wang},\ and\ \citenamefont {Zhang}}]{guo2013high}%
  \BibitemOpen
  \bibfield  {author} {\bibinfo {author} {\bibfnamefont {H.}~\bibnamefont {Guo}}, \bibinfo {author} {\bibfnamefont {T.}~\bibnamefont {Yang}}, \bibinfo {author} {\bibfnamefont {P.}~\bibnamefont {Tao}}, \bibinfo {author} {\bibfnamefont {Y.}~\bibnamefont {Wang}}, \ and\ \bibinfo {author} {\bibfnamefont {Z.}~\bibnamefont {Zhang}},\ }\bibfield  {title} {\enquote {\bibinfo {title} {High pressure effect on structure, electronic structure, and thermoelectric properties of mos2},}\ }\href@noop {} {\bibfield  {journal} {\bibinfo  {journal} {Journal of Applied Physics}\ }\textbf {\bibinfo {volume} {113}} (\bibinfo {year} {2013})}\BibitemShut {NoStop}%
\bibitem [{\citenamefont {Ponc{\'e}}\ \emph {et~al.}(2021)\citenamefont {Ponc{\'e}}, \citenamefont {Macheda}, \citenamefont {Margine}, \citenamefont {Marzari}, \citenamefont {Bonini},\ and\ \citenamefont {Giustino}}]{ponce2021first}%
  \BibitemOpen
  \bibfield  {author} {\bibinfo {author} {\bibfnamefont {S.}~\bibnamefont {Ponc{\'e}}}, \bibinfo {author} {\bibfnamefont {F.}~\bibnamefont {Macheda}}, \bibinfo {author} {\bibfnamefont {E.~R.}\ \bibnamefont {Margine}}, \bibinfo {author} {\bibfnamefont {N.}~\bibnamefont {Marzari}}, \bibinfo {author} {\bibfnamefont {N.}~\bibnamefont {Bonini}}, \ and\ \bibinfo {author} {\bibfnamefont {F.}~\bibnamefont {Giustino}},\ }\bibfield  {title} {\enquote {\bibinfo {title} {First-principles predictions of hall and drift mobilities in semiconductors},}\ }\href@noop {} {\bibfield  {journal} {\bibinfo  {journal} {Physical Review Research}\ }\textbf {\bibinfo {volume} {3}},\ \bibinfo {pages} {043022} (\bibinfo {year} {2021})}\BibitemShut {NoStop}%
\bibitem [{\citenamefont {Ganose}\ \emph {et~al.}(2021)\citenamefont {Ganose}, \citenamefont {Park}, \citenamefont {Faghaninia}, \citenamefont {Woods-Robinson}, \citenamefont {Persson},\ and\ \citenamefont {Jain}}]{ganose2021efficient}%
  \BibitemOpen
  \bibfield  {author} {\bibinfo {author} {\bibfnamefont {A.~M.}\ \bibnamefont {Ganose}}, \bibinfo {author} {\bibfnamefont {J.}~\bibnamefont {Park}}, \bibinfo {author} {\bibfnamefont {A.}~\bibnamefont {Faghaninia}}, \bibinfo {author} {\bibfnamefont {R.}~\bibnamefont {Woods-Robinson}}, \bibinfo {author} {\bibfnamefont {K.~A.}\ \bibnamefont {Persson}}, \ and\ \bibinfo {author} {\bibfnamefont {A.}~\bibnamefont {Jain}},\ }\bibfield  {title} {\enquote {\bibinfo {title} {Efficient calculation of carrier scattering rates from first principles},}\ }\href@noop {} {\bibfield  {journal} {\bibinfo  {journal} {Nature communications}\ }\textbf {\bibinfo {volume} {12}},\ \bibinfo {pages} {2222} (\bibinfo {year} {2021})}\BibitemShut {NoStop}%
\bibitem [{\citenamefont {Bhattacharyya}, \citenamefont {Pandey},\ and\ \citenamefont {Singh}(2014)}]{bhattacharyya2014effect}%
  \BibitemOpen
  \bibfield  {author} {\bibinfo {author} {\bibfnamefont {S.}~\bibnamefont {Bhattacharyya}}, \bibinfo {author} {\bibfnamefont {T.}~\bibnamefont {Pandey}}, \ and\ \bibinfo {author} {\bibfnamefont {A.~K.}\ \bibnamefont {Singh}},\ }\bibfield  {title} {\enquote {\bibinfo {title} {Effect of strain on electronic and thermoelectric properties of few layers to bulk mos2},}\ }\href@noop {} {\bibfield  {journal} {\bibinfo  {journal} {Nanotechnology}\ }\textbf {\bibinfo {volume} {25}},\ \bibinfo {pages} {465701} (\bibinfo {year} {2014})}\BibitemShut {NoStop}%
\bibitem [{\citenamefont {Min}\ \emph {et~al.}(2016)\citenamefont {Min}, \citenamefont {Park}, \citenamefont {Wallace}, \citenamefont {Cho},\ and\ \citenamefont {Hong}}]{min2016reduction}%
  \BibitemOpen
  \bibfield  {author} {\bibinfo {author} {\bibfnamefont {K.-A.}\ \bibnamefont {Min}}, \bibinfo {author} {\bibfnamefont {J.}~\bibnamefont {Park}}, \bibinfo {author} {\bibfnamefont {R.~M.}\ \bibnamefont {Wallace}}, \bibinfo {author} {\bibfnamefont {K.}~\bibnamefont {Cho}}, \ and\ \bibinfo {author} {\bibfnamefont {S.}~\bibnamefont {Hong}},\ }\bibfield  {title} {\enquote {\bibinfo {title} {Reduction of fermi level pinning at au--mos2 interfaces by atomic passivation on au surface},}\ }\href@noop {} {\bibfield  {journal} {\bibinfo  {journal} {2D Materials}\ }\textbf {\bibinfo {volume} {4}},\ \bibinfo {pages} {015019} (\bibinfo {year} {2016})}\BibitemShut {NoStop}%
\bibitem [{\citenamefont {Kang}\ \emph {et~al.}(2014)\citenamefont {Kang}, \citenamefont {Liu}, \citenamefont {Sarkar}, \citenamefont {Jena},\ and\ \citenamefont {Banerjee}}]{kang2014computational}%
  \BibitemOpen
  \bibfield  {author} {\bibinfo {author} {\bibfnamefont {J.}~\bibnamefont {Kang}}, \bibinfo {author} {\bibfnamefont {W.}~\bibnamefont {Liu}}, \bibinfo {author} {\bibfnamefont {D.}~\bibnamefont {Sarkar}}, \bibinfo {author} {\bibfnamefont {D.}~\bibnamefont {Jena}}, \ and\ \bibinfo {author} {\bibfnamefont {K.}~\bibnamefont {Banerjee}},\ }\bibfield  {title} {\enquote {\bibinfo {title} {Computational study of metal contacts to monolayer transition-metal dichalcogenide semiconductors},}\ }\href@noop {} {\bibfield  {journal} {\bibinfo  {journal} {Physical Review X}\ }\textbf {\bibinfo {volume} {4}},\ \bibinfo {pages} {031005} (\bibinfo {year} {2014})}\BibitemShut {NoStop}%
\bibitem [{\citenamefont {Gong}\ \emph {et~al.}(2014)\citenamefont {Gong}, \citenamefont {Colombo}, \citenamefont {Wallace},\ and\ \citenamefont {Cho}}]{gong2014unusual}%
  \BibitemOpen
  \bibfield  {author} {\bibinfo {author} {\bibfnamefont {C.}~\bibnamefont {Gong}}, \bibinfo {author} {\bibfnamefont {L.}~\bibnamefont {Colombo}}, \bibinfo {author} {\bibfnamefont {R.~M.}\ \bibnamefont {Wallace}}, \ and\ \bibinfo {author} {\bibfnamefont {K.}~\bibnamefont {Cho}},\ }\bibfield  {title} {\enquote {\bibinfo {title} {The unusual mechanism of partial fermi level pinning at metal--mos2 interfaces},}\ }\href {https://pubs.acs.org/doi/epdf/10.1021/nl403465v} {\bibfield  {journal} {\bibinfo  {journal} {Nano letters}\ }\textbf {\bibinfo {volume} {14}},\ \bibinfo {pages} {1714--1720} (\bibinfo {year} {2014})}\BibitemShut {NoStop}%
\bibitem [{\citenamefont {Komsa}\ and\ \citenamefont {Krasheninnikov}(2015)}]{komsa2015native}%
  \BibitemOpen
  \bibfield  {author} {\bibinfo {author} {\bibfnamefont {H.-P.}\ \bibnamefont {Komsa}}\ and\ \bibinfo {author} {\bibfnamefont {A.~V.}\ \bibnamefont {Krasheninnikov}},\ }\bibfield  {title} {\enquote {\bibinfo {title} {Native defects in bulk and monolayer mos 2 from first principles},}\ }\href@noop {} {\bibfield  {journal} {\bibinfo  {journal} {Physical Review B}\ }\textbf {\bibinfo {volume} {91}},\ \bibinfo {pages} {125304} (\bibinfo {year} {2015})}\BibitemShut {NoStop}%
\bibitem [{\citenamefont {Li}\ \emph {et~al.}(2022)\citenamefont {Li}, \citenamefont {Chen}, \citenamefont {Wang},\ and\ \citenamefont {Li}}]{li2022conductive}%
  \BibitemOpen
  \bibfield  {author} {\bibinfo {author} {\bibfnamefont {X.-D.}\ \bibnamefont {Li}}, \bibinfo {author} {\bibfnamefont {N.-K.}\ \bibnamefont {Chen}}, \bibinfo {author} {\bibfnamefont {B.-Q.}\ \bibnamefont {Wang}}, \ and\ \bibinfo {author} {\bibfnamefont {X.-B.}\ \bibnamefont {Li}},\ }\bibfield  {title} {\enquote {\bibinfo {title} {Conductive mechanism in memristor at the thinnest limit: The case based on monolayer boron nitride},}\ }\href@noop {} {\bibfield  {journal} {\bibinfo  {journal} {Applied Physics Letters}\ }\textbf {\bibinfo {volume} {121}} (\bibinfo {year} {2022})}\BibitemShut {NoStop}%
\bibitem [{\citenamefont {Ruban}\ \emph {et~al.}(1997)\citenamefont {Ruban}, \citenamefont {Hammer}, \citenamefont {Stoltze}, \citenamefont {Skriver},\ and\ \citenamefont {N{\o}rskov}}]{ruban1997surface}%
  \BibitemOpen
  \bibfield  {author} {\bibinfo {author} {\bibfnamefont {A.}~\bibnamefont {Ruban}}, \bibinfo {author} {\bibfnamefont {B.}~\bibnamefont {Hammer}}, \bibinfo {author} {\bibfnamefont {P.}~\bibnamefont {Stoltze}}, \bibinfo {author} {\bibfnamefont {H.~L.}\ \bibnamefont {Skriver}}, \ and\ \bibinfo {author} {\bibfnamefont {J.~K.}\ \bibnamefont {N{\o}rskov}},\ }\bibfield  {title} {\enquote {\bibinfo {title} {Surface electronic structure and reactivity of transition and noble metals},}\ }\href@noop {} {\bibfield  {journal} {\bibinfo  {journal} {Journal of Molecular Catalysis A: Chemical}\ }\textbf {\bibinfo {volume} {115}},\ \bibinfo {pages} {421--429} (\bibinfo {year} {1997})}\BibitemShut {NoStop}%
\bibitem [{\citenamefont {Schnur}\ and\ \citenamefont {Gro{\ss}}(2010)}]{schnur2010strain}%
  \BibitemOpen
  \bibfield  {author} {\bibinfo {author} {\bibfnamefont {S.}~\bibnamefont {Schnur}}\ and\ \bibinfo {author} {\bibfnamefont {A.}~\bibnamefont {Gro{\ss}}},\ }\bibfield  {title} {\enquote {\bibinfo {title} {Strain and coordination effects in the adsorption properties of early transition metals: A density-functional theory study},}\ }\href@noop {} {\bibfield  {journal} {\bibinfo  {journal} {Physical Review B}\ }\textbf {\bibinfo {volume} {81}},\ \bibinfo {pages} {033402} (\bibinfo {year} {2010})}\BibitemShut {NoStop}%
\bibitem [{\citenamefont {Bardeen}(1947)}]{bardeen1947surface}%
  \BibitemOpen
  \bibfield  {author} {\bibinfo {author} {\bibfnamefont {J.}~\bibnamefont {Bardeen}},\ }\bibfield  {title} {\enquote {\bibinfo {title} {Surface states and rectification at a metal semi-conductor contact},}\ }\href@noop {} {\bibfield  {journal} {\bibinfo  {journal} {Physical review}\ }\textbf {\bibinfo {volume} {71}},\ \bibinfo {pages} {717} (\bibinfo {year} {1947})}\BibitemShut {NoStop}%
\bibitem [{\citenamefont {Darling}(1991)}]{darling1991defect}%
  \BibitemOpen
  \bibfield  {author} {\bibinfo {author} {\bibfnamefont {R.~B.}\ \bibnamefont {Darling}},\ }\bibfield  {title} {\enquote {\bibinfo {title} {Defect-state occupation, fermi-level pinning, and illumination effects on free semiconductor surfaces},}\ }\href@noop {} {\bibfield  {journal} {\bibinfo  {journal} {Physical Review B}\ }\textbf {\bibinfo {volume} {43}},\ \bibinfo {pages} {4071} (\bibinfo {year} {1991})}\BibitemShut {NoStop}%
\bibitem [{\citenamefont {Heine}(1965)}]{heine1965theory}%
  \BibitemOpen
  \bibfield  {author} {\bibinfo {author} {\bibfnamefont {V.}~\bibnamefont {Heine}},\ }\bibfield  {title} {\enquote {\bibinfo {title} {Theory of surface states},}\ }\href@noop {} {\bibfield  {journal} {\bibinfo  {journal} {Physical Review}\ }\textbf {\bibinfo {volume} {138}},\ \bibinfo {pages} {A1689} (\bibinfo {year} {1965})}\BibitemShut {NoStop}%
\bibitem [{\citenamefont {Louie}\ and\ \citenamefont {Cohen}(1975)}]{louie1975self}%
  \BibitemOpen
  \bibfield  {author} {\bibinfo {author} {\bibfnamefont {S.~G.}\ \bibnamefont {Louie}}\ and\ \bibinfo {author} {\bibfnamefont {M.~L.}\ \bibnamefont {Cohen}},\ }\bibfield  {title} {\enquote {\bibinfo {title} {Self-consistent pseudopotential calculation for a metal-semiconductor interface},}\ }\href@noop {} {\bibfield  {journal} {\bibinfo  {journal} {Physical Review Letters}\ }\textbf {\bibinfo {volume} {35}},\ \bibinfo {pages} {866} (\bibinfo {year} {1975})}\BibitemShut {NoStop}%
\bibitem [{\citenamefont {Hasegawa}\ and\ \citenamefont {Sawada}(1983)}]{hasegawa1983electrical}%
  \BibitemOpen
  \bibfield  {author} {\bibinfo {author} {\bibfnamefont {H.}~\bibnamefont {Hasegawa}}\ and\ \bibinfo {author} {\bibfnamefont {T.}~\bibnamefont {Sawada}},\ }\bibfield  {title} {\enquote {\bibinfo {title} {On the electrical properties of compound semiconductor interfaces in metal/insulator/semiconductor structures and the possible origin of interface states},}\ }\href@noop {} {\bibfield  {journal} {\bibinfo  {journal} {Thin Solid Films}\ }\textbf {\bibinfo {volume} {103}},\ \bibinfo {pages} {119--140} (\bibinfo {year} {1983})}\BibitemShut {NoStop}%
\bibitem [{\citenamefont {Choi}(2018)}]{choi2018strain}%
  \BibitemOpen
  \bibfield  {author} {\bibinfo {author} {\bibfnamefont {M.}~\bibnamefont {Choi}},\ }\bibfield  {title} {\enquote {\bibinfo {title} {Strain-enhanced p doping in monolayer mos 2},}\ }\href@noop {} {\bibfield  {journal} {\bibinfo  {journal} {Physical Review Applied}\ }\textbf {\bibinfo {volume} {9}},\ \bibinfo {pages} {024009} (\bibinfo {year} {2018})}\BibitemShut {NoStop}%
\bibitem [{\citenamefont {Velicky}\ \emph {et~al.}(2020)\citenamefont {Velicky}, \citenamefont {Rodriguez}, \citenamefont {Bousa}, \citenamefont {Krayev}, \citenamefont {Vondracek}, \citenamefont {Honolka}, \citenamefont {Ahmadi}, \citenamefont {Donnelly}, \citenamefont {Huang}, \citenamefont {Abruna} \emph {et~al.}}]{velicky2020strain}%
  \BibitemOpen
  \bibfield  {author} {\bibinfo {author} {\bibfnamefont {M.}~\bibnamefont {Velicky}}, \bibinfo {author} {\bibfnamefont {A.}~\bibnamefont {Rodriguez}}, \bibinfo {author} {\bibfnamefont {M.}~\bibnamefont {Bousa}}, \bibinfo {author} {\bibfnamefont {A.~V.}\ \bibnamefont {Krayev}}, \bibinfo {author} {\bibfnamefont {M.}~\bibnamefont {Vondracek}}, \bibinfo {author} {\bibfnamefont {J.}~\bibnamefont {Honolka}}, \bibinfo {author} {\bibfnamefont {M.}~\bibnamefont {Ahmadi}}, \bibinfo {author} {\bibfnamefont {G.~E.}\ \bibnamefont {Donnelly}}, \bibinfo {author} {\bibfnamefont {F.}~\bibnamefont {Huang}}, \bibinfo {author} {\bibfnamefont {H.~D.}\ \bibnamefont {Abruna}},  \emph {et~al.},\ }\bibfield  {title} {\enquote {\bibinfo {title} {Strain and charge doping fingerprints of the strong interaction between monolayer mos2 and gold},}\ }\href@noop {} {\bibfield  {journal} {\bibinfo  {journal} {The journal of physical chemistry letters}\ }\textbf {\bibinfo {volume} {11}},\ \bibinfo {pages} {6112--6118} (\bibinfo {year}
  {2020})}\BibitemShut {NoStop}%
\bibitem [{\citenamefont {Michail}\ \emph {et~al.}(2016)\citenamefont {Michail}, \citenamefont {Delikoukos}, \citenamefont {Parthenios}, \citenamefont {Galiotis},\ and\ \citenamefont {Papagelis}}]{michail2016optical}%
  \BibitemOpen
  \bibfield  {author} {\bibinfo {author} {\bibfnamefont {A.}~\bibnamefont {Michail}}, \bibinfo {author} {\bibfnamefont {N.}~\bibnamefont {Delikoukos}}, \bibinfo {author} {\bibfnamefont {J.}~\bibnamefont {Parthenios}}, \bibinfo {author} {\bibfnamefont {C.}~\bibnamefont {Galiotis}}, \ and\ \bibinfo {author} {\bibfnamefont {K.}~\bibnamefont {Papagelis}},\ }\bibfield  {title} {\enquote {\bibinfo {title} {Optical detection of strain and doping inhomogeneities in single layer mos2},}\ }\href@noop {} {\bibfield  {journal} {\bibinfo  {journal} {Applied Physics Letters}\ }\textbf {\bibinfo {volume} {108}} (\bibinfo {year} {2016})}\BibitemShut {NoStop}%
\end{thebibliography}%
%=============================================================
%\clearpage
%\appendix \label{Supplementary_Information}
%\section{ Supplementary Information}

\clearpage
\pagebreak
\widetext
\begin{center}
\textbf{\large Supplemental Material: \\
Nature of point defects in monolayer MoS$_2$ and the Mo$_2$/(111)-Au heterojunction\\}

Roozbeh Anvari and Wennie Wang\\
\rm McKetta Department of Chemical Engineering, University of Texas at Austin 
\end{center}
%%%%%%%%%% Merge with supplemental materials %%%%%%%%%%
%%%%%%%%%% Prefix a "S" to all equations, figures, tables and reset the counter %%%%%%%%%%
\setcounter{equation}{0}
\setcounter{figure}{0}
\setcounter{table}{0}
\setcounter{section}{0}
\setcounter{page}{1}
\makeatletter
\renewcommand{\theequation}{S\arabic{equation}}
\renewcommand{\thefigure}{S\arabic{figure}}
\renewcommand{\thetable}{S\arabic{table}}
\renewcommand{\thesection}{S\arabic{section}}
\renewcommand{\bibnumfmt}[1]{[S#1]}
\renewcommand{\citenumfont}[1]{S#1}

\section{Point defects in free-standing monolayer MoS$_2$}
% ======== Figure defect configs =========
%-----------------------
\begin{figure}[h!]
    \centering
    \subfloat[\label{simple_defect_struct}]{%
    \includegraphics[width=.8\columnwidth] 
      {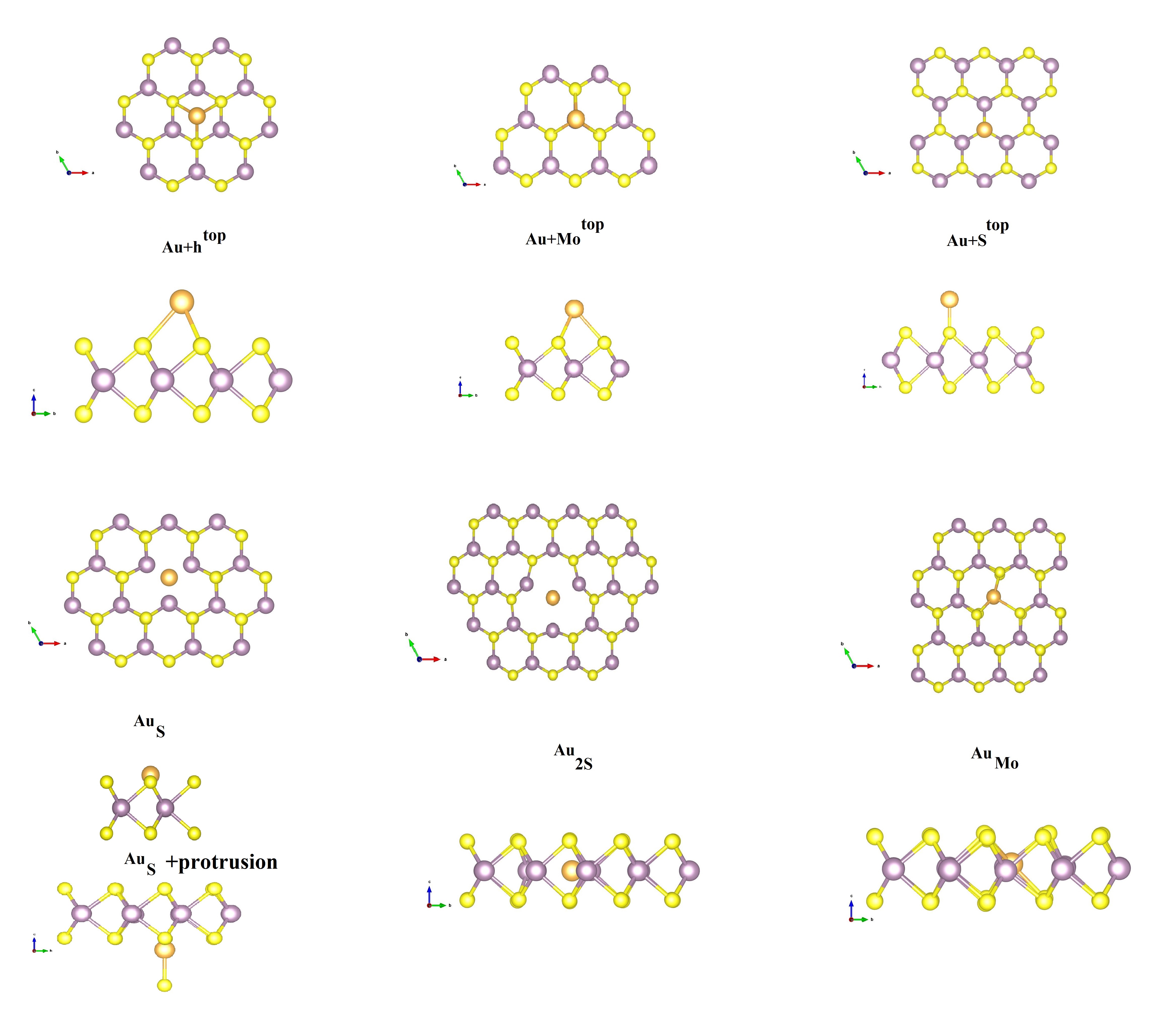}
} 
     \caption{ Top and side-view of the studied stand-alone structures. The first two rows correspond to the top and side view of Au-adatom absobed on top of the h-site ($\rm Au+h^{top}$), Mo-top site ($\rm Au+Mo^{top}$), and S-top site ($\rm Au_S$). The third and fourth rows correspond to the top and side view of ($\rm Au_S$) after protrusion, Au adsorbed into sulfur-divacancy ($\rm Au_{2S}$), and Mo-vacancy ($\rm Au_{Mo}$).  }
     \label{simple_defect_struct}
\end{figure} 

\begin{figure}[h!]
 \centering
    \subfloat[\label{hetero_coordinates_a}]{%
    \includegraphics[width=.35\columnwidth] 
      {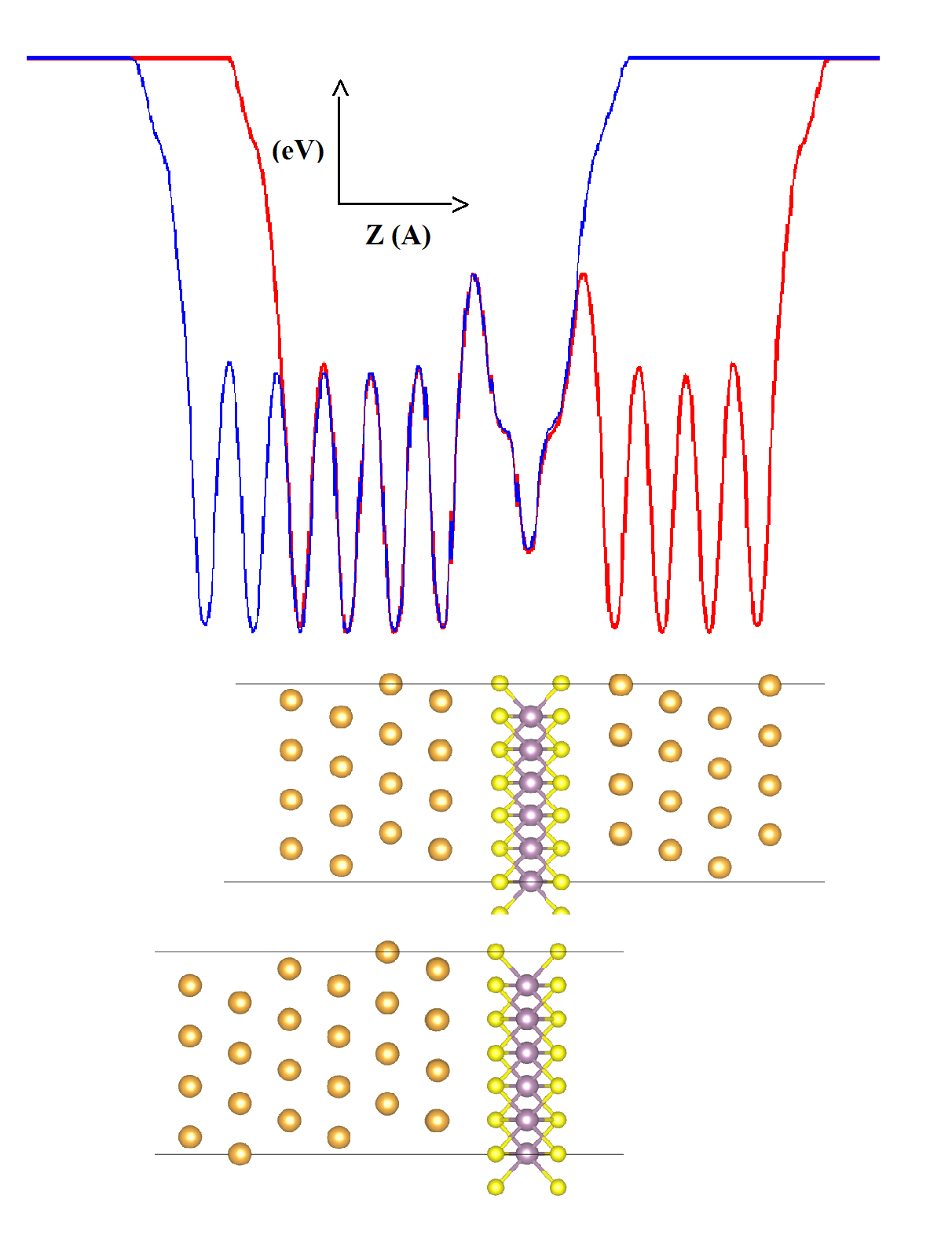}
      }
     \subfloat[\label{hetero_coordinates_b}]{  
    \includegraphics[width=.4\columnwidth]
      {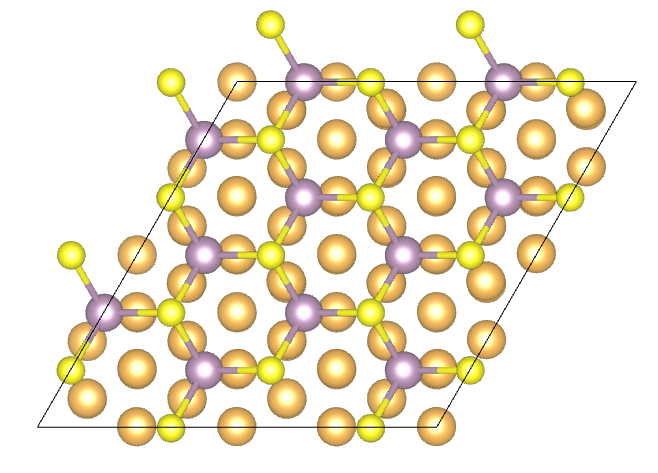}
} 
     \caption{ (a) Side view of the Au(111)-MoS$_2$-Au(111) and Au(111)-MoS$_2$ structures, and the corresponding electrostatic potential relative to the vacuum level. (b) Top-view of the reconstructed $\rm (2\sqrt{3} \times 2\sqrt{3})-MoS_2$ layer. 
     }
     \label{hetero_coordinates}
\end{figure}

Figure~\ref{hetero_coordinates} shows the fully relaxed atomic coordinates of the vertical (111)Au-MoS$_2$-(111)Au device, and the half (111)Au-MoS$_2$ system from the side and top.
The calculated formation energy of a single sulfur vacancy ($\rm V_S$) in (111)Au-MoS$_2$-(111)Au and (111)Au-MoS$_2$ heterostructures is 3.33 eV and 3.86 eV,respectively. 
The normal distance between S of MoS$_2$ and Au of the (111)-Au electrode is 2.897 $\rm \AA$ (2.912 and 2.916 $\rm \AA$) for the one (two)-sided heterostructure. 
As depicted in Figure \ref{hetero_coordinates_a}, the calculated electrostatic potential with respect to the vacuum level indicates that the interlayer distances, and the corresponding charge distribution are approximately identical between the full (111)Au-MoS$_2$-(111)Au heterostructure and the (111)Au-MoS$_2$ half system. 
Due to spatial and electrostatic symmetry of the (111)Au-MoS$_2$-(111)Au, we have focus on the characteristics of the half system MoS$_2$/(111)Au in this study. 

%   , system preview ,  fig S2 

% FIG/TAB native point defect formation energy   fig s3 

\begin{figure}[h!]
 \centering
    \includegraphics[width=.4\columnwidth] 
      {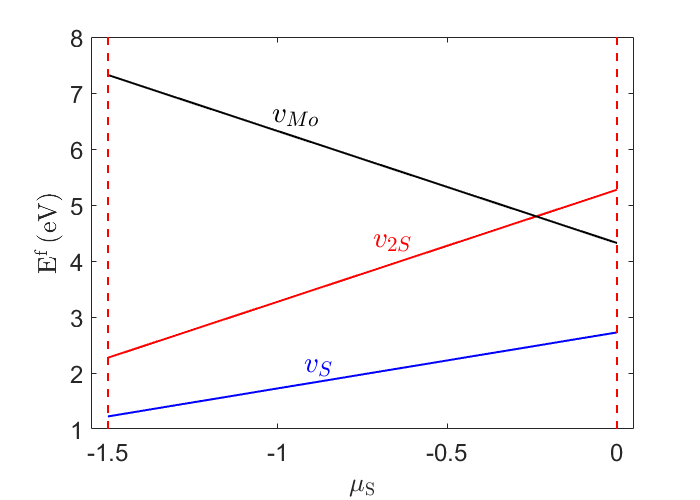}
     \caption{ Defect formation energy of the $\rm V_S$, $\rm V_{2S}$ and $\rm V_{Mo}$ vacancy defects in stand-alone (no substrate) $\rm MoS_2$ as a function of the chemical potential of sulfur. Chemical potentials are obtained from equilibrium of monolayer-$\rm MoS_2$ with bulk Mo (solid) and solid phase S. Left and right boundaries correspond to the sulfur-poor and sulfur-rich environments, respectively. 
     }
     \label{simple_defect_energ_vac}
\end{figure} 

\begin{table}
    \centering
    \begin{tabular}{|c|c|c|c||c|c|c|c|}

        \hline
        structure            & $E^f$ & $E^{ads}$& $E^b$ & $E^{f}$ & $E^{ads}$   & REF.\cite{komsa2015native}   & REF. \cite{theory_tumino2020nature}\\
         \hline 
        $\rm v_S$            & 2.72 & &  &2.799 &   & {2.7}  & { 2.81}  \\
        $\rm v_{2S}$         & 5.27 & & & &   &   &   { 5.44}    \\
        $\rm v_{Mo}$         & {4.32} & & & &   & { 3.6}  &   { 2.22}    \\
        \hline
        $\rm Au+pr, h{top}$  & {2.}7 &  2.830 & & & 2.64   &   &\\
        $\rm Au+pr, S{top}$  & {2.7} &  2.831 & & & 2.69   &   &\\
        $\rm Au+pr, Mo{top}$ & {2.8} &  2.925 & & & 2.76   &   &\\
        \hline
         $\rm Au_{S}$& {3.55} &  {0.95} &{-1.87 } & &   &   &\\
         $\rm Au_{2S}$&{10.65}  & 5.498 &{2.86}  & &\\
         $\rm Au_{Mo}$&{4.17}  & {-0.026} &{-2.95}  & &   &    &\\
        \hline   
    \end{tabular}
    \caption{Formation energy of vacancies ($\rm E^f$), adsorption energy ($E^{ads}$), and binding energy ($\rm E^b$) of Au+vacancy defect complexes (all in eV) in freestanding $\rm (6x6)-MoS_2$ supercell with no strain. The last 2 columns are calculated using Quantum Espresso for freestanding $\rm (4x4)-MoS_2$ supercell. Values are tabulated for sulfur-rich conditions.} 
    \label{table_stal_energies}
\end{table}

%==============================================================================
%  Electronic structure of simple defects- BIG FIGURE
%==============================================================================
%   Fig S5 
\begin{figure}[!htb]
 \centering
    \subfloat[\label{native_defect_band_dos_cond_a}]{%
    \includegraphics[width=.6\columnwidth] 
      {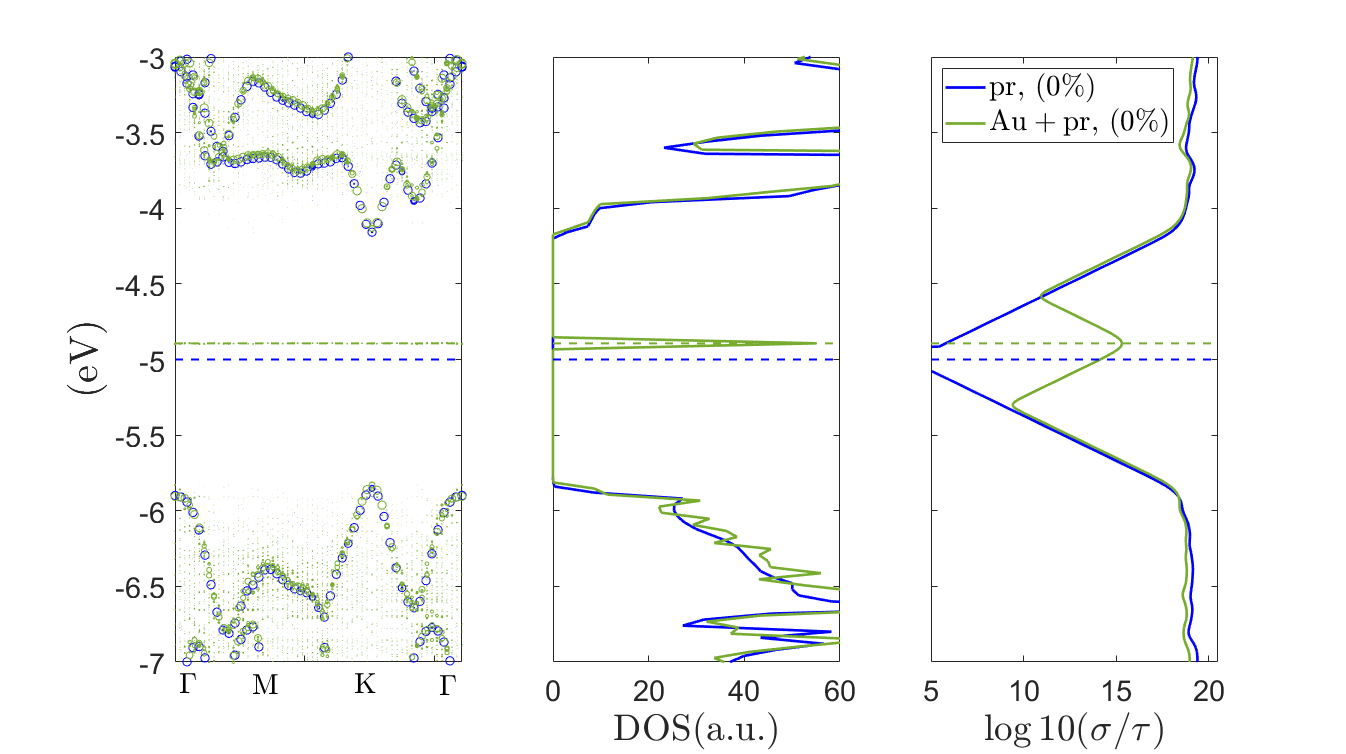}
      }
     \hfill 
     \subfloat[\label{native_defect_band_dos_cond_SI_b}]{  
    \includegraphics[width=.6\columnwidth]
      {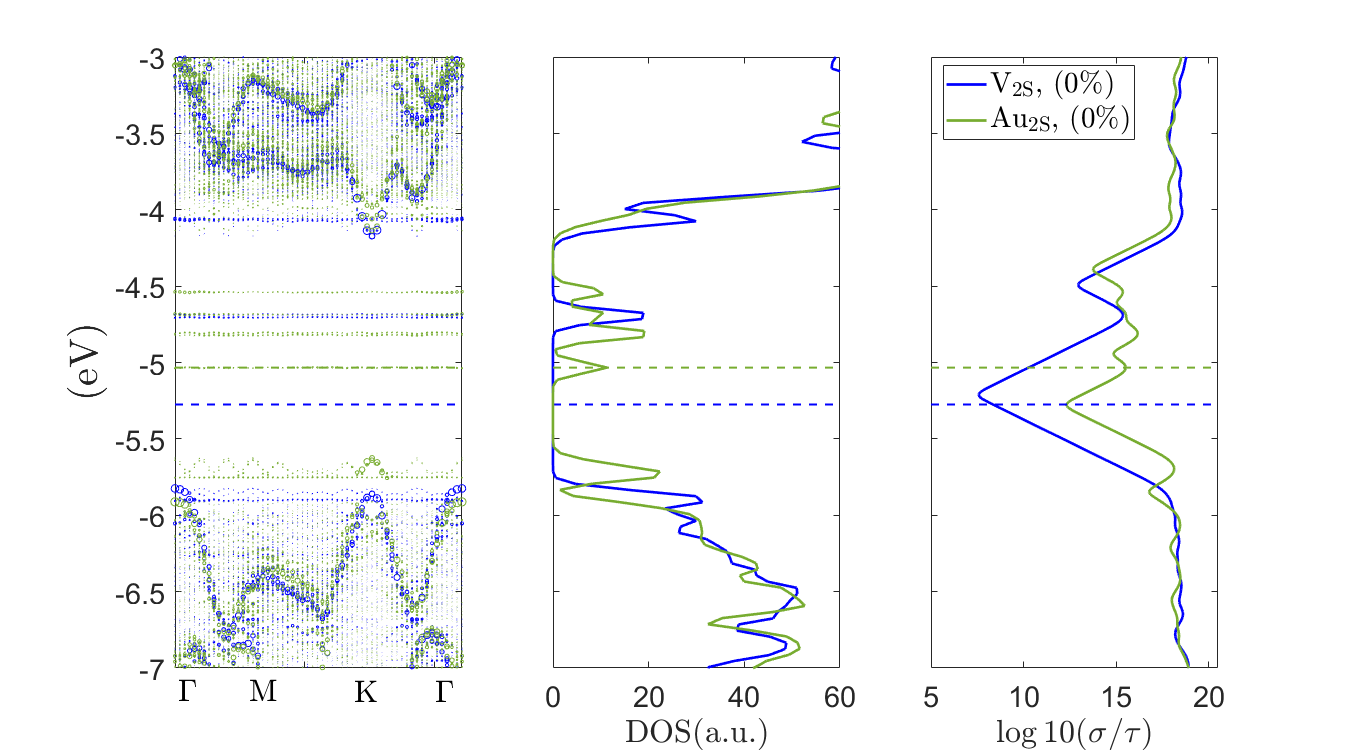}
     }
  \hfill
 \subfloat[\label{native_defect_band_dos_cond_SI_c}]{  
    \includegraphics[width=.6\columnwidth]
      {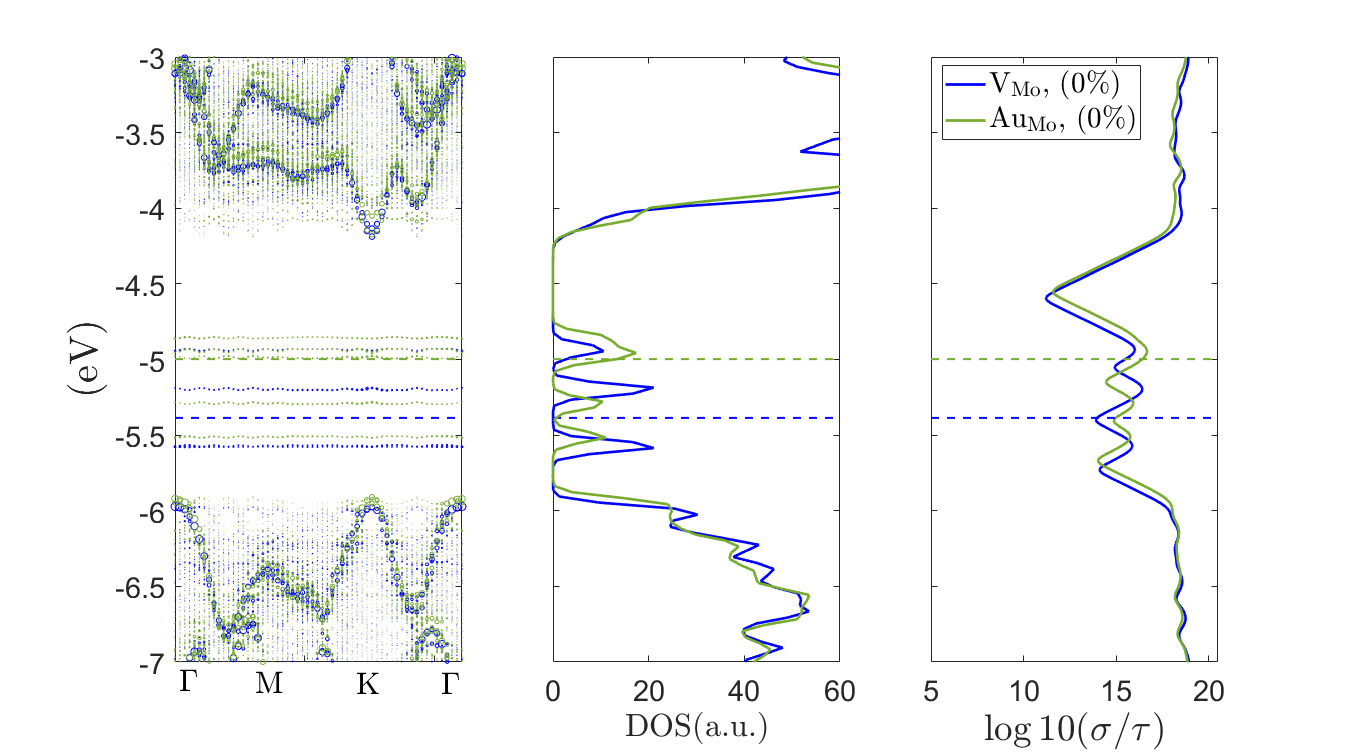}
     }
     \caption{ Unfolded bandstructure (left), DOS (middle) and normalized electrical conductivity (right) of (6x6)  supercell $\rm MoS_2$ comparing the (a) pristine monolayer and Au adsorbed on the pristine monolayer, (b) sulfur divacancy and Au adsorbed on the sulfur divacancy, (c) molydenum vacancy and Au adsorbed on the molybdenum vacancy, in blue and green, respectively. The dotted lines indicate the computed Fermi level. 
     }
     \label{native_defect_band_dos_cond_SI}
     
\end{figure} 

%==============================================================================
%  summary of band-edges of simple defects, no strain   , Fig S4 
\begin{figure}[!htb]
 \centering
    \subfloat[\label{simple_defect_band_edges}]{%
    \includegraphics[width=.75\columnwidth] 
      {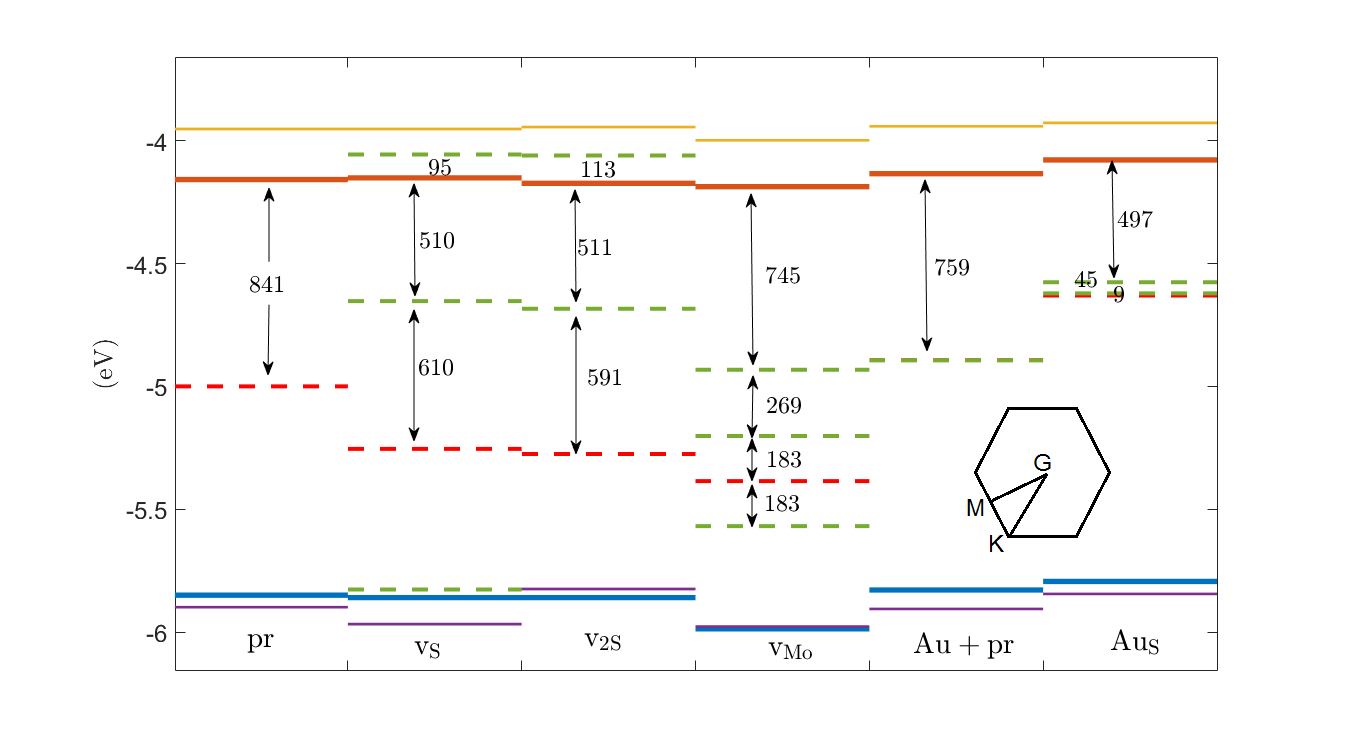}
       }
     \caption{Energy diagram of the band edges of the free-standing monolayer without defects, with vacancies alone, and with Au+defect complexes. Solid orange and blue lines correspond to the conduction band minima and valence band maxima at the K-point, respectively. Solid gold and purple lines correspond to the conduction band minima and valence band maxima at the satellite valley next to CBM and $\rm \Gamma$-point, respectively. Dashed red and dashed green lines correspond to the Fermi-level and defect levels, respectively. Numbers correspond to the energy difference between the levels indicated, in units of meV. As discussed in the main text, CBM and VBM of structure with zero strain are located at K-point. The valley that is presented in this figure is located along the K-$\rm \Gamma$ high symmetry line.
     } 
     \label{simple_defect_band_edges}
\end{figure} 

%=====================
% Tables: normalized conductivity
%===================

\begin{table}[!htb]
    \centering
    \begin{tabular}{c|  c   c }
    \hline
           & $\rm \sigma_{xx}$ &   $\rm \sigma_{zz} $ \\
\hline           
$\rm V_S$       &  7.851 &   5.53  \\
$\rm V_{2S}$  &  8.345   &   6.16 \\
$\rm V_{Mo}$   &  13.97   &   10.44  \\
$\rm Au+pr$    &  15.31   &   11.17    \\
$\rm Au_S$     &  15.57   &   11.53    \\  
$\rm Au_{2S}$  &  15.52   &   11.47    \\
$\rm Au_{Mo}$  &  16.42   &   12.91  \\
\hline
    \end{tabular}
    \caption{Calculated normalized conductivity ($\rm \log_{10}(\sigma/\tau)$) using the constant time relaxation approximation and the Boltzmann transport equation of native defects of free-standing MoS$_2$ at the computed Fermi level.}
    \label{table_point_def_cond}
\end{table}
 \FloatBarrier
%------------------------------------------------------
% Fig S6 
\newpage
\section{Effect of strain in free-standing MoS$_2$}

%  DOS of simple defects ------------------------------------------------------
\begin{figure}[!htb]
 \centering
    \subfloat[\label{stal_strain_bands_a}]{%
    \includegraphics[width=.5\columnwidth] 
      {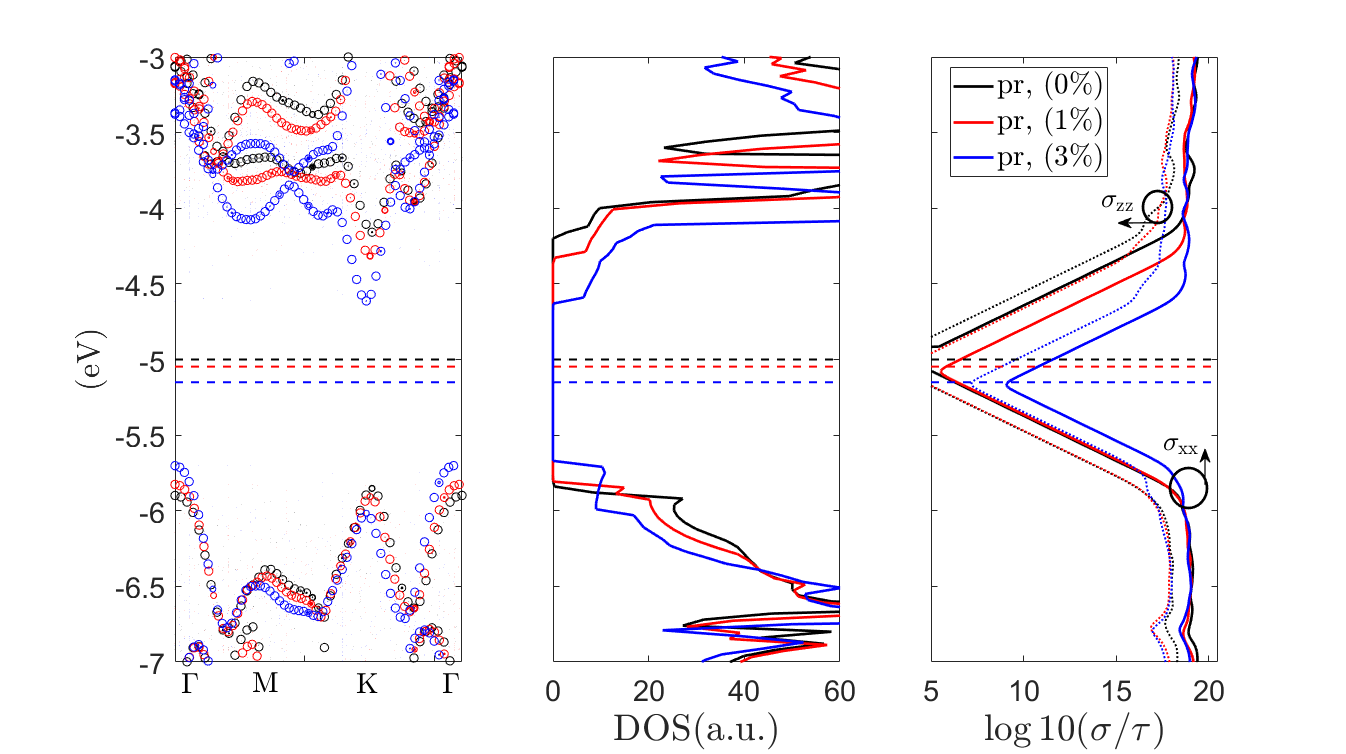}
      }
      \hfill
     \subfloat[\label{stal_strain_bands_b}]{  
    \includegraphics[width=.55\columnwidth]
      {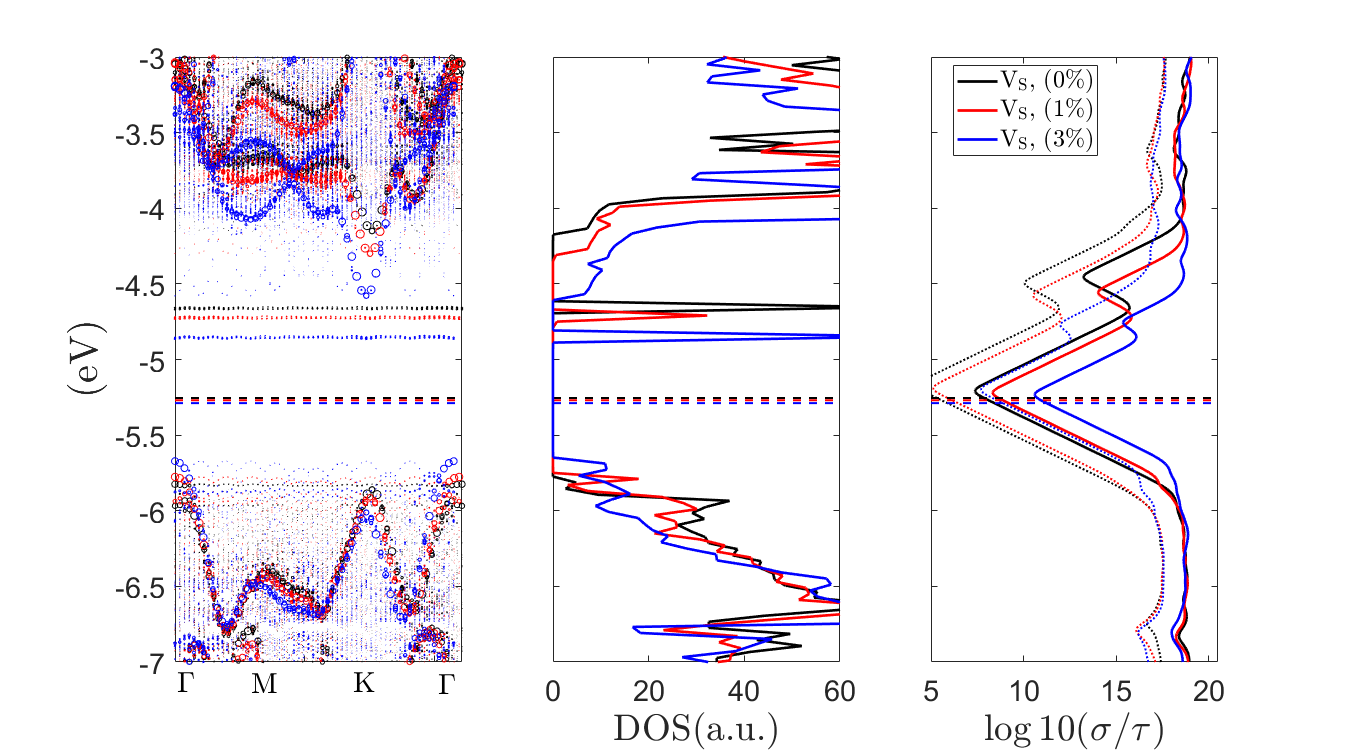}
     }
      \hfill
    \subfloat[\label{stal_strain_bands_c}]{%
    \includegraphics[width=.55\columnwidth] 
      {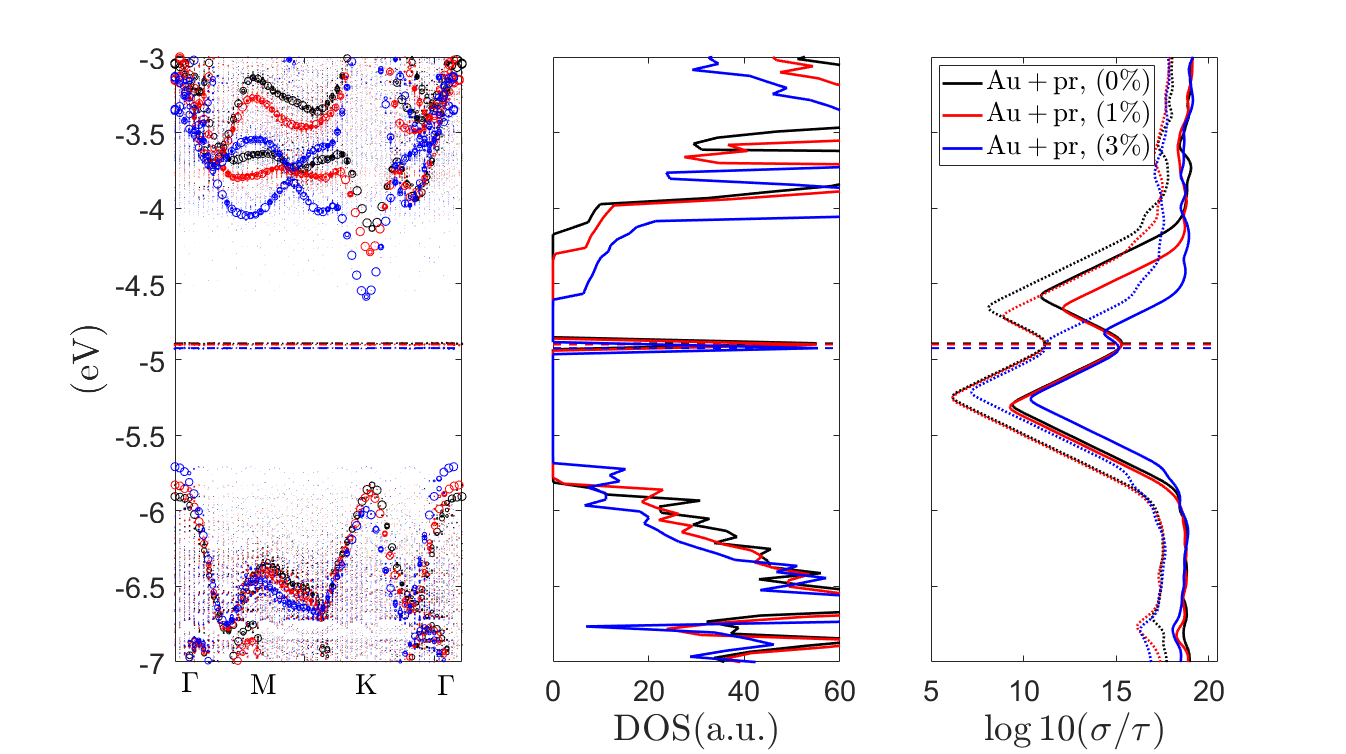}
      }
       \hfill
     \caption{ Effect of biaxial tensile strain on the electronic structure and conductivity of (a) pristine MoS$_2$, (b) MoS$_2$ with a sulfur vacancy, and (c) pristine MoS$_2$ with an Au adatom adsorbed.}
     \label{stal_strain_bands_SI}
\end{figure} 

%=================================================================
% FiG s7 
%-----------------------
\begin{figure}[!t]
    \centering
    \subfloat[]{\includegraphics[width=0.45\linewidth] 
      {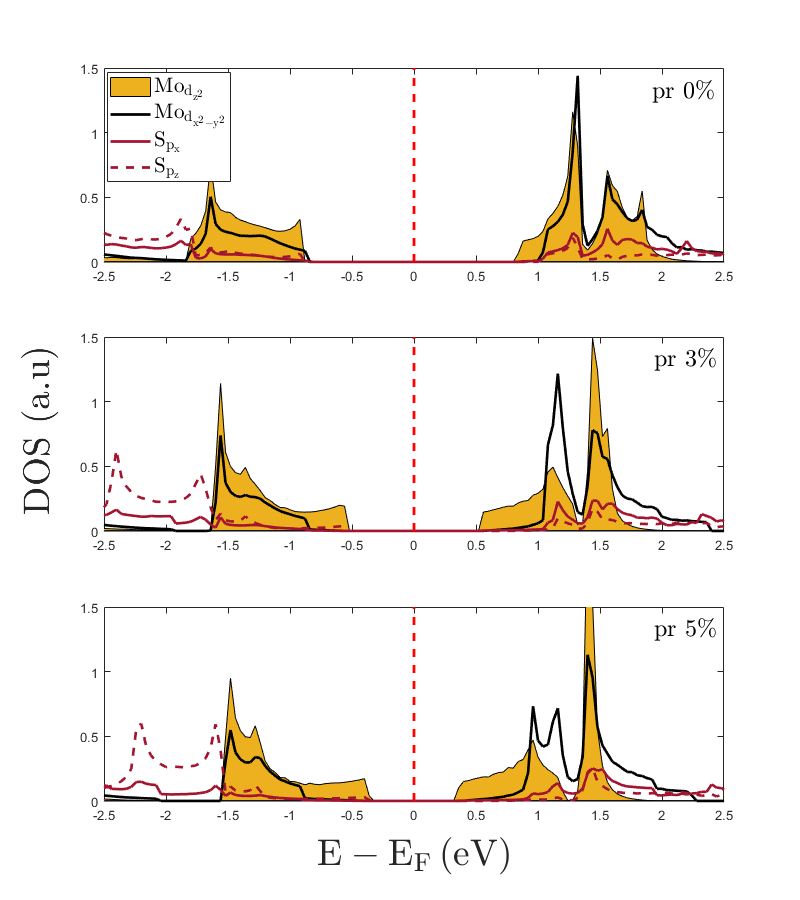}
      \label{fig_stal_pdos_strain_a}}
       \subfloat[\label{fig_stal_pdos_strain_b}]{%
    \includegraphics[width=0.45\linewidth] 
      {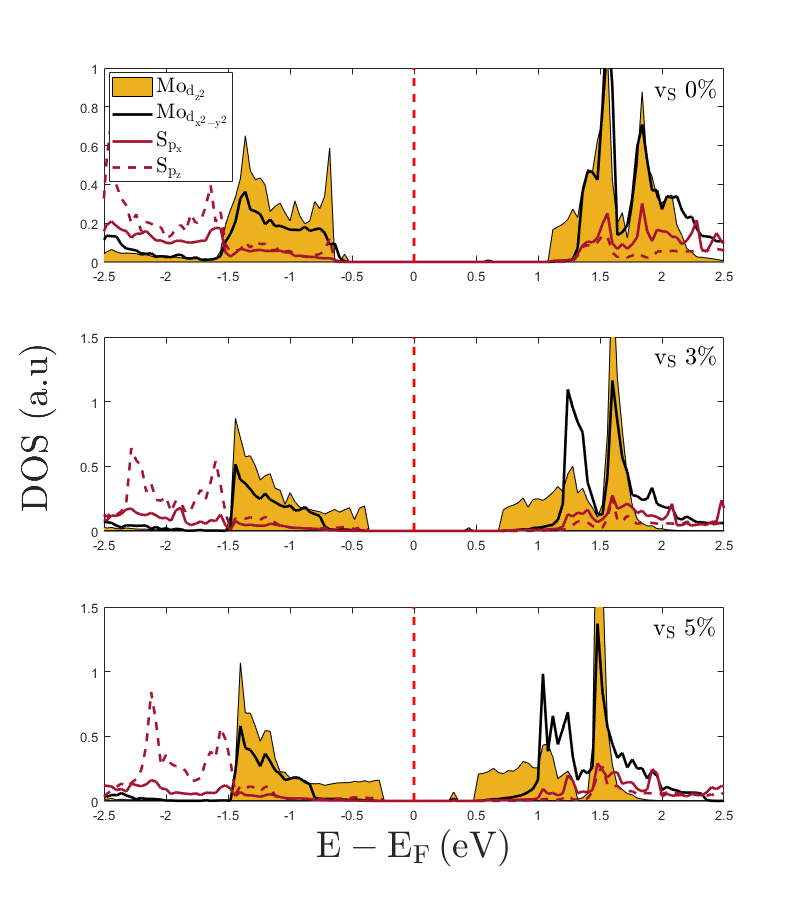}
      } 
      \hfill
   \subfloat[\label{fig_stal_pdos_strain_c}]{%
    \includegraphics[width=0.45\linewidth]  
      {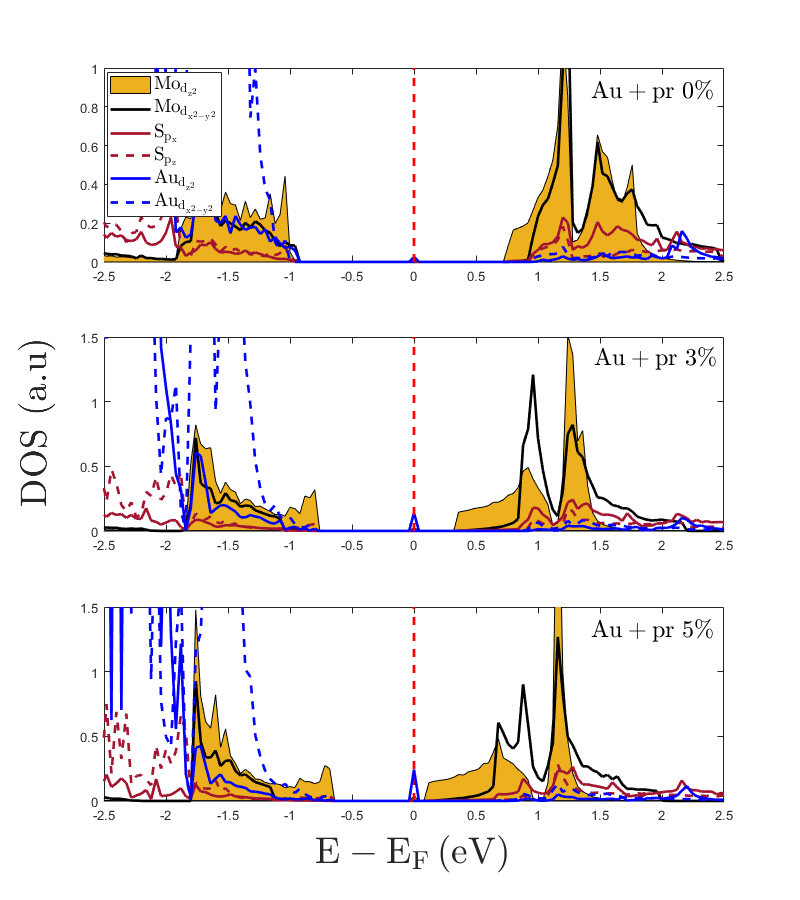}
         }     
     \caption{Changes of the PDOS of free-standing MoS$_2$ structures with biaxial tensile strain. Changes of net charge accumulated on the Au adatom with strain indicates that Au is hybridizing differently with the host MoS$_2$ in Au+pr and $\rm Au_S$ structures. PDOS analysis indicates that the hybridization of Au in Au+pr is mainly $d_{z^2}$, while the main contribution of Au in $\rm Au_S$ is $d_{x^2-y^2}$.}
     \label{fig_stal_pdos_strain}
     \vspace{-10 pt}  
\end{figure}

%=======================
%% changes of band edges, energies shifted with respect to the  Fermi level 
%% fig s8 
%\begin{figure}[!htb]
% \centering
%    \subfloat[\label{fig_SI_strain_band}]{%
%    \includegraphics[width=.7\columnwidth] 
%      {"./fig_draft4/stal_band_edges_no_elec".png}
%       }
%
%     \caption{ Effect of strain on the band edges of freestanding pristine and single sulfur %vacancy structure (top left), Au + pr (middle), and AuS (top right) structures. Dashed blue %and red lines indicate the defect induced midgap state, and Fermi-level, respectively. Center %of the d-band is shown by green and yellow (vS) colors (color inline). Our calculations show %a downshift of the d-band center with increasing tensile strain  }
%     \label{fig_SI_strain_band}
%\end{figure} 

% FiG 
%-----------------------
% fig s10 
\begin{figure}[h!]
    \centering
    \subfloat[\label{fig_stal_unfold_pr_projected}]{%
    \includegraphics[width=.7\columnwidth] 
      {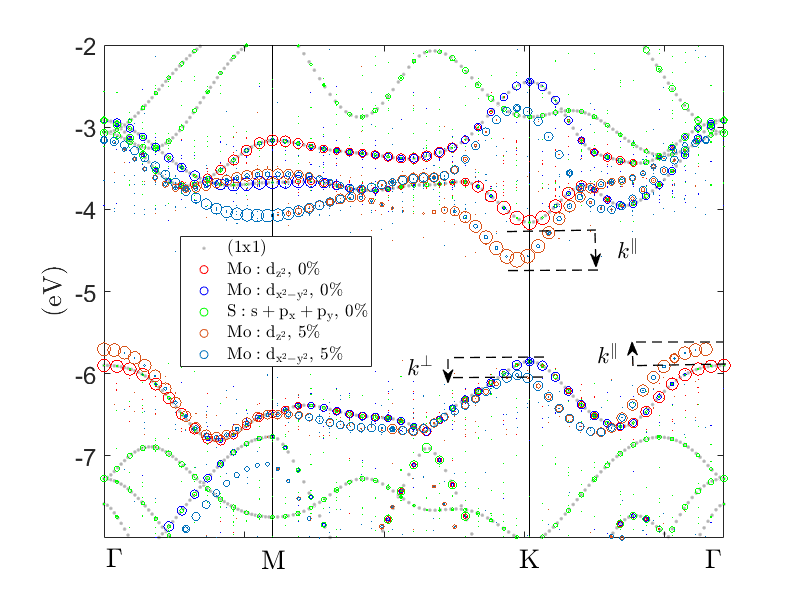}
} 
     \caption{ Unfolded band structure of free standing (6x6)-MoS$_2$ structure under 0 and 5$\%$ tensile strain. The radius of the circles represent the magnitude of the spectral weight obtained by unfolding. Energy levels are relative to vacuum. Arrows indicates the direction of shift in energy with increasing tensile strain. 
     }
     \label{fig_stal_unfold_pr_projected}
\end{figure} 

% fig s9 
\begin{figure}[!htb]
 \centering
    \subfloat[\label{fig_SI_strain_band_edge_2}]{%
    \includegraphics[width=.9\columnwidth] 
      {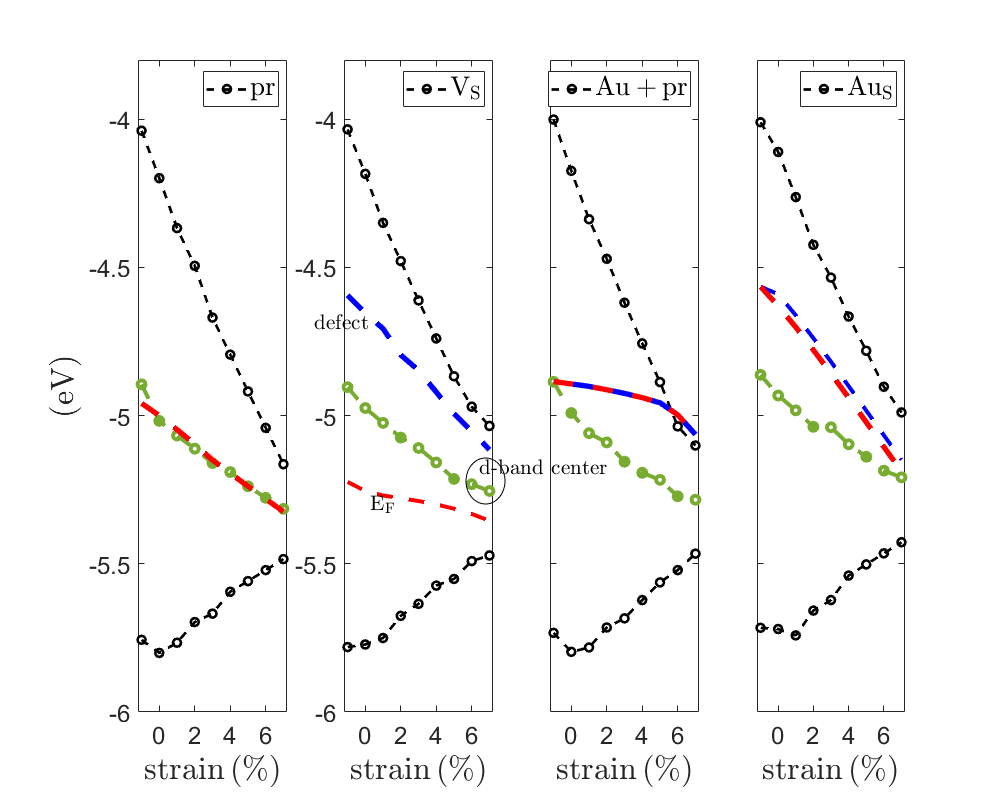}
       }

     \caption{ Effect of strain on the band edges of freestanding structures.  }
     \label{fig_SI_strain_band_edge_2}
\end{figure}

%================================================

\begin{table}
    \centering
    \begin{tabular}{| c |  c  |  c  |  c  |  c  |}
          \hline
 strain $\%$& $\rm pr$& $\rm v_S$& $\rm Au+pr$& $\rm Au_S$\\
       \hline
 -1  & -4.958   &  -5.223  &  -4.884  &  -4.564  \\
  0 & -4.999   &  -5.254  &  -4.892  &  -4.630 \\
  1 & -5.047   &  -5.269  &  -4.901  &  -4.701 \\
  2 & -5.096   &  -5.278  &  -4.912  &  -4.778 \\
  3 & -5.150   &  -5.288  &  -4.924  &  -4.858 \\
  4 & -5.194   &  -5.299  &  -4.939  &  -4.939 \\
  5 & -5.238   &  -5.314  &  -4.956  &  -5.022 \\
  6 & -5.281   &  -5.331  &  -4.997  &  -5.105 \\
  7 & -5.323   &  -5.354  &  -5.063  &  -5.188 \\
        \hline
    \end{tabular}
    \caption{ Changes of workfunction (in units of eV) of freestanding MoS$_2$ structures with biaxial strain.}
    \label{tab:my_label}
\end{table}

%-------------------------------
% Effect of strain 
% FiG s11
%-----------------------
\begin{figure}[h!]
    \centering
    \subfloat[\label{fig_}]{%
    \includegraphics[width=.8\columnwidth] 
      {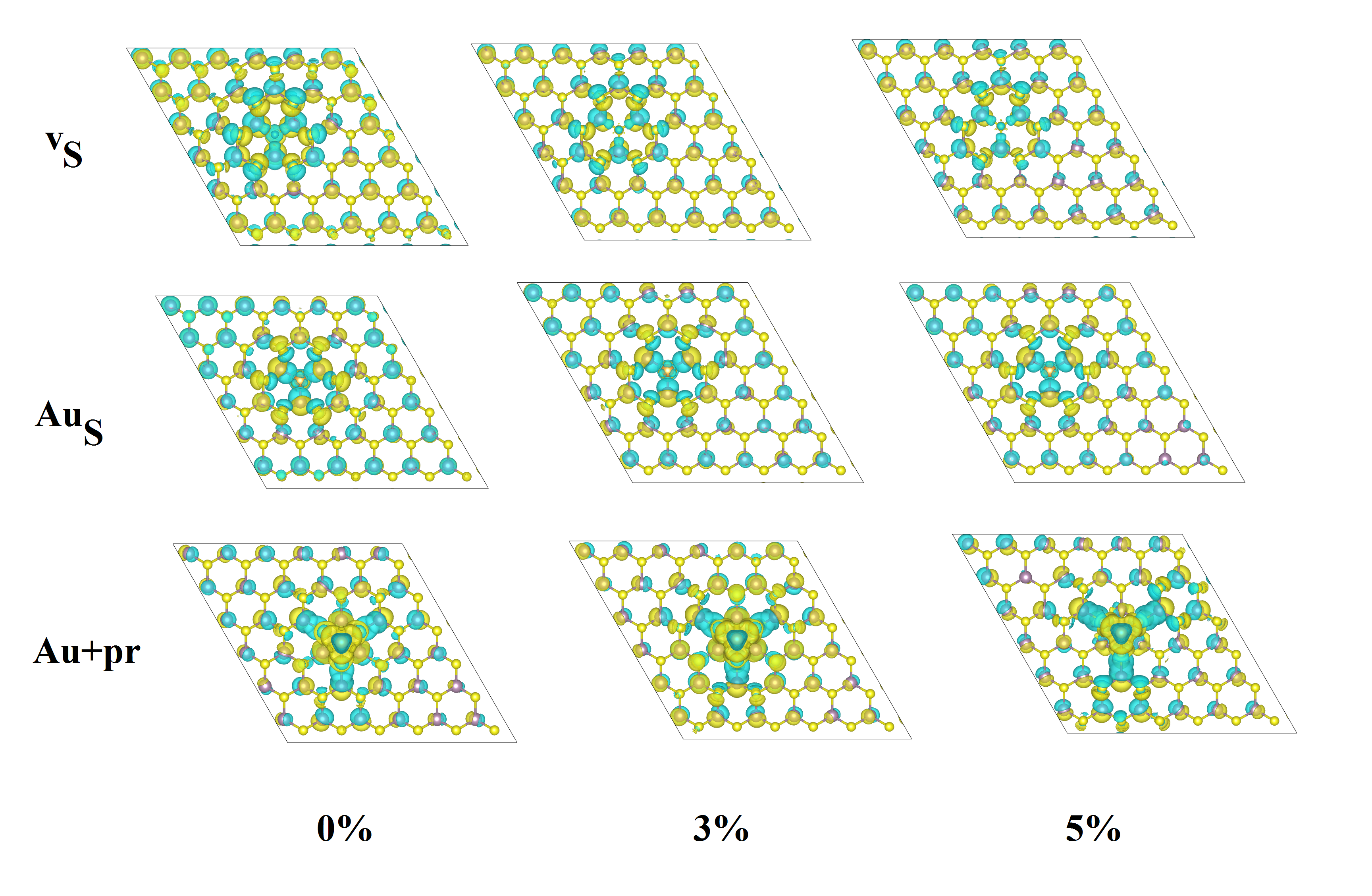}
 }     
   
      \subfloat[\label{fig_}]{%
    \includegraphics[width=1.0\columnwidth] 
      {"./fig_draft4/STAL_charge_diff_2".png}
  
} 
     \caption{(a) Charge difference profiles for free standing MoS$_2$ structures. Green represents charge depletion, yellow represents charge accumulation. (a) top view and (b) side view . Blue represents charge accumulation, red represents charge depletion. } 
     \label{fig_}
\end{figure} 
%===========================================================

%============================================================
% STAL : Atomic corrdinates + strain 

% Atomic coordinates and d-orbital charges of Au+pr
     % Au + pr 

%=====================================================================

%=====================================================

\begin{table}
    \centering
    \begin{tabular}{| c | c | c | c | c | c |}
    \hline
    $ \rm strain $  &   $ \rm  d_S-d_{Mo}$     & $ \rm  d_S-d_{Mo}$    &  $ \rm d_{Au}-d_{Mo}$    &    $ d_{z^2}$ &  $ d_{x^2-y^2}$   \\
    $\%$& $\AA$ & $\AA$& $\AA$& $\rm q/e$&$\rm q/e$  \\
    \hline
   -1 &   -1.5968  &  1.5963 &   3.7014 &  -0.0009 &   0.0007 \\ 
    0 &   -1.5860  &  1.5860 &   3.6703 &        0 &        0 \\
    1 &  -1.5763   & 1.5747  &  3.6376  &  0.0010  & -0.0012 \\
    2 &  -1.5651   & 1.5664  &  3.6132  &  0.0019  & -0.0021 \\
    3 &  -1.5558   & 1.5563  &  3.5786  &  0.0029  & -0.0031 \\
    4 &  -1.5471   & 1.5465  &  3.5453  &  0.0039  & -0.0048 \\
    5 &  -1.5395   & 1.5369  &  3.5063  &  0.0048  & -0.0060 \\
    6 &  -1.5300   & 1.5291  &  3.4595  &  0.0062  & -0.0071 \\
    7 &  -1.5201   & 1.5217  &  3.4070  &  0.0078  & -0.0081 \\
    \hline
    \end{tabular}
    \caption{ (columns 2--4) Effect of biaxial strain on the interlayer distance of the plane of S and Au to that of Mo atoms of the freestanding (6x6)-$\rm Au+pr$ structure consisting of an Au adatom on a pristine monolayer of MoS$_2$; (columns 5--6) the corresponding change in the charge population of the $ d_{3z^2-r^2}$ and $ d_{x^2-y^2}$ orbitals of the Au-adatom of this structure with respect to those of $0\%$ strain. }
    \label{table_stal_strain_charge_Au}
\end{table}

%=====================================================================
\begin{table}
    \centering
    \begin{tabular}{| c | c | c | c | c | c |}
    \hline
    $ \rm strain $  &   $ \rm  d_S-d_{Mo}$     & $ \rm  d_S-d{Mo}$    &  $ \rm d_{Au}-d{Mo}$    &    $d_{3z^2-r^2}$ &  $d_{x^2-y^2}$   \\
    $\%$& $\AA$ & $\AA$& $\AA$& $\rm q/e$&$\rm q/e$  \\
    \hline
   -1.0000 &  -1.5957 &   1.5962 &   2.0291  & -0.0018 &   0.0001  \\
         0 &  -1.5865 &   1.5839 &   2.0099  &       0 &        0  \\
    1.0000 &  -1.5758 &   1.5745 &   1.9856  &  0.0014 &   0.0002  \\
    2.0000 &  -1.5666 &   1.5640 &   1.9650  &  0.0025 &  -0.0006  \\
    3.0000 &  -1.5556 &   1.5559 &   1.9479  &  0.0038 &  -0.0013  \\
    4.0000 &  -1.5463 &   1.5470 &   1.9206  &  0.0053 &  -0.0023  \\
    5.0000 &  -1.5369 &   1.5385 &   1.9047  &  0.0070 &  -0.0032  \\
    6.0000 &  -1.5288 &   1.5297 &   1.8821  &  0.0089 &  -0.0041  \\
    7.0000 &  -1.5197 &   1.5223 &   1.8631  &  0.0109 &  -0.0049  \\
    \hline
    \end{tabular}
       \caption{(columns 2--4) Effect of biaxial strain on the interlayer distance of the plane of S and Au to that of Mo atoms of the freestanding (6x6)-$\rm Au_S$ structure consisting of an Au adatom adsorbed on a sulfur vacancy site; (columns 5--6) the corresponding change in the charge population of the $d_{3z^2-r^2}$ and $d_{x^2-y^2}$ orbitals of the Au-adatom of this structure with respect to those of $0\%$ strain. }
    \label{table_stal_strain_charge_AuS}
\end{table}

\begin{table}[!htb]
     \centering
     \begin{tabular}{c   |     c         c         c       c}
     \hline
    $\rm \epsilon \%$   & pr              &  $\rm V_S$            & $\rm Au+pr$         & $\rm Au_S$         \\          
    \hline
   -1  &    -          &   18.46     &  35.23      & 35.42  \\
    0  &    -          &   18.90     &  35.19      & 35.93  \\
    1  &  13.41        &   20.46     &  35.18      & 36.44  \\
    2  & 17.39        &   23.33     & 34.84      & 36.87  \\
    3  &  21.13       &   25.80     & 34.79      & 37.48  \\
    4  &   24.70       &   28.21     & 34.78      & 37.99  \\
    5  &   28.26       &   30.53     & 37.68      & 38.53 \\
    6  &   31.17       &   32.72     & 40.15      & 38.03  \\
    7  &   34.12      &   35.14     & 41.34      & 39.57 \\
    \hline
     \end{tabular}
     \caption{Calculated normalized conductivity ($\rm \log_{10}(\sigma/\tau)$) of freestanding MoS$_2$ under biaxial strain at the computed Fermi-level, ($\pm \epsilon$ and 0 correspond to tensile/compressive and no strain, respectively) calculated using constant time relaxation approximation and the Boltzmann transport equation.}
     \label{STAL_strain_conductivity}
 \end{table}

\FloatBarrier

%============================================================
%============================================================
%============================================================
%                   Heterostructure 
%============================================================
%---------------------------------%---------------------------------
%Mo   Au_ads    S_top   S_bot  (111)Au_surf   
%0    -2.04     1.53    -1.45   -4.10
%0    -2.07     1.58    -1.45   -4.10
%0     1.86     1.49    -1.54   -4.72
%0    -2.04     1.54    -1.51   -4.36
%0     1.82     1.52    -1.40   -4.37
%0    -2.02     1.54    -1.46   -4.26

\newpage
\section{Effect of strain on heterostructure of MoS$_2$/(111)-Au}

% FiG , subs , coordinate, charge transfer 
%-----------------------  , Fig 7 
\begin{figure}[h!]
    \centering
    \subfloat[\label{fig_subs_all_a}]{%
    \includegraphics[width=.99\columnwidth] 
      {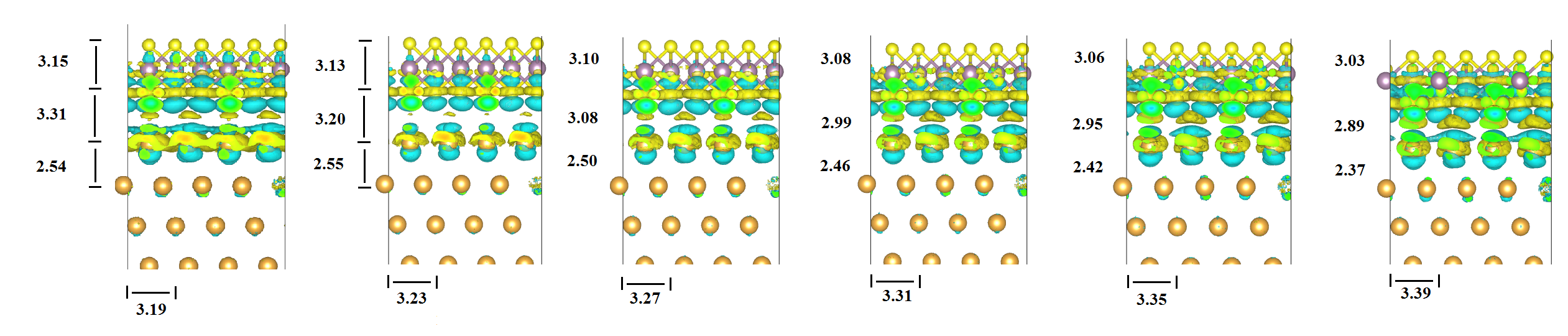}
    }  
 
\subfloat[\label{fig_subs_all_b}]{%
    \includegraphics[width=.99\columnwidth] 
      {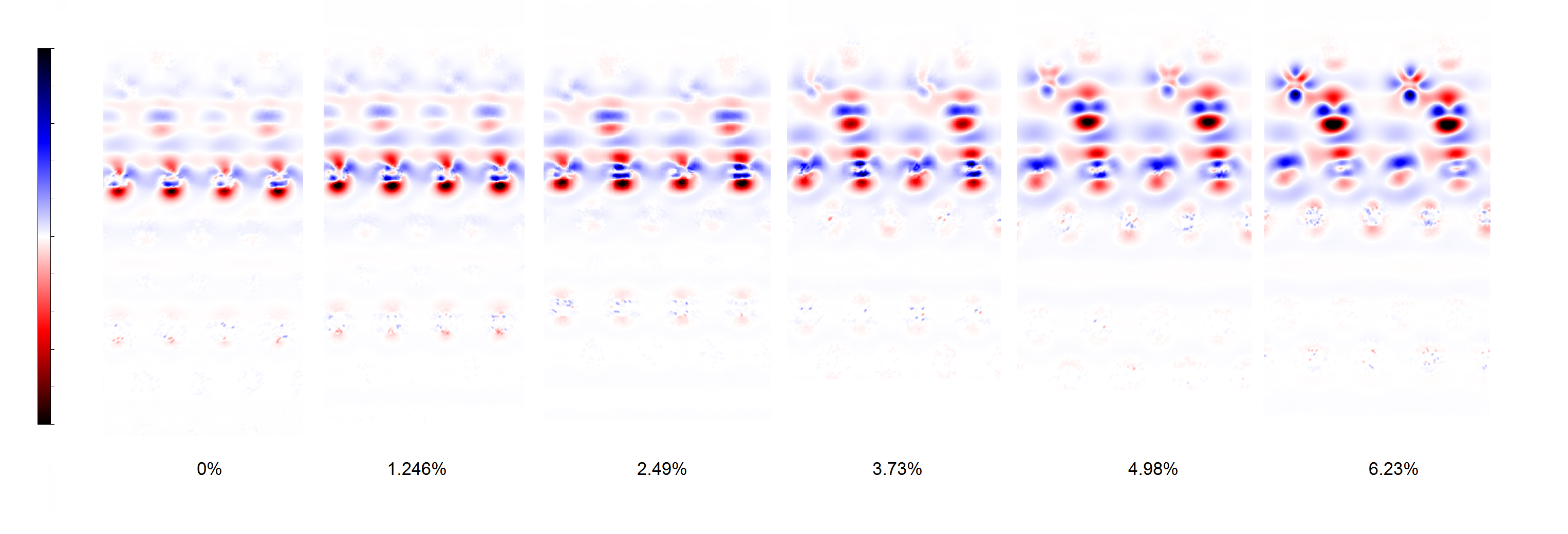}
    }
 
    \subfloat[\label{fig_subs_all_c}]{%
    \includegraphics[width=.4\columnwidth] 
      {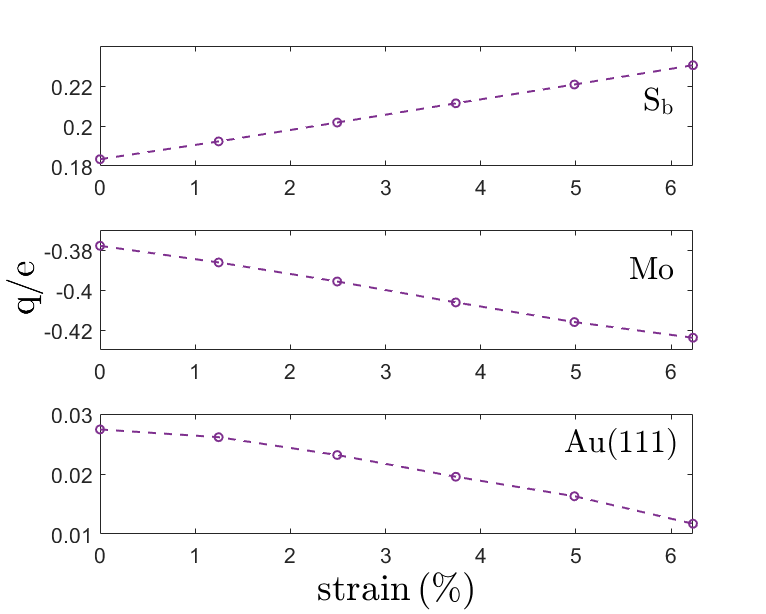}
}
\subfloat[\label{SI_fig_subs_all_d}]{%
    \includegraphics[width=.6\columnwidth] 
      {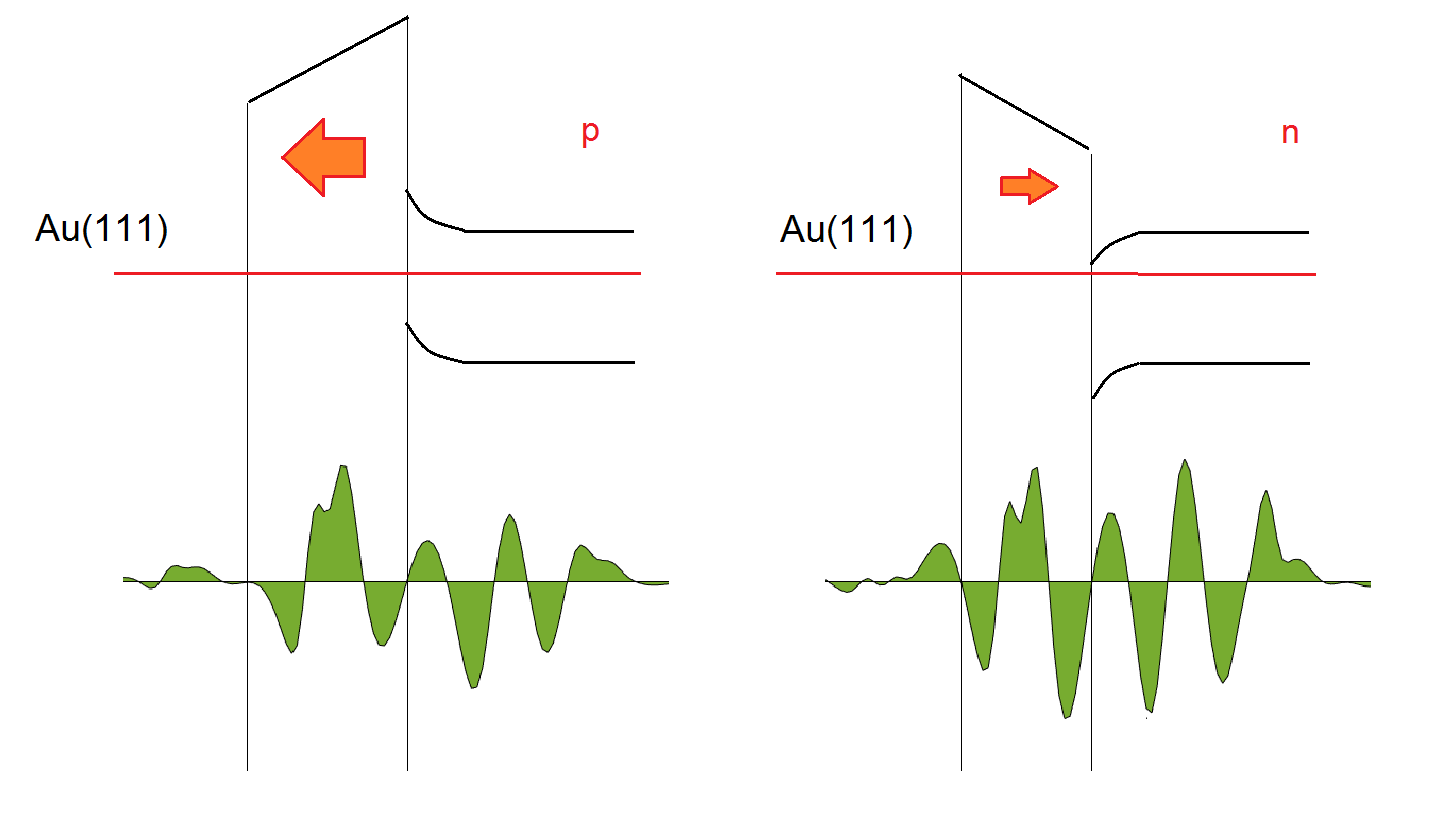}
    }

\caption{(a) Charge difference plot and interlayer distances due to biaxial strain caused by lattice mismatch of the top $\rm MoS_2$ layer with the (111)-Au substrate. The charge difference is obtained by $\rm \rho_{heterostructure}-\rm \rho_{Au(111)}-\rm \rho_{MoS_2}$. Biaxial tensile strain increases from most left structure with top layer at 0$\%$ strain (supercell consists of lattice coordinates of stand-alone $\rm MoS_2$), to the most right structure with top layer at 6.23$\%$ strain (lattice coordinates of (111)-Au). (b) Changes of the lattice vector constants and interlayer distances with increasing strain. (c) Mulliken charge analysis for the interfacial atoms forming the Mo-S-Au bonds. (d)Schematic (dimensions not to scale) demonstrates the effect of charge accumulation/depletion on the direction of interfactial dipole moment and corresponding changes of barrier height, and local band bending in the semiconductor. Variation of the interlayer distance and strain across the heterostructure cause local regions with different charge doping.  
}
     \label{SI_fig_subs_all}
\end{figure}

\begin{table}[h!]
\begin{center}
\begin{tabular}{| c| c | c |c |c |c |}
  \hline
  structure & Mo  & $\rm Au_{ads}$  &  $\rm  S_{top}$  & $\rm S_{bot}$ & $\rm (111)Au_{surf}$    \\ 
  \hline
  $\rm v_{Au}^{Mo} + Au_{2S}^{bot}$ & 0   & -2.04   &  1.53   & -1.45 &  -4.10\\ 
    \hline
  $\rm v_{Au}^{S}  + Au_{2S}^{bot}$ &0   & -2.07   &  1.58   & -1.45 &  -4.1 \\ 
    \hline
  $\rm Au_{2S}^{top}$ & 0   &  1.86   &  1.49   & -1.54 &  -4.72\\ 
    \hline
  $\rm Au_{2S}^{bot}$ &0   & -2.04   &  1.54   & -1.51 &  -4.36 \\ 
    \hline
  $\rm Au_{S}^{top}$ &0   &  1.82   &  1.52   & -1.40 &  -4.37 \\  
    \hline
  $\rm Au_{S}^{bot}$ & 0   & -2.02   &  1.54   & -1.46 &  -4.26 \\ 
      \hline
\end{tabular}

 \caption{Atomic coordinates of Au+vacancy defect complexes in the $\rm MoS_2$/(111)-Au heterostructure with lattice coordinates of the substrate (6.23\% biaxial strain). Coordinates are taken relative to the plane of Mo-atoms. In all cases, $Au^{top}$ is located approximately 0.36 \AA above the plane of $\rm S^{top}$ atoms,  $\rm Au^{bot}$ is located approximately 0.5 below the $\rm S^{bot}$ atoms; the interlayer distance between $\rm MoS_2$ and (111)-Au is 2.9 \AA.}
\label{table_hetero_coordinates}
\end{center} 
\end{table}  

%-------------------------------------------------------------------
\begin{table}
    \centering
    \begin{tabular}{| c | c  c | c  c| c  c|| c  c|}
      \hline
       structure  & 0 $\%$  &    & 1.24 $\%$  &    & 6.23 $\%$  &   &  6.23 $\%$ , 2 elect. & \\
      \hline   
       $\rm v_{S}$                        & 3.95 &       & 5.43    &    & 3.86 &        &  3.34&   \\
       $\rm Au_{S}^{bot}$                 &      & -0.71 &     &  -2.14  &      & -0.77    &  & \\
       $\rm Au_{S}^{top}$                 &      &       &     &    &      & 0.081    &  & \\
       $\rm Au_{S}^{bot}+v_{Au}^{S}$     &      &       &     &  -0.77  &      & -0.50    &  & \\
       $\rm Au_{S}^{bot}+v_{Au}^{Mo}$    &      & 0.58 &     &    &      & -0.48     &  &\\
      \hline  %-------------------------------------------------------------------------------------
       $\rm v_{2S}$                       & 8.38 &       &  8.87   &    & 8.03   &          & 7.01 & \\
       $\rm Au_{2S}^{bot}$                &      & -1.21 &     & 0.092   &        &-1.10     &  &\\
       $\rm Au_{2S}^{top}$                &      &       &     &    &        & -0.065    &  & \\
       $\rm Au_{2S}^{bot}+v_{Au}^{S}$    &      &       &     &    &        & -0.662    &  & \\
       $\rm Au_{2S}^{bot}+v_{Au}^{Mo}$   &      & -0.41 &     &    &        & -0.756    &  & \\
       $\rm Au_{2S}^{top}+v_{Au}^{Mo}$   &      &       &     &    &        & 0.252    &  & \\
       \hline  
    \end{tabular}
    \caption{ Calculated formation energies of vacancies (left side of each column) and adsorption energy of Au+vacancy defect complexes in the $\rm MoS_2$/(111)-Au heterostructure under 0 $\%$, 1.24$\%$, 6.23$\%$ biaxial strain. The last column corresponds to the Au(111)/MoS$_2$/Au(111) structure under  6.23 $\%$ tensile strain. The notation $\rm V^{X}_{Au}$ corresponds to a Au vacancy on the surface of the electrode under atom X=$\rm {Mo, S}$ of the MoS$_2$. }
    \label{table_hetero_energies}
\end{table}

% FiG , Hetero, DOS , all S , all Mo  
%-----------------------
% fig s13 
\begin{figure}[h!]
    \centering
    \subfloat[\label{fig_het_dos_all_S}]{%
    \includegraphics[width=.8\columnwidth] 
      {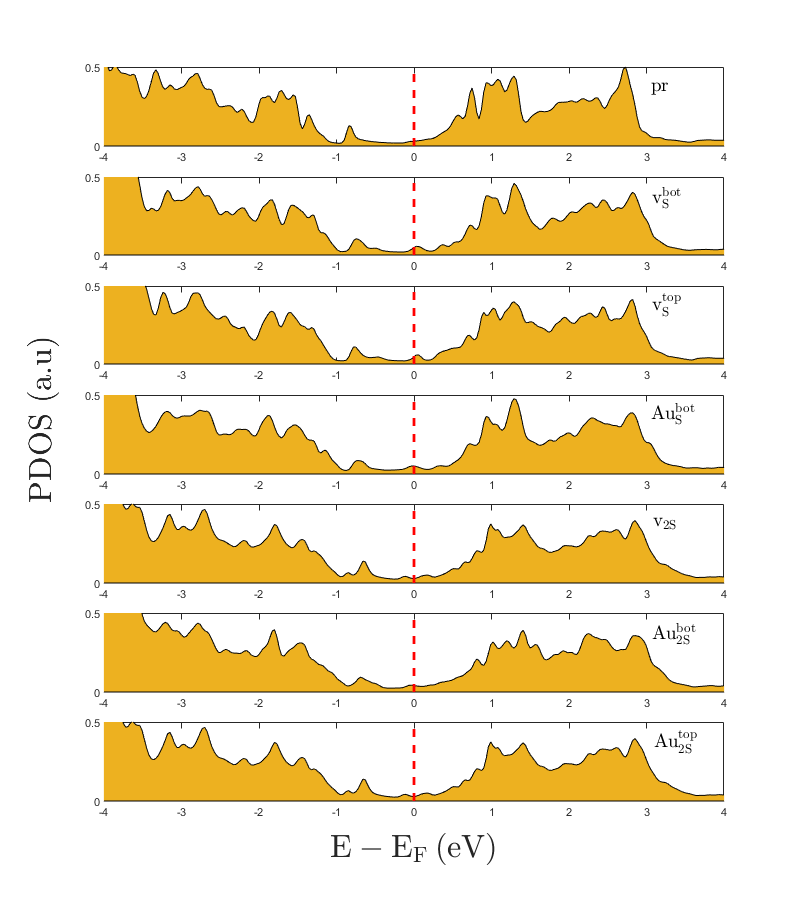}
} 
     \caption{ Electronic structure (PDOS) of bottom S atom in various Au+defect structures in the $\rm MoS_2$/(111)-Au heterostructure complex.  }
     \label{fig_het_dos_all_S}
\end{figure} 

%========================
% FiG s13 
%-----------------------
\begin{figure}[h!]
    \centering
    \subfloat[\label{fig_subs_dos_S}]{%
    \includegraphics[width=.55\columnwidth] 
      {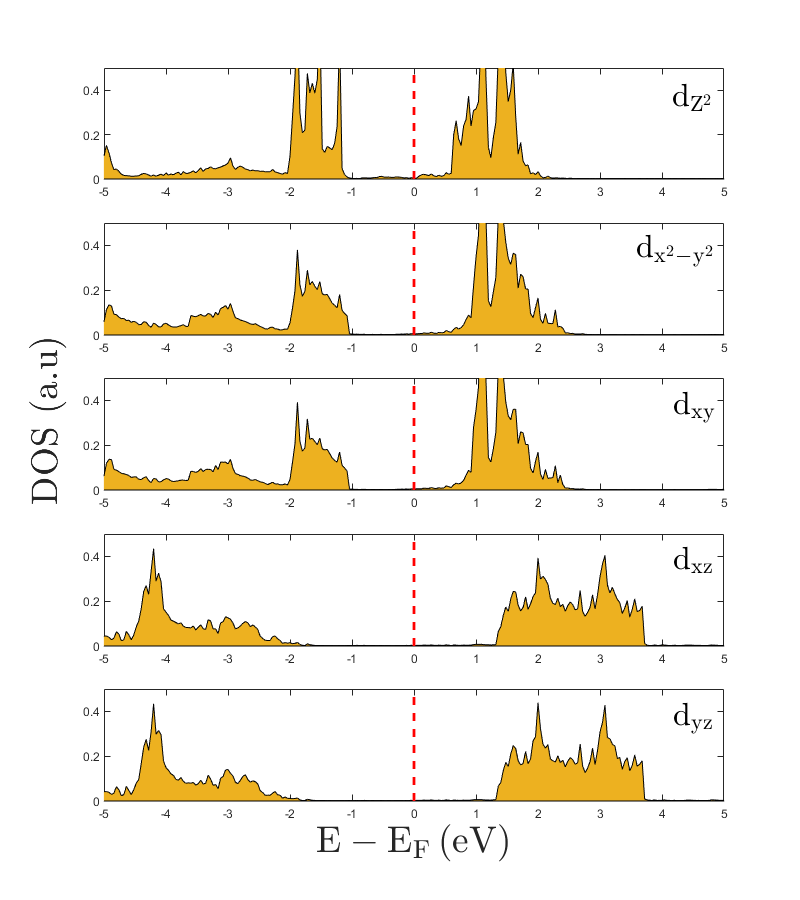}
} 

\subfloat[\label{fig_subs_dos_S}]{%
    \includegraphics[width=.55\columnwidth] 
      {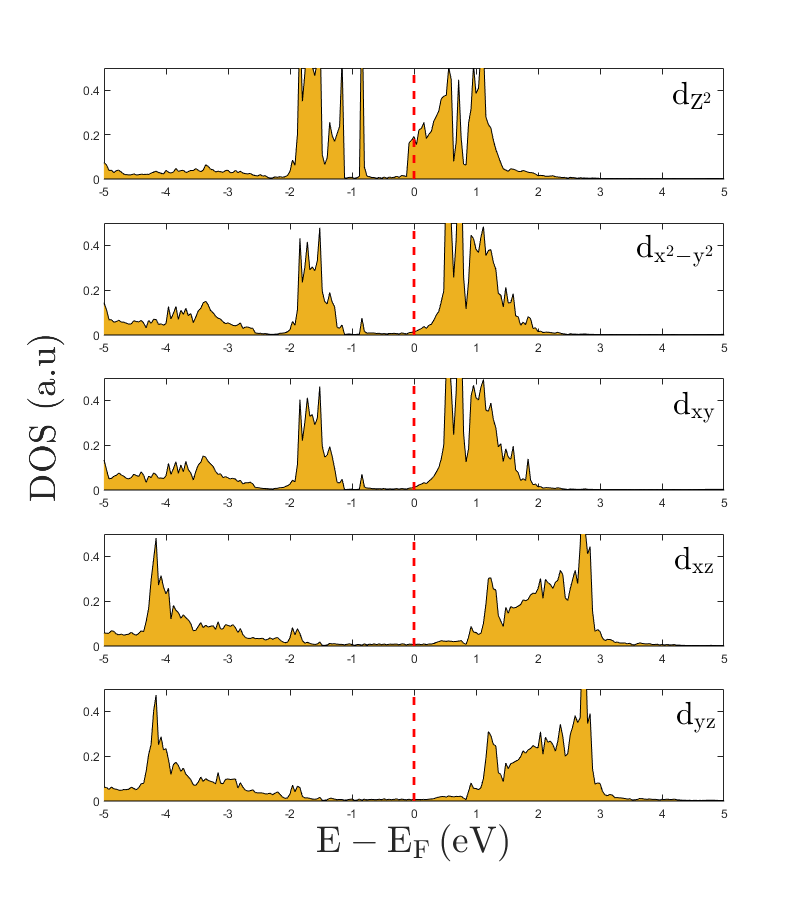}
} 

     \caption{Changes of d-orbitals of Mo in the $\rm MoS_2$/(111)-Au heterostructure under (a) 0$\%$ and (b) {6.23$\%$} biaxial tensile strain. As discussed in the main text, the $d$-bands of the Mo shift downward with increasing strain. The band edges and gap energy levels are dominated by $d_{z^2}$  and $d_{x^2-y^2}$  orbitals. }
     \label{fig_subs_dos_S}
\end{figure} 
% ===========================

% ===========================
% FiG , subs, dos , d Mo 
%-----------------------
\begin{figure}[h!]
    \centering
    \subfloat[\label{fig_subs_dos_a}]{%
    \includegraphics[width=.5\columnwidth] 
      {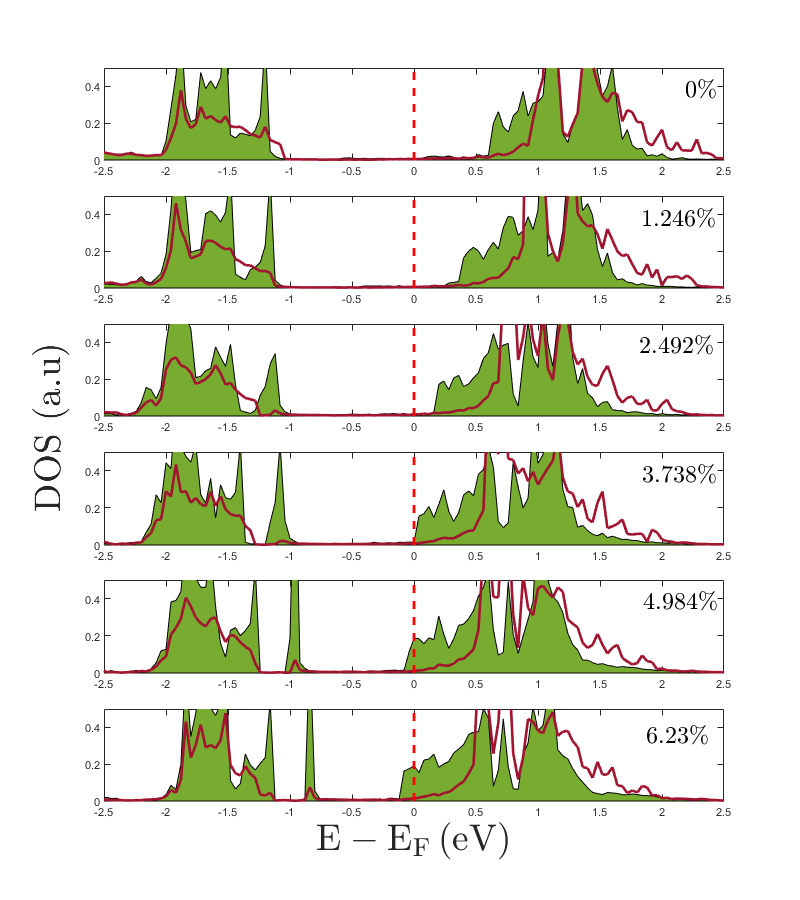}
} 
    \subfloat[\label{fig_subs_dos_b}]{%
    \includegraphics[width=.5\columnwidth] 
      {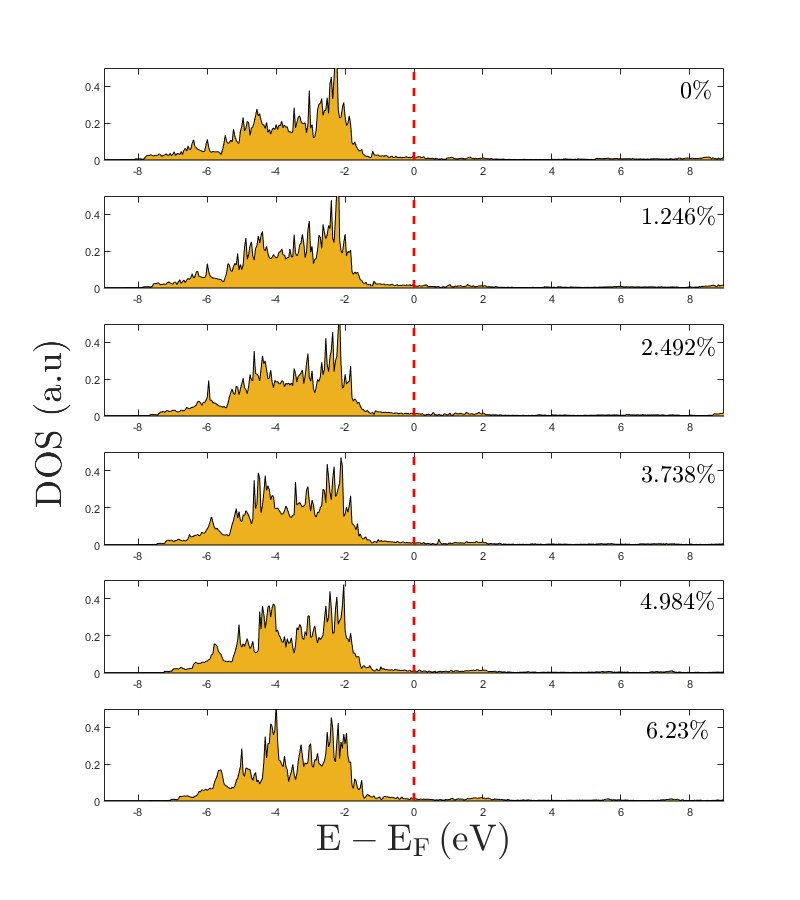}
} 
\hfill
    \subfloat[\label{fig_subs_dos_c}]{%
    \includegraphics[width=.5\columnwidth] 
      {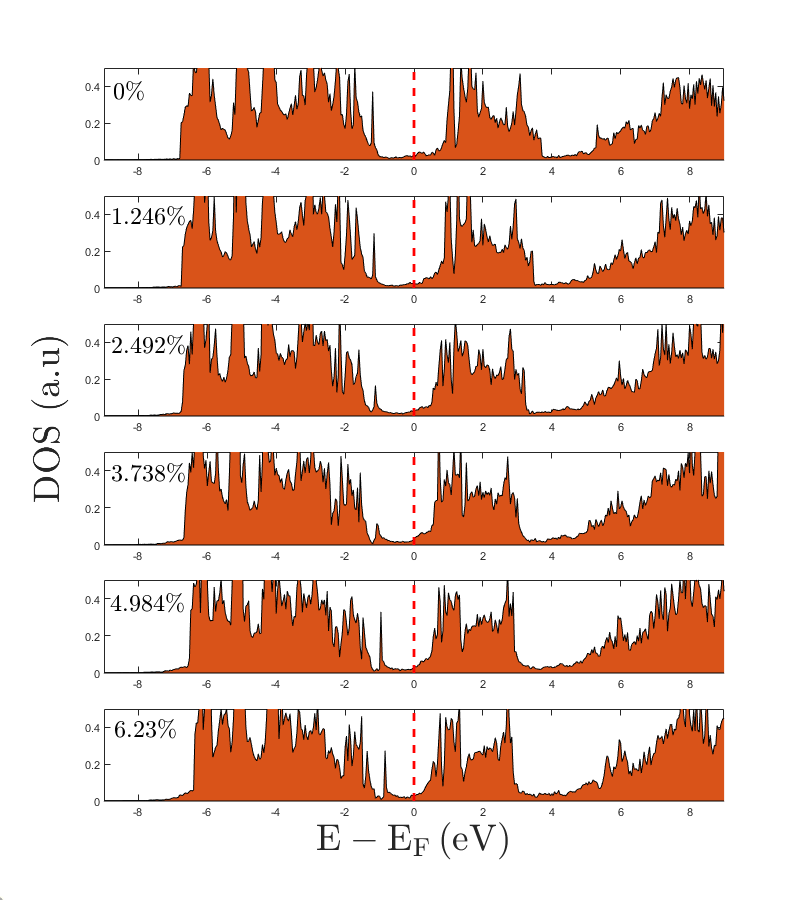}
}%

     \caption{  Combined effect of heterostructure formation and strain on the $d_{z^2}$ (green) $d_{x^2-y^2}$  (red) orbital of Mo (a), $d_{z^2}$-band of interfacial Au or the electrode (b) and all bands of interfacial S (c) in $\rm MoS_2$/(111)-Au heterostructure. Legend corresponds to the biaxial tensile strain in $MoS_2$, and the Fermi-level is indicated by red dashed line. The center of the d-band of Mo  significantly downshifts with increasing strain. On the contrary, the width and location of peaks of bands of interfacial Au and S atoms are not affected as significantly. }
     \label{fig_subs_dos_}
\end{figure}

Here, we describe the change in the electronic structure of the Mo$_2$/(1110)-Au heterostructure with strain.
Similar to freestanding MoS$_2$, the VBM and CBM contain by Mo:${d_{x^2-y^2}}$ and Mo:${d_{z^2}}$ character, respectively, under no strain.
The peak of the Mo:${d_{x^2-y^2}}$ states that is located 200 meV below VBM eventually shifts up and splits from the VBM in biaxially strained structures. 
In the $\rm V_S$ structure, the defect level induced by $\rm V_S$ is located approximately 0.4 eV below the CBM. 
This
defect level mainly consists of S orbitals marginally mixing with Mo orbitals.
The peak of this defect level gets slightly closer to the Fermi level by 40 meV as the $\rm Au_S$ structure forms and  
mainly consists of S and Au with a marginally contribution from Mo. %The peak is located approximately 360 meV below CBM.
The general trends are the same for $\rm Au_S$+$\rm V_{2S}$ structure. 
%The defect level (peak) in $\rm V_{2S}$ structure is located approximately 0.2 eV above the Fermi level. The band gap of this structure is approximately 1.4 eV. The defect level mainly consists of S hybridizing with Au(111), with a marginal contribution from Mo orbitals. 
As the $\rm Au_{2S}$ defect complex forms, multiple peaks 
arise in the band gap, located at -0.56, -0.12, 0.12 and 0.68 eV, relative to the Fermi level. 
%The defect levels located at 0.12 and 0.68 eV have a significant contribution from S and Mo orbitals, while those located at lower energy levels are approximately metallic.

% FiG 
%-----------------------
\begin{figure}[h!]
       \subfloat[\label{fig_pdos_het_0_defect_a}]{%
    \includegraphics[width=.5\columnwidth] 
      {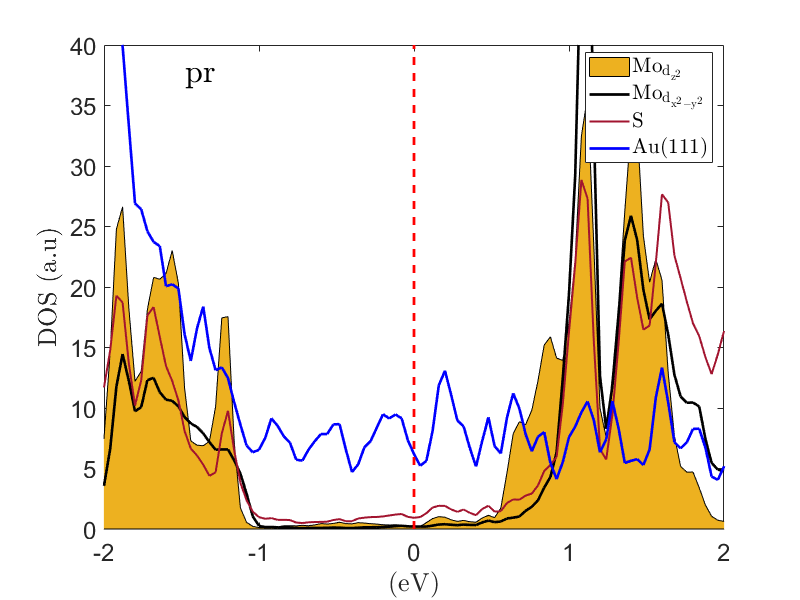}
     }
     \hfill
     \subfloat[\label{fig_pdos_het_0_defect_b}]{%
    \includegraphics[width=.5\columnwidth] 
    {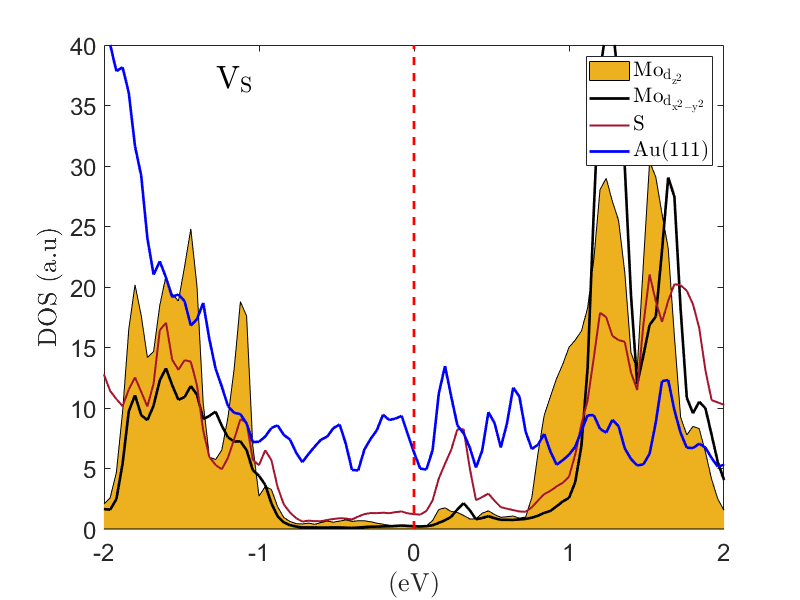}
     }
     \subfloat[\label{fig_pdos_het_0_defect_c}]{%
    \includegraphics[width=.5\columnwidth] 
      {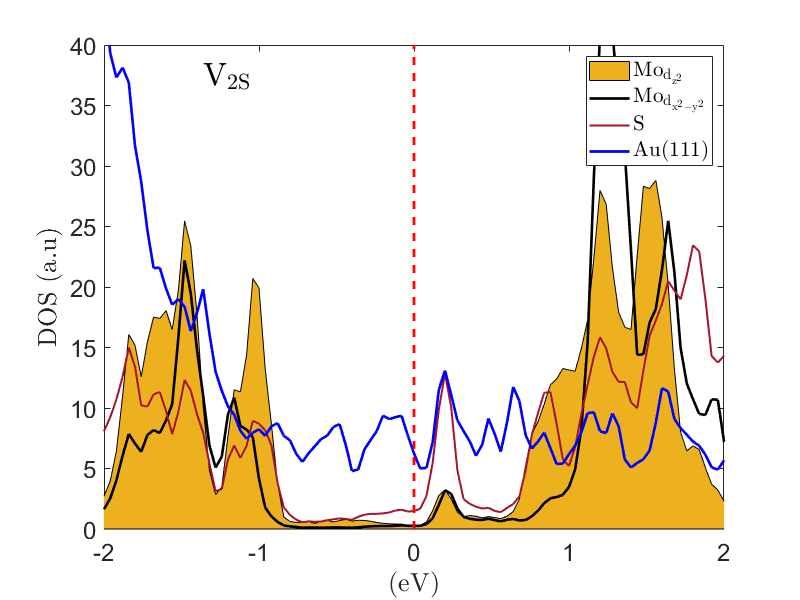}
     }
      \hfill
      \subfloat[\label{fig_pdos_het_0_defect_d}]{%
    \includegraphics[width=.5\columnwidth] 
      {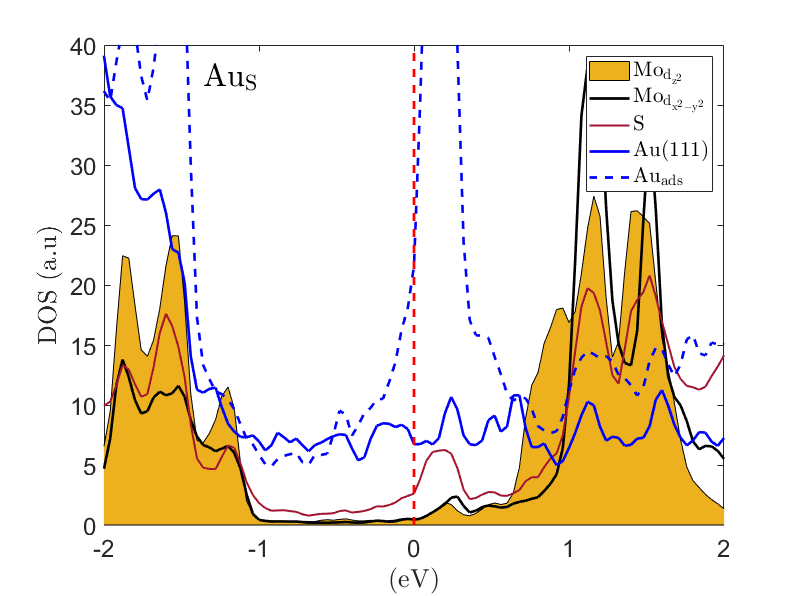}
     }
     \subfloat[\label{fig_pdos_het_0_defect_e}]{%
    \includegraphics[width=.5\columnwidth] 
      {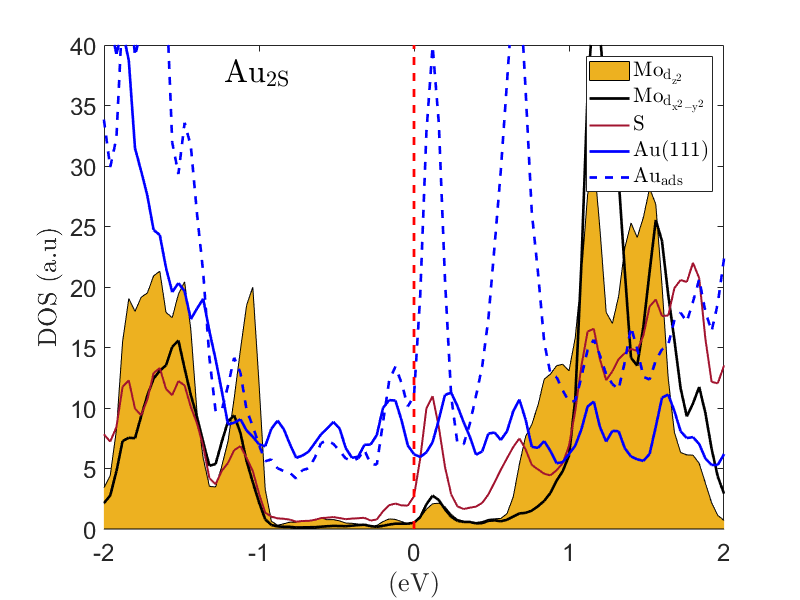}
     }
     \caption{ PDOS of defect complexes in MoS$_2$/Au(111) heterostructure under 0 $\%$ biaxial strain. PDOS of Mo, S and $\rm Au_{ads}$ are multiplied by 50, 50 and 100, respectively for ease of eye. }
     \label{fig_pdos_het_0_defect_}
\end{figure}

% FiG 
%-----------------------
\begin{figure}[h!]
       \subfloat[\label{fig_pdos_het_1p2_defect_a}]{%
    \includegraphics[width=.5\columnwidth] 
      {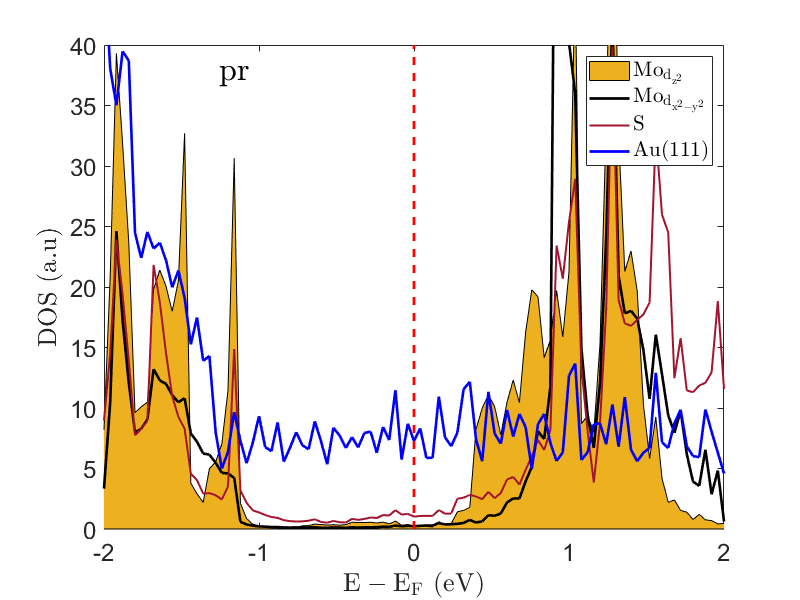}
     }
     \hfill
     \subfloat[\label{fig_pdos_het_1p2_defect_b}]{%
    \includegraphics[width=.5\columnwidth] 
    {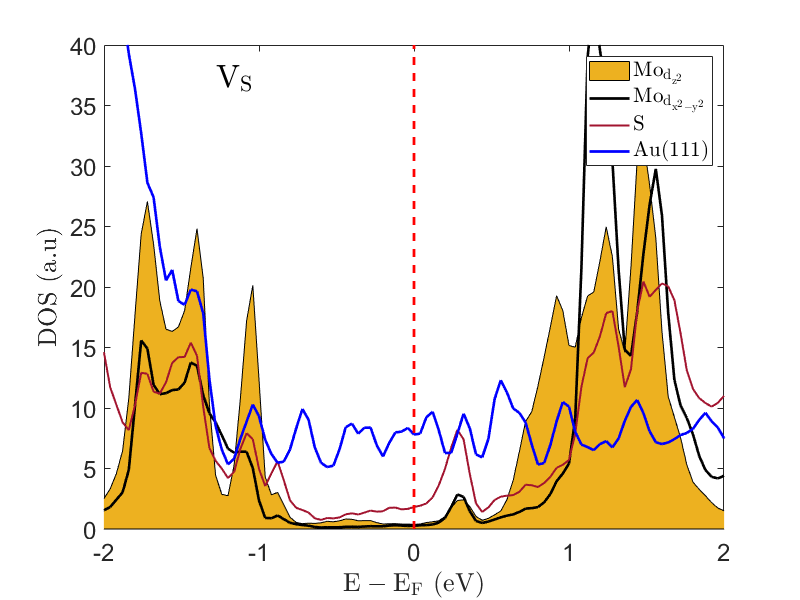}
     }
     \subfloat[\label{fig_pdos_het_1p2_defect_c}]{%
    \includegraphics[width=.5\columnwidth] 
      {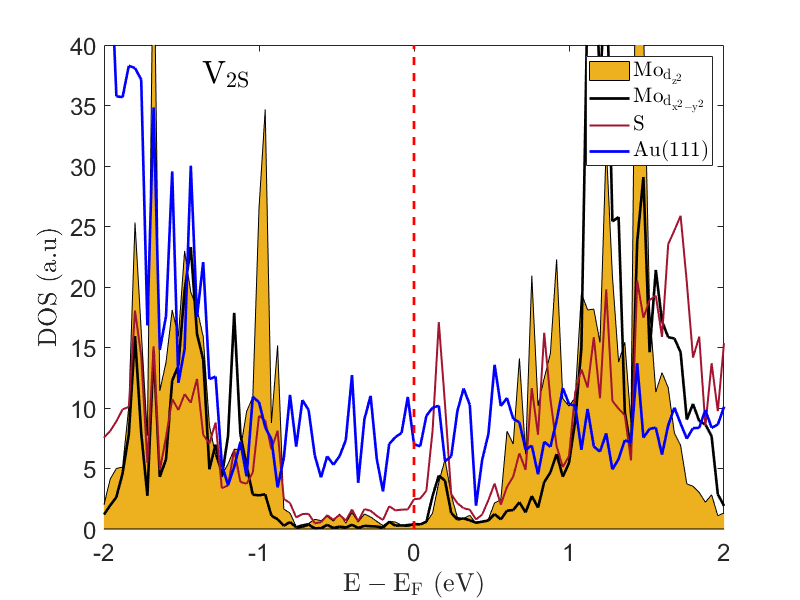}
     }
      \hfill
      \subfloat[\label{fig_pdos_het_1p2_defect_d}]{%
    \includegraphics[width=.5\columnwidth] 
      {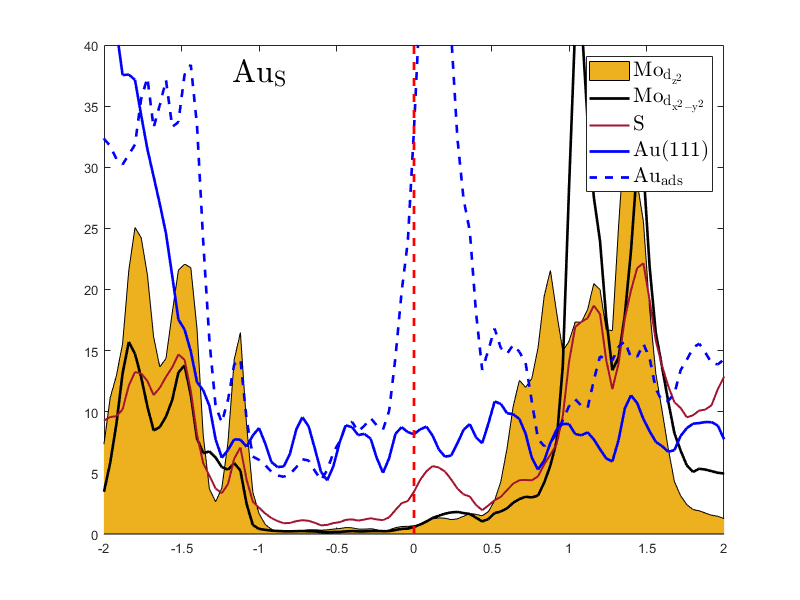}
     }
     \subfloat[\label{fig_pdos_het_1p2_defect_e}]{%
    \includegraphics[width=.5\columnwidth] 
      {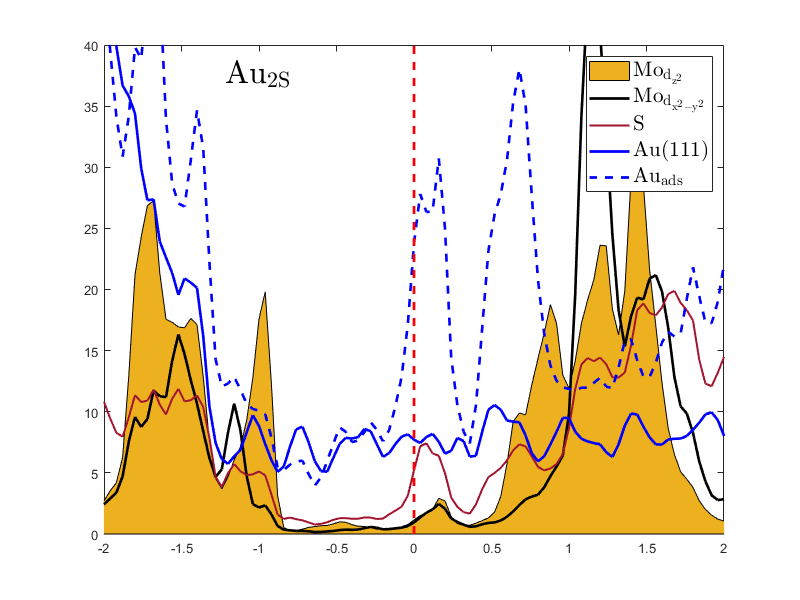}
     }
     \caption{ PDOS of defect complexes in MoS$_2$/Au(111) heterostructure under 1.24$\%$ biaxial strain. PDOS of Mo, S and $\rm Au_{ads}$ are multiplied by 50, 50 and 100, respectively for ease of eye. }
     \label{fig_pdos_het_1p2_defect_}
\end{figure}

% FiG 
%-----------------------
\begin{figure}[h!]
       \subfloat[\label{fig_pdos_het_5_defect_a}]{%
    \includegraphics[width=.5\columnwidth] 
      {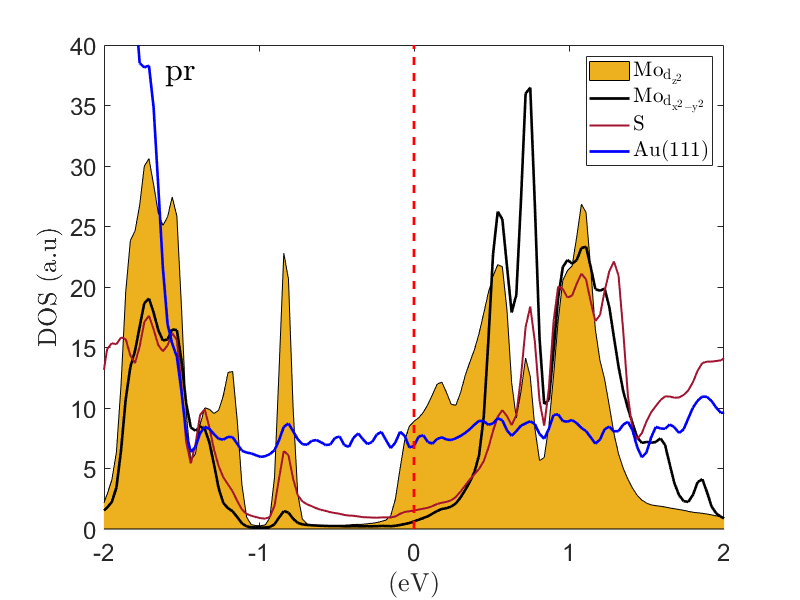}
     }
     \hfill
     \subfloat[\label{fig_pdos_het_5_defect_b}]{%
    \includegraphics[width=.5\columnwidth] 
    {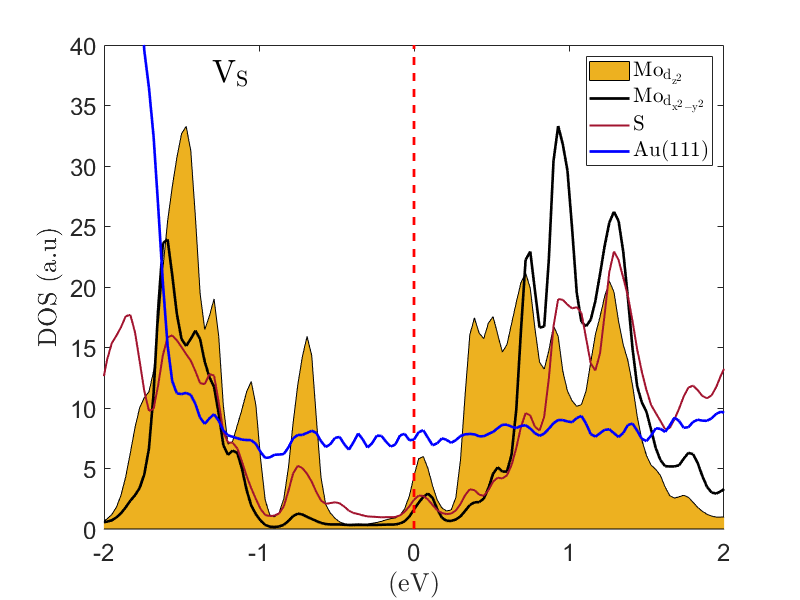}
     }
     \subfloat[\label{fig_pdos_het_5_defect_c}]{%
    \includegraphics[width=.5\columnwidth] 
      {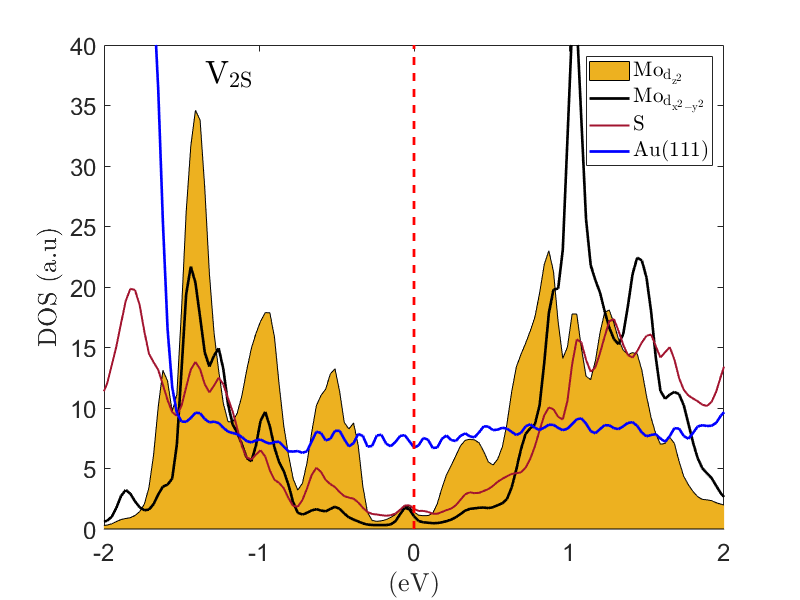}
     }
      \hfill
      \subfloat[\label{fig_pdos_het_5_defect_d}]{%
    \includegraphics[width=.5\columnwidth] 
      {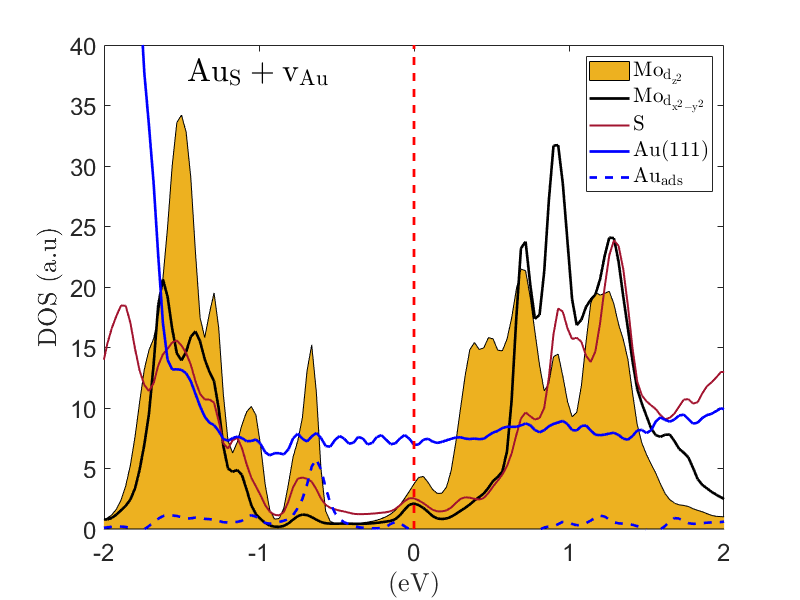}
     }
     \subfloat[\label{fig_pdos_het_5_defect_e}]{%
    \includegraphics[width=.5\columnwidth] 
      {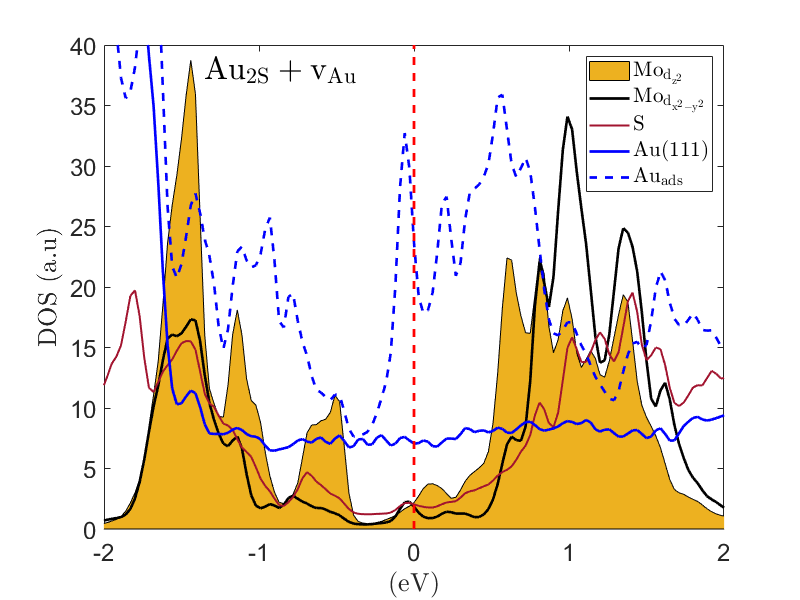}
     }
     \caption{ PDOS of defect complexes in MoS$_2$/Au(111) heterostructure under 6.23 $\%$ biaxial strain. PDOS of Mo, S and $\rm Au_{ads}$ are multiplied by 50, 50 and 100, respectively for ease of eye. }
     \label{fig_pdos_het_5_defect_}
\end{figure}

%===========================================
\begin{figure}[h!]
       \subfloat[\label{subs_shift_charge_5p8}]{%
    \includegraphics[width=.5\columnwidth] 
      {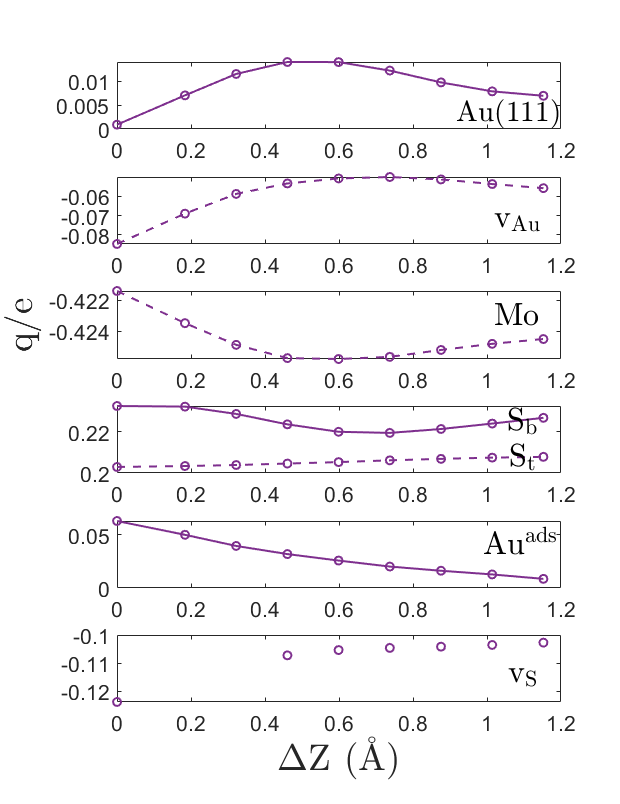}
     }
     \hfill
  
     \caption{  Changes of the Mulliken charge of atoms with interlayer distance in the MoS$_2$/Au(111) heterostructure under 6.23$\%$ tensile strain. }
     \label{subs_shift_charge_5p8}
\end{figure}

%===========================================

\end{document}